# Superhydrides on the way to ambient pressure: weak localization and persistent X-ray photoconductivity in BaSiH$_8$

Dmitrii V. Semenok[1,†,*], Di Zhou[1,*,†], Sven Luther[2], Toni Helm[2], Hirokazu Kadobayashi[3], Yuki Nakamoto[4], Katsuya Shimizu[4], Kirill S. Pervakov[5], Andrei V. Sadakov[5], Oleg A. Sobolevskiy[5], Vladimir M. Pudalov[5,6], Simone Di Cataldo[7], Roman Lucrezi[8], Lilia Boeri[7], Michele Galasso[9], Frederico G. Alabarse[10], Ivan A. Troyan[11,12], and Viktor V. Struzhkin[12,13,*]

[1] *Center for High Pressure Science & Technology Advanced Research, Bldg. #8E, ZPark, 10 Xibeiwang East Rd, Haidian District, Beijing, 100193, China*
[2] *Hochfeld-Magnetlabor Dresden (HLD-EMFL) and Würzburg-Dresden Cluster of Excellence ctd.qmat, Helmholtz-Zentrum Dresden-Rossendorf, 01328 Dresden, Germany*
[3] *Japan Synchrotron Radiation Research Institute (JASRI), 1-1-1 Kouto, Sayo, Hyogo 679-5198, Japan*
[4] *Center for Science and Technology under Extreme Conditions, Graduate School of Engineering Science, Osaka University, 1-3 Machikaneyama, Toyonaka, Osaka 560-8531, Japan*
[5] *V. L. Ginzburg Center for High-Temperature Superconductivity and Quantum Materials, Moscow, 119991 Russia*
[6] *National Research University Higher School of Economics, Moscow 101000, Russia*
[7] *Dipartimento di Fisica, Sapienza Università di Roma, Piazzale Aldo Moro 5, 00185 Roma, Italy*
[8] *Department of Materials and Environmental Chemistry, Stockholm University, SE-10691 Stockholm, Sweden*
[9] *Department of Electrotechnology, Faculty of Electrical Engineering, Czech Technical University in Prague, Czech Republic*
[10] *Elettra - Sincrotrone Trieste S.C.p.A., Strada Statale 14 - km 163,5 in AREA Science Park, 34149 Basovizza, Trieste, Italy*
[11] *A.V. Shubnikov Institute of Crystallography of the Kurchatov Complex of Crystallography and Photonics (KKKiF), 59 Leninsky Prospekt, Moscow 119333, Russia*
[12] *Shanghai Advanced Research in Physical Sciences (SHARPS), 68 Huatuo Rd, Bldg 3, Pudong, Shanghai 201203, P.R. China*
[13] *Center for High Pressure Science and Technology Advanced Research (HPSTAR), Bldg. 6, 1690 Cailun Rd., Pudong, Shanghai 201203, P.R. China*

*Corresponding authors: Dmitrii V. Semenok (dmitrii.semenok@hpstar.ac.cn), Di Zhou (di.zhou@hpstar.ac.cn), and Viktor V. Struzhkin (viktor.struzhkin@hpstar.ac.cn).

†These authors contributed equally.

## Abstract

Reducing the stabilization pressure of superhydrides represents one of the most important challenges in hydrogen-saturated compound chemistry. Moving in this direction, we studied the Ba-Si-H system at 0 – 142 GPa using transport measurements, $^1$H nuclear magnetic resonance, single-crystal and powder X-ray diffraction in the temperature range of 4–317 K. We synthesized the previously predicted cubic BaSiH$_8$ at pressures of 18–31 GPa. Remarkably, we demonstrate that BaSiH$_8$ remains stable upon decompression to ambient conditions and can be recovered from the diamond anvil cell. Obtained Ba-Si polyhydrides exhibit metallic and superconducting properties ($T_c$ = 9 K, $B_{c2}(0) \approx$ 13-16 T) at 142 GPa. However, at pressures below 50 GPa, these hydrides behave as degenerate semiconductors (bandgap < 0.4 meV) or poor metals with weak electron localization, negative magnetoresistance, photovoltaic effect, and persistent photoconductivity in the X-ray and visible range. Our work demonstrates the high-pressure synthesis of Ba-Si polyhydrides that remain stable upon decompression to ambient conditions, overcoming a critical bottleneck in superhydride chemistry and establishing a foundation for practical applications in hydrogen storage.

## 1. Introduction

The field of superhydrides physics and chemistry began its rapid expansion approximately a decade ago from the discovery of superconductivity in H$_3$S[1]. For much of that time, the prevailing consensus was that these compounds could only exist at high pressures exceeding 100 GPa (1 Mbar)[2]. While these conditions enable remarkable high-temperature superconductivity (for instance, in LaH$_{10}$ [3], ThH$_{10}$ [4], YH$_{6-9}$ [5,6]) and exceptional hydrogen storage capacities (for example, in SrH$_9$ [7], IH$_{27}$ [8], LaH$_{11-12}$ [3], and BaH$_{12}$ [9]), recent studies on molecular polyhydrides have begun to challenge this paradigm. Polyhydrides stable at significantly lower pressures have now been identified in numerous systems, including K-H (KH$_9$ [10]), Rb-H (RbH$_9$ [11,12]), Cs-H (CsH$_{15+}$ [11], CsH$_{25}$ [13]), and others. Notably, the stabilization pressure for certain polyhydrides has dropped below 10-20 GPa [10-12]. Such compounds can be "pumped" with hydrogen at high pressure of several ten thousand atmospheres, and then remain stable for a long time after the load is removed. This reduction in pressure requirements suggests that working with diamond anvil cells (DACs) may no longer be necessary, paving the way for large-volume synthesis in hydraulic presses or using non-diamond anvils.

However, the pursuit of superconducting hydrides faces a distinct challenge. High-$T_c$ superconductivity is typically observed only above 80–100 GPa[14-17]. Furthermore, depressurizing samples below 70 GPa often results in the failure of diamond anvils, complicating the recovery of samples. Consequently, the synthesis of a high-$T_c$ superconducting hydride below 60–70 GPa remains challenging. Bridging this gap is critical for transitioning superhydrides from laboratory curiosities to industrial applications.

A promising strategy to address this is the targeted search for ternary polyhydrides that stabilize at moderate pressures, significantly below 50 GPa. Theoretical models and experiments indicate that LaBeH$_8$ [15], LuBeH$_8$ [18], and the recently proposed BaSiH$_8$ [19] and SrSiH$_8$ [19] should exhibit high critical temperatures in the pressure range of 5 to 100 GPa. Motivated by the prediction that BaSiH$_8$ can be synthesized from the barium monosilicide (BaSi) — a precursor stable at ambient conditions[20] — we experimentally investigated Ba-Si-H system.

Here, we report the successful synthesis of cubic BaSiH$_8$ at a remarkably low pressure of 18 GPa. Most notably, this compound is very stable, retaining its structure and high hydrogen content (80 mol%, 4.6 wt%) even upon full decompression to ambient pressure (0 GPa). Powder X-ray diffraction reveals a small broadening of the 111 reflection of BaSiH$_8$, indicative of a long-period structural distortion and a likely enlargement of the unit cell relative to the idealized cubic model. While the high-pressure phase exhibits superconductivity below 9 K at 142 GPa, the system undergoes a metal-to-insulator transition near 40 GPa, displaying degenerate semiconductors or poor metal behavior characterized by a small electronic gap (< 0.4 meV). This we confirm by electrical-transport and NMR measurements. Furthermore, BaSiH$_x$ samples exhibit negative magnetoresistance with signatures of localization ($k_Fl$~1), photovoltaic effect, and pronounced persistent photoconductivity in the X-ray range, that shows the potential for applications such as radiation detectors. A complete list of experiments performed is given in Supporting Table S1. Our work opens the way for the synthesis and study of metallic polyhydrides at pressures significantly below 50 GPa.

## 2. Results

*2.1 Synthesis of precursor (BaSi)*

Starting material (BaSi) for high-pressure synthesis was obtained by cold mechanochemical synthesis: grinding for 1 hour an equimolar mixture of Ba and Si in a ball mill under argon atmosphere. According to the EDS and XRD analysis (Figures 1a, b, e, and Supporting Figure S4), the Ba:Si ratio is close to 1:1 in the final product, but some variations of the elements concentration has been found. XRD shows that the sample includes two main Ba-Si phases[20], *Cmcm*-BaSi, and *P*4$_2$/*mnm*-Ba$_3$Si$_4$, as well as impurity phases: *Pnma*-Ba$_2$Si, *Fd*$\bar{3}$*m*-Si (under pressure it goes to Si-V [21], Supporting Figure S7), and *Pnma*-Ba$_2$SiO$_4$ [22](result of oxidation), found by single-crystal XRD (Supporting Figure S14, Table S10). The reaction product is a black powder that self-ignites in air and slowly oxidizes when stored in an argon glovebox.

Attempts to synthesize *cmcm*-BaSi using high-temperature approach in an inert metal or SiO$_2$ ampoules were unsuccessful due to the high vapor pressure of barium and its reactivity at high temperatures. Therefore, despite the fact that the starting material was not single-phase, but contained an average ratio of Ba and Si close to equimolar, we used a mechanochemical reaction to synthesize the precursor. Laser synthesis of hydrides in a diamond anvil cell in any case mixes the elements. The deficiency of barium is likely due to its viscosity and adhesion to the inner surface of the mill.

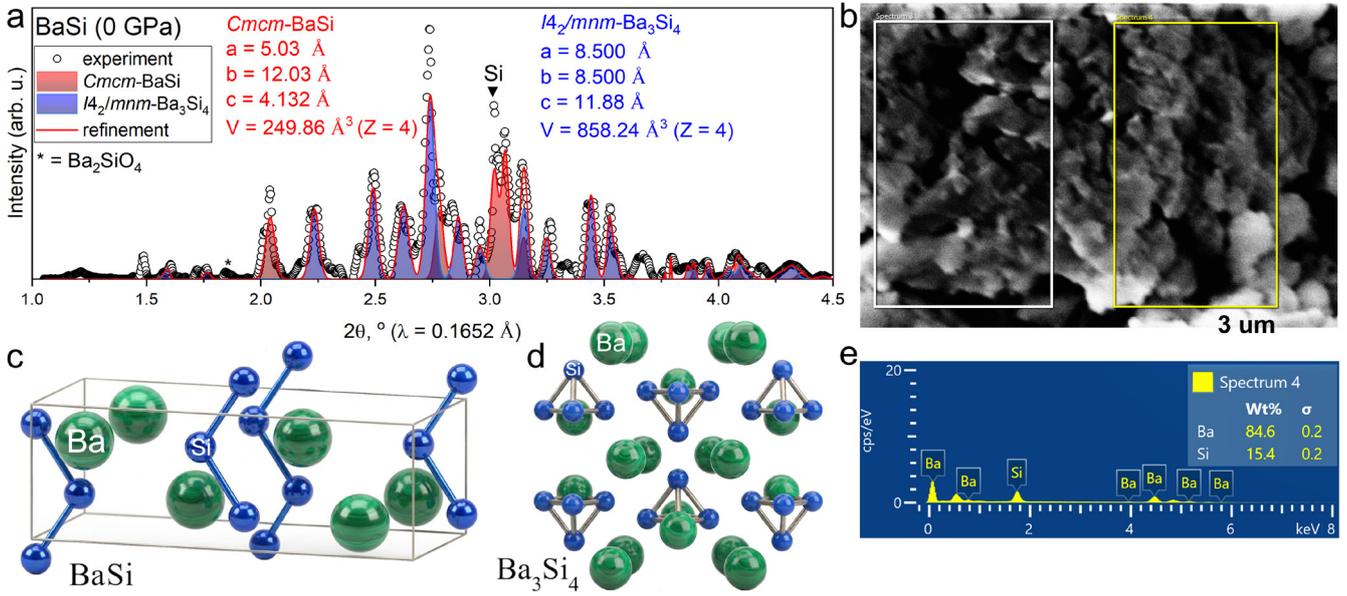

**Figure 1. Structural characterization and phase composition of BaSi at ambient pressure.** (a) X-ray diffraction pattern at 0 GPa showing coexistence of two main phases: orthorhombic *Cmcm*-BaSi (red shading, a = 5.03 Å, b = 12.03 Å, c = 4.132 Å, V = 249.86 Å$^3$, Z = 4) and tetragonal *I4$_2$/mnm*-Ba$_3$Si$_4$ (blue shading, a = b = 8.500 Å, c = 11.88 Å, V = 858.24 Å$^3$, Z = 4). Open circles represent experimental data, and solid lines show Le Bail refinement fit. Minor Ba$_2$SiO$_4$ impurity phase marked with asterisks. (b) Scanning electron microscopy image showing the surface morphology of the sample, scale bar is 3 μm. (c) Crystal structure of orthorhombic *Cmcm*-BaSi phase showing Ba atoms (large green spheres) and Si atoms (small blue spheres) with unit cell outlined. (d) Crystal structure of tetragonal *I4$_2$/mnm*-Ba$_3$Si$_4$ phase displaying Ba atoms (large green spheres) and Si$_4$ tetrahedral clusters (small blue spheres connected by bonds). (e) Energy-dispersive X-ray spectroscopy (EDS) analysis that shows elemental composition of 84.6 wt% Ba and 15.4 wt% Si, which is close to 1:1 Ba:Si atomic ratio, with characteristic X-ray peaks for Ba and Si labeled across 0–8 keV energy range.

## 2.2 High-pressure synthesis and stability of BaSiH$_8$

Reproducible synthesis of cubic BaSiH$_8$ was achieved in two experiments: in DAC BS-1 (Figures 2a, b) at 31 GPa, and in DAC BS-2 at 18 GPa (Figure 3). Before loading, at ambient pressure and in high-pressure DACs, the sample was an opaque material, presumably a metal or narrow-bandgap semiconductor (Figure 2a, d). After laser heating of the BaSi precursor at 31 GPa in DAC BS-1 with an ammonia borane (AB), used as a hydrogen source and pressure transmitting media, the formation of $Fm\bar{3}m$-BaSiH$_8$ was observed (Figure 2b). This compound is similar to $Fm\bar{3}m$-UH$_8$ [23] in composition and volume, the main diffraction reflections come from the heavy atom sublattice (Ba or U), but these compounds differ in the structure of the hydrogen sublattice. In addition to the cubic BaSiH$_8$, we detected another phase, probably, hexagonal *hP*-BaSiH$_x$, refined as $P6_3/mmc$ (Supporting Figure S10) with a = 4.420 Å, c = 7.430 Å, V = 62.84 Å$^3$/Ba or 31.42 Å$^3$/(Ba,Si). This phase accompanies cubic BaSiH$_8$ in many other samples (for instance, DAC BS-4, Supporting Figure S9), has a similar unit cell volume and, probably, a similar hydrogen content (x ~ 8). For comparison, the previously studied BaH$_4$[24] has a volume of about 40–42 Å$^3$/Ba at 30 GPa, which is 20 Å$^3$ less than in the BaSiH$_8$ in terms of Ba atoms.

In both experiments, sample decompression was carried out to the full opening of the DAC, and the synthesized material particle retains its opaque appearance (Figure 2e, h) and the characteristic cubic diffraction pattern even at 0 GPa (Figure 2f), with possibly slight distortions of the cubic structure and broadening of some reflections (e.g., 111, most likely due to long-period distortion). This suggests that near the low-pressure stability limit, BaSiH$_8$ does not decompose abruptly but instead forms a distorted superstructure in which hydrogen partially reorganizes into molecular or covalent configurations. Formation of the BaSiH$_8$ was observed in a significant part of the sample since XRD experiments were performed at SSRF and SPring-8, with an X-ray beam sizes of about 8×12 um or larger.

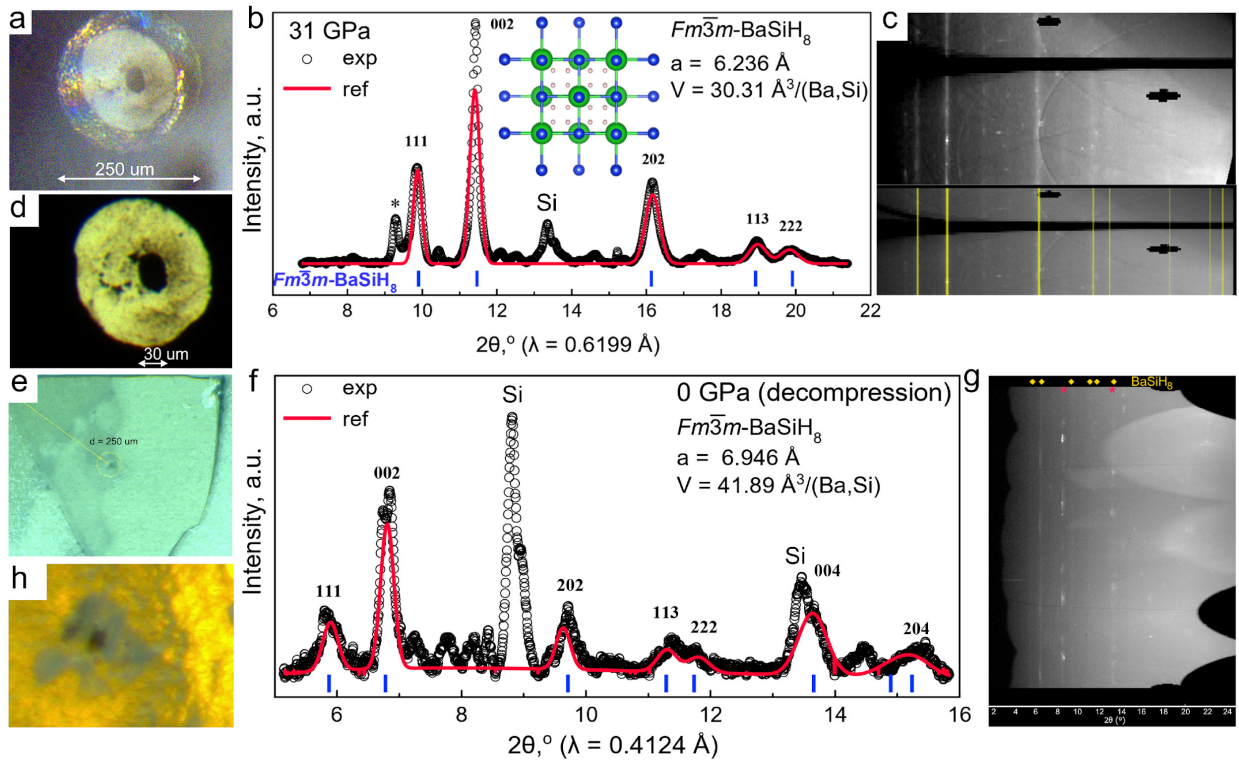

**Figure 2.** Synthesis and structural characterization of BaSiH$_8$ under pressure in DAC BS-1. (a) Optical photo of the BaSi sample in the reflected light at 31 GPa before laser heating. Scale bar: 250 μm. (b) Experimental X-ray diffraction pattern and Le Bail refinement at 31 GPa after laser heating of the BaSi/AB sample, showing the main phase assigned to $Fm\bar{3}m$-BaSiH$_8$ with a lattice parameter a = 6.236 Å and volume $V$ = 30.31 Å$^3$/(Ba, Si). The observed Bragg reflections are indexed; the main impurity is likely $Ia\bar{3}$-Si (Supporting Table S9) and $hP$-BaSiH$_x$ (Supporting Figure S10). An unexplained diffraction peak is marked with an asterisk. Measurements were done at SSRF. (c) Corresponding raw diffraction image used for the integration. Yellow lines correspond to the XRD peaks of BaSiH$_8$. (d) Optical photo of the BaSi sample in the transmitted light at 31 GPa before laser heating. (e) Optical photo of the BaSiH$_8$ in the reflected light at 0 GPa after full decompression and opening of the DAC BS-1. (f) Experimental XRD pattern and Le Bail refinement after decompression of BaSiH$_8$ to ambient pressure, retaining cubic $Fm$-3$m$ structure with expanded lattice (a = 6.946 Å, $V$ = 41.89 Å$^3$/(Ba, Si)). Peak positions of the reference phase are indicated. Measurements were taken at SPring-8. (g) Corresponding raw diffraction image used for the integration. (h) Optical photo of the BaSiH$_8$ in the transmitted light at 0 GPa after full decompression and opening of the DAC.

Distribution of the XRD reflections intensity remains constant after opening of the DAC, indicating that the ordered true ternary $Fm\bar{3}m$ sublattice of heavy atoms, Ba and Si is preserved even at atmospheric pressure. Namely, the 111 reflection is significantly weaker than that of 002 and is close in intensity to the 202 reflection at both 31 GPa and 0 GPa. An important argument in favor of BaSiH$_8$ formation is the very good agreement between the experimental and calculated unit cell volume performed with an anharmonic correction (see Figure 3c, Supporting Table S2). It is interesting to note that, according to single crystal analysis data, cubic γ-Si$_3$N$_4$ is also present in the DAC BS-2 reaction mixture. This silicon nitride is formed as a by-product during the reaction of Si or BaSi with NH$_3$BH$_3$ at high temperature (Supporting Table S8).

The next experiment at DESY, conducted in 2024, involved laser heating of DAC BS-2 at a pressure of 18 GPa (Figure 3). Powder XRD analysis confirmed the formation of cubic BaSiH$_8$ and its stability upon decompression to 0 GPa (Figure 3a). The synthesis also pointed to the formation of a byproduct, another cubic hydride with a lower hydrogen content, likely BaSiH$_{4-x}$ (Figure 3b). We considered several structural possibilities for this byproduct. Cubic BaSiH$_6$ has only a slightly smaller volume than BaSiH$_8$ and cannot explain the difference of ΔV = 3–4 Å$^3$/(Ba,Si). The best result is achieved for pseudocubic BaSiH$_4$ (Figure 3c), but its calculated volume is still slightly larger than the experimental data, indicating a slightly lower hydrogen content, close to BaSiH$_{4-x}$ with x ≈ 0.5. The distribution of BaSiH$_{4-x}$ throughout the sample volume virtually mirrors that of BaSiH$_8$ (Figure 3d, e). Decompression of the DAC BS-2 to atmospheric pressure (0 GPa) leads to a significant decrease in the BaSiH$_{4-x}$ abundance below 3.5 GPa, and likely distortion of its structure. This is particularly noticeable in the splitting of the 111 reflection, while BaSiH$_8$ with a larger cell volume retains a virtually perfect cubic structure of heavy atoms (Figures 3a, f, Supporting Figure S6).

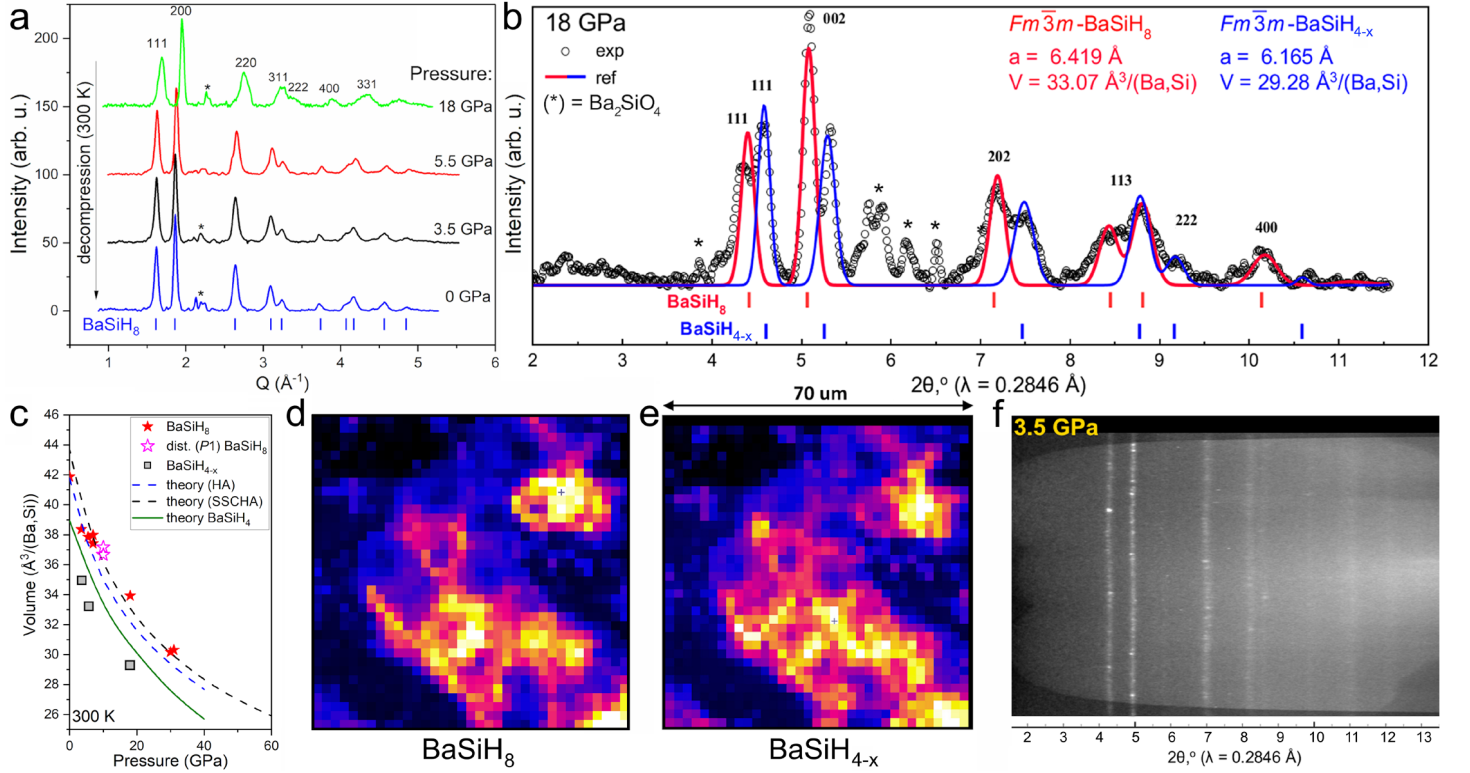

**Figure 3.** Low-pressure stability of BaSiH$_8$ after synthesis at 18 GPa. (a) Decompression of DAC BS-2 from 18 GPa to 0 GPa at 300 K. No significant broadening of the 111 reflection is observed with decompression, as it is the case for BaSiH$_6$ due to the reduction of the lattice symmetry to $C2/c$. (b) Experimental diffraction pattern at 18 GPa and Le Bail refinement of the parameters of two cubic phases: BaSiH$_8$ and, probably, BaSiH$_6$. The sample also contained a significant admixture of $Pnma$-Ba$_2$SiO$_4$ (marked with asterisks *) due to the oxidation of BaSi during storage. (c) Experimental volume $V(P)$ of the unit cell of BaSiH$_8$ (red stars) and BaSiH$_{4-x}$ (grey squares) compared to the theoretically predicted unit-cell volumes in the harmonic (HA) and anharmonic (SSCHA) approximations. (d, e) Color map of the spatial distribution of BaSiH$_8$ and BaSiH$_{4-x}$ in the BS-2 sample. Lighter colors correspond to higher 111 reflection intensity. (f) Example of a raw X-ray diffraction image from the DAC BS-2 at 3.5 GPa used for the pattern integration. Even at low pressure, a clear cubic XRD pattern is maintained.

Synchrotron diffraction of the sample in the DAC BS-3, prepared for electrical transport measurements, exhibiting superconducting properties at 142 GPa (Figure 4), was not studied. XRD from another electric cell, DAC BS-4, was studied at 48 GPa in the range of 51 to 317 K. The diffraction pattern can mainly be explained by a mixture of two phases, hexagonal and cubic ($cF$-BaSiH$_x$, V = 24.54 Å$^3$/(Ba,Si)), and does not undergo any changes upon cooling to 51 K, indicating the absence of phase transitions in the discovered Ba-Si polyhydrides (Supporting Figure S8). The sample of DAC BS-4 is a classic illustration [25] of the degree of non-uniformity of chemical synthesis observed during laser heating: we examined a high-density XRD map of DAC BS-4 (50×50 pixels, 2 μm step, 2500 XRD images) and found that within 2-6 μm the diffraction pattern changes significantly, which corresponds to the temperature gradient at the focus of the IR laser with a diameter of 2-3 μm.

*2.3 Possible formation of other BaSi hydrides*

In addition to the main cubic BaSiH$_8$ phase, identified in DACs BS-1 and BS-2, we systematically observed the formation of hexagonal phases ($hP$) with similar hydrogen content (BaSiH$_{\approx 8}$) in samples of DACs BS-1 and BS-4 (Supporting Figures S9-S10, Table S4). In the DAC BS-4 at 48 GPa, two primary phases — cubic and hexagonal — are also detected; however, their unit cell volume is 2–3 Å$^3$/(Ba,Si) lower than expected for the BaSiH$_8$ composition at this pressure.

Synthesis at 10 GPa in DAC BS-6 results in the formation of a distorted pseudocubic structure, $P1$-BaSiH$_{6-8}$ (Supporting Figures S13, S15), which closely resembles the previously studied pseudocubic $C2/c$-BaSiH$_6$ [26]. The later structure can also be refined as $P1$ [26]. The unit cell volumes of BaSiH$_6$ and BaSiH$_8$ are very similar, making it difficult to reliably determine the hydrogen content from volume data alone. Alongside this $P1$ structure, other low-symmetry hydrides with lower hydrogen content also form (Supporting Figures S12, S16-S18). All samples contain impurities of Ba$_2$SiO$_4$ and Si.

*2.4 Transport measurements*

Transport studies were performed in several cells (DAC BS-3, 4, and 5) using a four-contact van der Pauw scheme in a DC current mode. Despite the prediction of outstanding superconducting properties of $Fm\bar{3}m$-BaSiH$_8$ ($T_c$ = 60 – 90 K in the range of 10 –100 GPa [19]), in the experiment we detect a superconducting (SC) transition only at rather low temperature of about 9 K at 142 GPa (DAC BS-3, Figure 4c), whereas harmonic calculations at this pressure predict $T_c$(AD) = 52 K (Supporting Figure S19). Judging by the multistage character of the temperature-dependent electrical resistance curve (Figure 4d), several superconducting phases are present in the BaSiH$_x$ sample BS-3. The detected SC transition linearly shifts in external magnetic fields, the critical value of which reaches $B_{c2}(0)$ = 13–16 T (inset in Figure 4c). Remarkable difference between the predicted and observed $T_c$ values can be explained by isomerization of polyhydrides with conversion of atomic hydrogen to the molecular one: 2HH → H$_2$, or by formation of covalent Si-H bonds resulting in stabilization of the molecular sublattice. This also explains the very high stability of such polyhydrides with respect to decompression to ambient pressure at room temperature.

At relatively low pressure (30 GPa, DAC BS-4), transport measurements of BaSiH$_x$ hydrides indicate their semiconducting properties, negative $dR/dT$, although the detected minimal indirect bandgap is small (< 1 meV, 5-10 K). Increasing the pressure from 30 GPa to 40 GPa (Figure 4g, h), and further to 48 GPa (Supporting Figure S22b), leads to metallization of the compound and its transformation into a poor metal, similar to the insulator-to-metal transition in solid hydrogen between 360 and 440 GPa[27]. For the bandgap estimation we assumed the simplest parallel connection of semiconducting BaSiH$_x$ grains with metallic ones that possess a weakly temperature-dependent electrical resistance. This is rather common for dirty metallic polyhydrides, with the resistance following:

$$\frac{1}{R(T)} = \frac{1}{R_0} + \frac{e^{-E_g/2kT}}{aT^c} \qquad (1)$$

where $R_0$, $a$, $c$, $E_g$ – are the fit parameters, and $T^c$ describes the temperature dependence of the effective density of states in the conduction band [28]. The most accurate interpolation of the experimental data is provided by a multiparameter fit, including both parallel and series connections of semiconducting and metallic grains. Although less accurate due to the large number of parameters, it nevertheless leads to the same conclusion about the small size (< 10 meV) of the band gap in semiconducting grains. The introduction of nonlinear terms $R(T) \propto T^2$ allows us to explain the inflection of the $R(T)$ curve above 250 K, which corresponds to thermally activated processes with changes in carrier concentration or the manifestation of interband electron-electron scattering.

No signs of superconducting transitions were observed. Given the fast change in the sign of the derivative $dR/dT$ when the pressure increases by 10 GPa (Figures 4f, g), we expect that synthesis of BaSiH$_x$ at about 60 GPa will yield a fully metallic sample. Indeed, resistance measurements applying the van der Pauw contact scheme in BaSiH$_x$ synthesized at 55 GPa (3 electrodes, 1 was broken) showed metallic behavior (see Supporting Figure S22).

The metallic ground state is supported by spin-echo $^1$H NMR data recorded for a DAC BS-NMR sample (Supporting Figures S26–S28). At 28 GPa, the proton spin-lattice relaxation time $T_1$ was 1.56 s at 150 K and 1.95 s at 100 K, rendering the $T \times T_1$ approximately constant (200–225 sK), which is typical for metals and degenerate semiconductors with a low density of states at the Fermi level $N(E_F)$. Good metals typically have $T_1$ in the millisecond range (e.g., 10–100 ms)[29], and hydrides with a hopping relaxation mechanism exhibit an exponential increase in $T_1$ with decreasing temperature [30]. Taking into account the $^1$H NMR data and the findings that at pressures above 40 GPa and temperatures of 100-150 K, BaSiH$_x$ samples have metallic $dR/dT > 0$ (Figure 4g), we conclude that the sample is a poor metal or a degenerate semiconductor with low $N(E_F)$, an order of magnitude less than in aluminum [31].

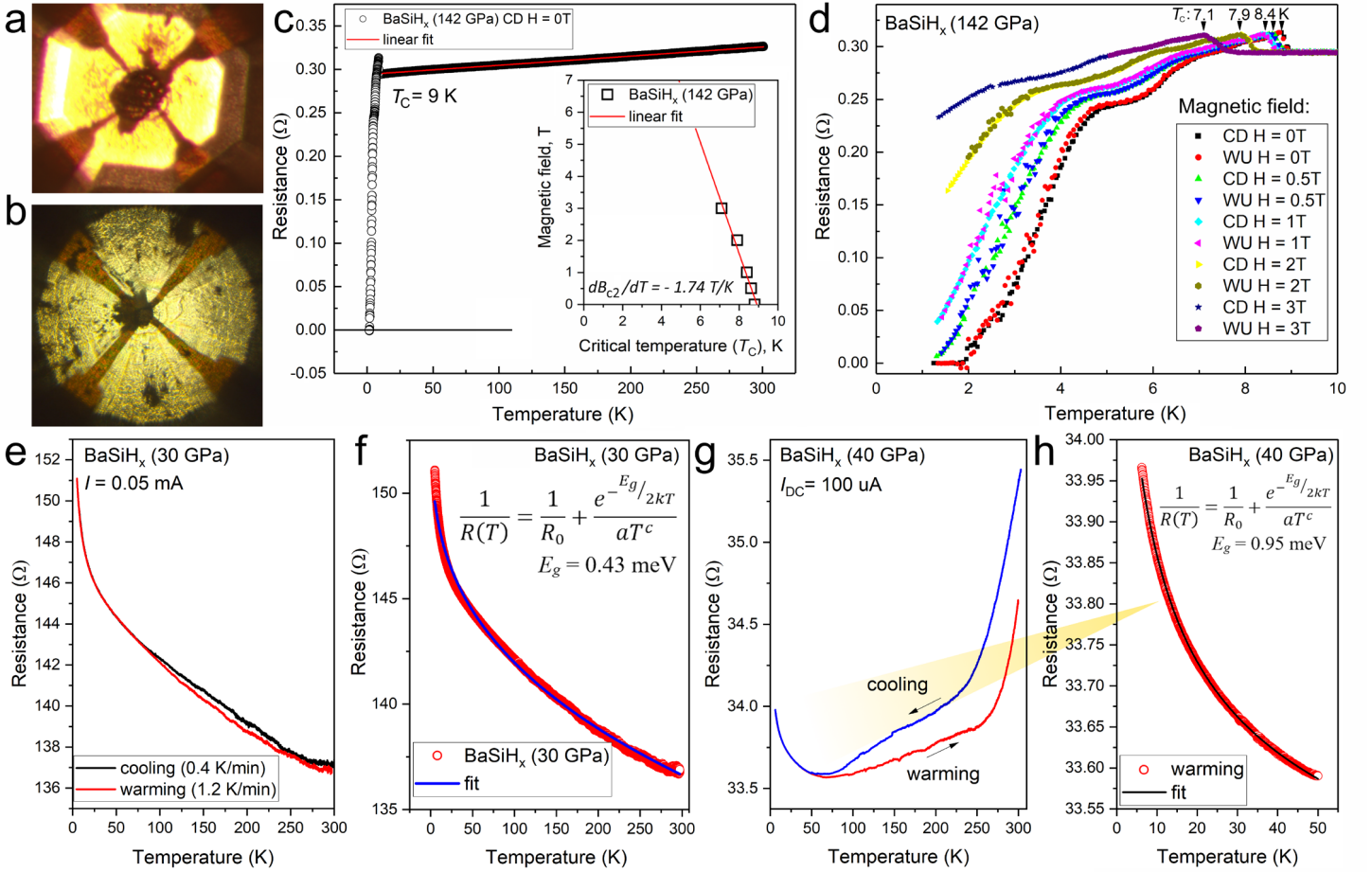

**Figure 4.** Transport properties of barium-silicon polyhydrides. (a, b) Photographs of culets of electrical DACs BS-4 (a) and BS-3 (b) after laser heating. Panel (a) shows partly transparent holes in the sample. (c) Temperature dependence of electrical resistance in zero magnetic field for BaSiH$_x$ sample in DAC BS-3 at 142 GPa. A superconducting transition is observed below 9 K, with resistance dropping to "zero" (below the noise level). Inset: magnetic field dependence of the onset critical temperature of SC, showing $dB_{c2}/dT |_{T=T_c} = -1.74$ T/K. (d) Series of resistive transitions in BaSiH$_x$ at 142 GPa measured during heating and cooling cycles in magnetic fields from 0 to 3 T, demonstrating the suppression of SC with increasing field. (e) Semiconducting behavior ($dR/dT < 0$) of BaSiH$_x$ sample in DAC BS-4 at 40 GPa tested with applied current $I_{DC}$ = 0.05 mA. (f) Fit of the temperature-dependent resistance using a parallel circuit model with a semiconducting shunt resistor, yielding an activation energy $\Delta E = E_g = 0.43 \pm 0.12$ meV. (g) Mostly metallic behavior ($dR/dT > 0$ above 50 K) of the same sample BaSiH$_x$ in DAC BS-4 synthesized at 40 GPa. Measurements were done with DC current $I_{DC}$ = 100 μA, showing a non-metallic trend only below 50 K. (h) Fit of the temperature-dependent resistance below 50 K using a parallel semiconducting shunt model yielding $\Delta E = E_g = 0.95$ meV.

The difference (hysteresis) observed at 40 GPa between cooling and warming cycles (Figure 4g) could indicate a phase transition in BaSiH$_x$ hydrides, but low-temperature X-ray diffraction study of DAC BS-4, performed in the range 51–317 K, indicates the absence of any pronounced changes in the heavy-atom sublattice over the entire temperature range (Supporting Figure S8). Still, we cannot completely exclude that the MIT transition above 40 GPa may be related with rearrangement of hydrogen atoms in the unit cell.

The absence of high-temperature superconductivity in Ba-Si hydrides is not surprising. Previous studies of barium hydrides have shown that, for example, cubic BaH$_{12}$ and other barium hydrides with lower hydrogen content (Ba$_4$H$_{23}$, BaH$_{\approx 6}$) do not exhibit pronounced superconducting properties: the maximum $T_c$ reaches 20 K at a pressure of about 140 GPa [9]. The same is for silicon hydrides: superconductivity has been reported in SiH$_4$ at high pressure and the temperature below $T_c$ = 17 K at 96 and 120 GPa [32]. Thus, if we combine these two systems (Ba-H and Si-H) that possess partially molecular hydrogen sublattices and exhibit low-$T_c$ superconductivity at pressures near or above 100 GPa, we can expect similar properties for BaSiH$_x$. Therefore, our observation of SC with a $T_c$ of only 9 K in DAC BS-3 (Fig. 4c) meets those expectations (Figure 4c). The special, step-like structure of the density of electron states $N(E)$ (see Fig. 3 in Ref. [19]) is also very sensitive to impurities, which can shift the Fermi level $E_F$ and reduce the $N(E_F)$ in BaSiH$_8$ several times, greatly reducing $T_c$. [33]

Theory predicts cubic BaSiH$_8$ to be thermodynamically stable above 140 GPa, and recoverable down to ambient pressure, with a $T_c$ of about 80 K. In this work, we synthesized several BaSiH$_x$ hydrides. Most of these are semiconducting. A superconducting phase forms at 142 GPa, but the $T_c$ is low. In a few cases, we found a cubic structure of BaSiH$_8$ that is recoverable at ambient pressure, but this phase is not superconducting, most likely due to structural distortions. In connection with the above, achieving $T_c$ of 80-100 K at pressures less than 50 GPa in the Ba-Si-H system is unlikely. The absence of high-$T_c$ superconductivity in BaSiH$_8$ is a disappointment, but we have a very reasonable explanation as to why that can help to guide future experiments.

*2.5 Weak localization in Ba-Si hydrides*

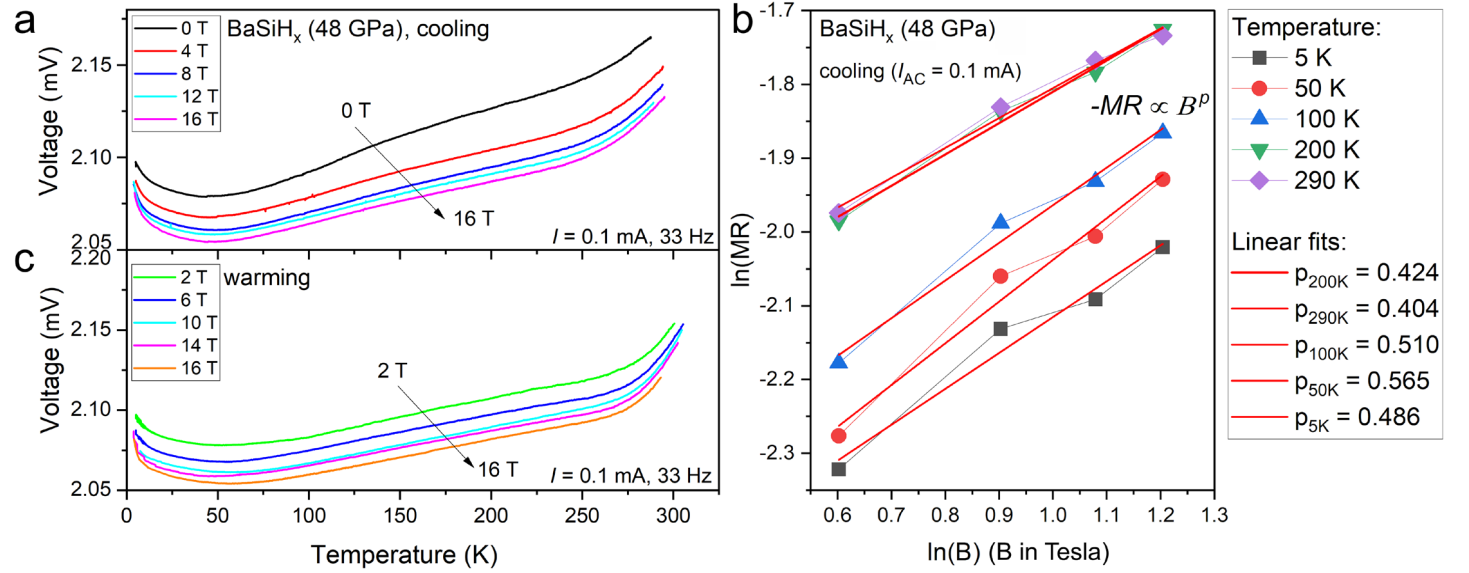

**Figure 5.** Negative magnetoresistance of BaSiH$_x$ at 48 GPa (DAC BS-4) in magnetic fields of 0-16 T over the temperature range from 5 to 300 K. (a) Temperature dependence of the real voltage drop across the sample (four-terminal van der Pauw circuit) in the AC measurement mode ($I_{AC}$ = 0.1 mA RMS, $f$ = 33 Hz) at a constant magnetic field (0–16 T) in the cooling mode. (b) The same, in the warming mode. (c) Linear fit of the dependence of magnetoresistance $MR = (R - R_0)/R_0$ on the magnetic field induction ($B$, in Tesla) in logarithmic coordinates at several selected temperature points (5, 50, 100, 200, and 290 K).

A remarkable property of the DAC BS-4 sample is its pronounced negative magnetoresistance (MR) reaching –2% change at 16 T, observed across the entire temperature range (Figures 5a, b, Supporting Figure S24). A negative MR has already been observed in several other polyhydrides (for instance, CeH$_{9-10}$ [34], La$_4$H$_{23}$ [35]) and in other materials under high pressure[36-39]. Plotted in logarithmic coordinates, $MR(B)$ exhibits a field dependence $|R - R_0|/R_0 \propto \sqrt{B}$ (Figure 5b), characteristic of weak localization in dirty metals and heavily doped semiconductors[40]. This phenomenon is originating from a disruption of the phase of electrons by the external magnetic field that leads to a localization of the charge carriers at defects in the hydrogen sublattice. Consequently, this reduces the electrical resistance of the sample. Considering equations in Refs. [41,42] for weak localization in the 3D case (Supporting Information, eqs. S5-S6), as well as $\Delta\sigma/\sigma_0 = -\Delta R/R_0$, we arrive at negative MR following

$$MR = \frac{\Delta R}{R_0} = -\frac{e^2}{2\pi^2 \hbar R_0}\sqrt{\frac{eB}{\hbar c}} f_{3D}\left(\frac{\hbar c}{4eBL_{Th}^2}\right) \propto \sqrt{B}, \quad (2)$$

where $L_{Th} = \sqrt{\tau_\varphi \tau_{tr} V_F^2/3}$ is the characteristic inelastic scattering distance. In high magnetic fields ($B$), the magnetoresistance is proportional to the square root of the magnetic induction. That is what we observed in the experiment: fits to Eq. (2) with $MR(B) \propto B^p$ at different temperatures yield an exponent $p \approx 0.5$ (Figure 5b).

It should be noted that a similar dependence $R \propto \sqrt{B}$ was also observed for a hopping transport [43] in some cases, for example, in thin films of $In_2O_{3-x}$ [44] and 2D heterojunctions GaAs/AlGaAs [45]. Thus, most likely $BaSiH_x$ is located on the border between hopping and diffusion transport mechanisms with a strong influence of localization effects. Other models of the hopping transport, for instance, tunneling in a random long-range potential [46,47], or hops in the presence of 3$^{rd}$ impurity [48] cannot explain the observed negative magnetoresistance over the entire range of magnetic fields and temperatures, as well as its positive temperature dependence.

Evaluation of the resistivity of the DAC BS-4 sample, having a shape close to square 72×72 μm$^2$ and thickness $t$ = 3.5 μm (assessment based on interference between diamond anvils), can be made using the symmetric van der Pauw formula: $\rho = \pi t R / \ln 2$, where $R \approx 34 - 35$ Ω at 40 GPa. It leads to ρ ≈ 5×10$^{-4}$ Ω·m. This value is ~10$^3$ times greater than in NiCr alloys and corresponds in order of magnitude to the specific resistance of overdoped semiconductors such as Si, Ge, indium-tin oxide, etc.

The Hall effect measurement for the BS-4 DAC shows (Supporting Figure S24d) that the Hall resistance $R_H$ = $V_H/IB \approx 0.021$ Ω/T. For a single type of carrier (electrons), this would result in a maximum carrier concentration $n_e^{max}$= 8.5×10$^{25}$ m$^{-3}$, which is characteristic of overdoped semiconductors in the vicinity of the metal-insulator transition. However, this is rarely observed in doped semiconductors due to the contribution of holes, which reduce the observed Hall effect and lead to an overestimation of the carrier concentration. More realistic estimates from magnetoresistance give $n_e$ = 3.6×10$^{24}$ m$^{-3}$, the Fermi energy $E_F$ = 8.6 meV, the Fermi velocity $V_F$ = 5.5×10$^4$ m/s, and mean free path $l_e$ = 9.7 Å. In this case, $k_F l_e \approx 0.46$ is close to unity, which allows us to speak of an important contribution of localization phenomena (weak or strong, depending on pressure) in $BaSiH_x$.

## 2.6 Persistent photoconductivity (PPC) under pressure

The rise in electrical resistance at low temperature and pressure (Figure 5) indicates the presence of a small bandgap in the band structure of some barium-silicon polyhydrides, typical for semiconductors. Semiconductors often exhibit pronounced photoconductivity[49], which can be effectively studied under pressure in high-pressure electrical DACs. In our case, an X-ray beam (25 keV, excitation time: 300 s, about 10$^{10}$ photons/s) was used to photoexcite the formation of electron-hole pairs in the $BaSiH_x$ sample under pressure of 48 GPa in the DAC BS-4. Experiment was conducted at the XPRESS beamline of the Elettra synchrotron, and was combined with simultaneous measurements of low-temperature X-ray diffraction (Figure 6). Almost immediately we discovered that the sample's response to X-ray irradiation was characterized by persistent photoconductivity (PPC, Figure 6), probably associated with a large number of structural defects (hydrogen vacancies) and grain boundaries.

The interaction of X-rays with a semiconductor result in the formation of electrons (e) and holes (h), and, at the first stage, a linear increase in conductivity (Figure 6b). However, fairly soon, charge carriers end up in various kinds of structural traps with an average activation energy $E_a$, and further photoconductivity is due to overcoming this activation barrier by thermal activation ($\propto e^{-E_a/kT}$). At the same time, recombination of electrons and holes in traps ($\propto n_e n_h e^{-E_a/kT}$) decreases the concentration of carriers. This leads to equations containing exponential functions of temperature

$$\frac{dn}{dt} = aIe^{-E_a/kT} - bn^2 e^{-E_a/kT}, \qquad (3)$$

where we assumed for simplicity that $n_e = n_h = n(t)$. The latter is the concentration of free charge carriers involved in the transport, $a$ – is the coefficient of illumination efficiency, $I$ – is the intensity of the X-ray or light beam, and $b$ – is the efficiency of carrier recombination during interaction of a free charge carrier with a bound one. For the illumination phase, eq. (3) leads to the following time dependence of the concentration of free charge carriers $n(t)$

$$\frac{n_{max} + n(t)}{n_{max} - n(t)} = exp\left(2\sqrt{abI}e^{\frac{-E_a}{kT}}t\right) = exp\left(\frac{2t}{\tau_1}\right), \quad (4)$$

where $\tau_1 = e^{E_a/kT}/\sqrt{abI}$ corresponds to the characteristic saturation time of photoconductivity ($\sigma = \sigma_0 + cn(t)$) under constant X-ray irradiation, $n_{max} = \sqrt{aI/b}$. Considering the large dark photoconductivity characteristic of narrow-bandgap semiconductors (Figure 6b), irradiation of the BaSiH$_x$ sample at 48 GPa does not lead to a significant drop in resistance, and the photoconductivity can be considered as a small correction to the initial resistance

$$R(t) = R_0 - c'\tanh\left(\frac{t}{\tau_1}\right). \quad (5)$$

At the initial stage ($t \ll \tau_1$) the drop in electrical resistance occurs linearly $\propto t/\tau_1$ (Figure 6b). We use eq. (5) to interpolate the data in the photoexcitation stage (exposure time: 300 s, Figure 6d). The interpolation indeed shows a significant increase of $\tau_1$ when the temperature decreases (Figure 6e), and the virtual absence of the resistance recovery at temperatures below 241 K for at least several hours (Figure 6d). This behavior, accompanied by prolonged existence of increased conductivity for many hours and days even after the light source is turned off, is called persistent photoconductivity, and is inherent in compounds such as ZnO, SnO$_2$, and GaAs [50].

Similar results, but with greater amplitude, were obtained using the visible range at room temperature (Figure 6c), including laser light of three wavelengths: 488 nm, 532 nm, and 660 nm. The characteristic times of photoconductivity excitation and resistance recovery at 300 K for visible light were: $\tau_1 \approx$ 2-20 s, and slightly longer $\tau_2 \approx$ 20-80 s, similar to what was obtained for the X-ray radiation. A linear component is evident in $R(t)$ both in resistance recovery and in the carrier generation stage under irradiation (Figure 6c). Like most semiconductors, BaSiH$_x$ exhibits a photovoltaic effect in the absence of a passing current and generates a microvolt photovoltage when illuminated with laser light (Figure 6f, Supporting Figure S25). Finally, application of strong 1-3 s pulses of 7 mA current to a pair of electrodes results in a sudden and irreversible increase in electrical resistance, possibly associated with electrochemical diffusion of hydrogen (Supporting Figure S23b).

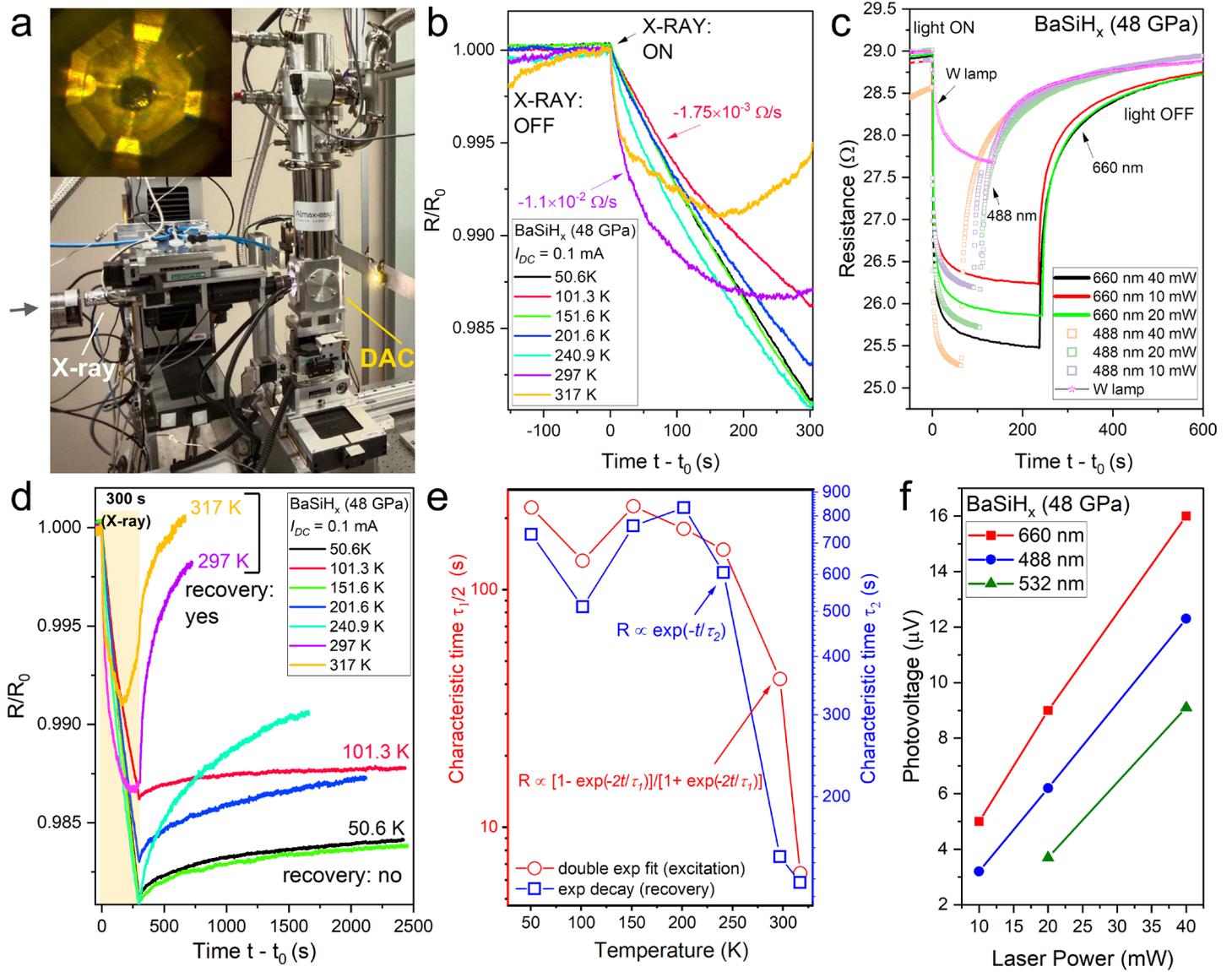

**Figure 6.** Photoconductivity and photovoltaic effect in BaSiH$_x$ at 48 GPa (DAC BS-4). (a) Experimental setup at the Elettra synchrotron's XPRESS beamline, highlighting the diamond anvil cell (DAC) in the cold finger cryostat, and the X-ray beam (25 keV) input direction. The inset shows an optical image of the BaSiH$_x$ sample with four electrodes sputtered on the DAC's culet. (b) Relative resistance (R/R$_0$) of BaSiH$_x$ as a function of time, demonstrating the transient photoconductivity response upon X-ray illumination (ON/OFF) at various temperatures from 50.6 K up to 317 K. Maximum and minimum rates of resistance change are indicated. (c) Photoconductivity of the sample, which occurs when BaSiH$_x$ is illuminated with visible light from a tungsten lamp, as well as laser light with a wavelength of 488 nm and 660 nm, and different powers (10-40 mW). (d) Long-term resistance recovery following X-ray exposure (total time: 300 s) at different temperatures, illustrating the presence or absence of recovery of electrical resistance back to its initial values for the time period of 30-40 min. (e) Temperature dependence of characteristic time constants, $\tau_1$ (excitation phase) and $\tau_2$ (recovery phase), derived from fitting the resistance dynamics. (f) Photovoltage on a sample (2-electrode circuit, Keithley 2182a nanovoltmeter) when applying a Π-shaped 10-second pulse of laser light of different wavelengths and power (see Supporting Figure S23).

## 3. Discussion

As noted in the previous work on the Ba-Si-H system[26], the immediate cause of the lack of metallic properties and pronounced SC in BaSiH$_6$ and BaSiH$_8$ is the distortion of the highly symmetric cubic structures with the destruction of H-H bonds and the formation of [SiH$_6$]$^{-2}$ octahedra. Such structures are semiconducting and more stable than the metallic, highly symmetric modifications of these compounds, with high predicted critical temperature of SC. Given the significant number of experiments conducted at various pressures, it is highly probable that the Ba-Si-H system does not exhibit high-temperature superconductivity, similar to the absence of such properties in the individual Ba-H and Si-H binary systems at pressures below 1 Mbar. The reason for this is that the metallic state of the hydrogen sublattice at pressure below 1 Mbar is unstable and tends to form a forbidden band in the spectrum of electronic states, which can be either a bandgap of a semiconductor (more advantageous) or a superconducting gap [51].

Despite numerous attempts to use high-pressure single-crystal diffractometry, we were only able to obtain good structural solutions for thermodynamically stable structures with a tendency to form large crystals: $Ba_2SiO_4$, $\gamma$-$Si_3N_4$, Si, and so on. We found that ternary Ba-Si hydrides crystallize poorly, and obtaining good single-crystal data directly from the polyhydrides is difficult, even with repeated laser heating cycles. This unfavorable property of hydrides distinguishes them from the behavior of most nitrides at high pressure and reduces the effectiveness of single-crystal diffractometry.

From a practical perspective, materials with PPC, such as B-doped diamond, GaAs, Se, CdZnTe, and $CsPbBr_3$, can be used to create long-term memories, as optical radiation sensors, and for creating neuromorphic devices with optical learning. Given their sensitivity to hard X-rays and potential sensitivity to neutron irradiation, the discovered effect can be used to create radiation leakage sensors for accelerator equipment, as well as cumulative radiation dose sensors (dosimeters). Such a sensor can be reset not only by heating but also by applying a high current or increasing pressure (tightening screws), which leads to metallization of the sample.

## 4. Conclusions

We studied the Ba-Si-H system using powder and single-crystal XRD at room and low temperatures, and performed electric transport measurements in magnetic fields over a wide pressure range (0–142 GPa). Along with several other phases, we synthesized previously predicted $BaSiH_8$, which remained stable during decompression even under ambient conditions. Despite predictions of high-temperature superconductivity in this compound, Ba-Si hydrides exhibit comparatively weak superconducting properties with a maximum $T_c$ of 9 K at a pressure of 142 GPa. We show that this can be explained by the preferential formation of semiconducting molecular phases, possibly containing $[SiH_6]^{-2}$ anions, rather than phases with a clathrate metallic hydrogen sublattice.

We have demonstrated that high-pressure electrical DACs with Ba-Si hydrides can be applied as sensors for detecting various types of radiation, including hard X-ray. Measurements in a magnetic field indicate a pronounced negative magnetoresistance ($\propto \sqrt{B}$) of $BaSiH_x$ samples, likely due to weak electron localization at structural defects. Overall, barium-silicon hydrides exhibit semiconducting properties, negative temperature coefficient of electrical resistance below 50 K, long spin-lattice relaxation time $T_1$, particularly noticeable at low temperatures (<100-150 K) and pressures below 30 GPa, characteristic of degenerate heavily doped semiconductors or bad metals with a low density of states at the Fermi level. Increasing the pressure to 50-60 GPa leads to metallization of Ba-Si polyhydrides.

**Data availability**

The authors declare that the main data supporting our findings of this study are contained within the paper and Supporting Information. All relevant data are available from the corresponding authors upon request.

**Code availability**

Quantum ESPRESSO code is free for academic use and available after registration on https://www.quantum-espresso.org/. The Vienna ab-initio Simulation Package (VASP) code is available after registration on https://www.vasp.at/. The USPEX code is free for academic use and can be obtained after registration on https://uspex-team.org/en/. To interpret complex diffraction patterns, we used python-based scripts for the postprocessing of USPEX calculations freely available on https://github.com/michelegalasso/xrpostprocessing.

**Acknowledgments**

D. S. and D. Z. thank the National Natural Science Foundation of China (NSFC, grant No. 12350410354) and Beijing Natural Science Foundation (grant No. IS23017) for support of this research. D.Z. thanks the China

Postdoctoral Science Foundation (Certificate No. 2023M740204) and financial support from HPSTAR. D.S. and D.Z. are grateful for financial support from the EMFL-ISABEL project. We acknowledge support from the Würzburg-Dresden Cluster of Excellence on Complexity, Topology and Dynamics in Quantum Matter–ctd.qmat (EXC 2147, Project No. 390858490), as well as the support of the HLD at HZDR, member of the European Magnetic Field Laboratory (EMFL). The high-pressure experiments were supported by the state assignment of the Kurchatov Complex of Crystallography and Photonics (KKKiF). XRF analysis was performed using the equipment of the Shared Research Center of the Kurchatov Complex of Crystallography and Photonics. We acknowledge the European Synchrotron Radiation Facility (ESRF) for the provision of synchrotron radiation facilities and Momentum Transfer for facilitating the measurements. Jakub Drnec is thanked for assistance and support in using beamline ID31. The measurement setup was developed with funding from the European Union's Horizon 2020 research and innovation program under the STREAMLINE project (grant agreement ID 870313). Measurements performed as part of the MatScatNet project were supported by OSCARS through the European Commission's Horizon Europe Research and Innovation programme under grant agreement No. 101129751. All authors thank the staff scientists of the SPring-8, Elettra, Xpress station (proposal No. 20245056) and the ESRF, station ID27 (proposal MA-5924), synchrotron radiation facilities for their help with X-ray diffraction measurements, especially Dr. Anna S. Pakhomova (ESRF), Dr. Saori Kawaguchi (SPring-8), Dr. Boby Joseph and Dr. Adrea Stolfa (Elettra). We also thank Dr. Takeshi Nakagawa (HPSTAR) for assistance with the X-ray diffraction studies at SPring-8 synchrotron research facility. We express our deep gratitude to Dr. Umbertoluca Ranieri (Centro de Fisica de Materiales, San Sebastián, Spain) and Dr. Dominique Laniel (University of Edinburgh, UK) for assistance with single crystal XRD.

**Contributions**

D.V.S., D.Z., I.A.T., M.G., A.V.S., O.A.S., H.K., F.G.-A., T.H., S.L., Y.N., and K.S.P. performed the experiments. S.D-C., R.L., M.G., and D.V.S. performed theoretical calculations for the paper. K.S.P. prepared starting BaSi intermetallics. I.A.T. prepared diamond anvil cells for electrical measurements. T.H., A.V.S., O.A.S., and V.M.P. performed the magneto transport experiments in fields up to 16 T and participated in the data processing and discussions. S.L. performed $^1$H NMR measurements. F.G.-A. performed low-temperature X-ray diffraction studies. D.V.S., D. Z., H. K., F.G.-A., Y. N. were responsible for the diffraction studies of the samples. D.V.S., V.M.P., K. S. and V.V.S. wrote the manuscript and supervised the project. All the authors discussed the results and offered useful inputs.

**Conflict of interest**

The authors declare that they have no conflict of interest

# SUPPORTING INFORMATION

# Superhydrides on the way to ambient pressure: weak localization and persistent X-ray photoconductivity in BaSiH$_8$

Dmitrii V. Semenok[1,†,*], Di Zhou[1,*,†], Sven Luther[2], Toni Helm[2], Hirokazu Kadobayashi[3], Yuki Nakamoto[4], Katsuya Shimizu[4], Kirill S. Pervakov[5], Andrei V. Sadakov[5], Oleg A. Sobolevskiy[5], Vladimir M. Pudalov[5,6], Simone Di Cataldo[7], Roman Lucrezi[8], Lilia Boeri[7], Michele Galasso[9], Frederico G. Alabarse[10], Ivan A. Troyan[11,12], and Viktor V. Struzhkin[12,13,*]

[1] *Center for High Pressure Science & Technology Advanced Research, Bldg. #8E, ZPark, 10 Xibeiwang East Rd, Haidian District, Beijing, 100193, China*
[2] *Hochfeld-Magnetlabor Dresden (HLD-EMFL) and Würzburg-Dresden Cluster of Excellence ctd.qmat, Helmholtz-Zentrum Dresden-Rossendorf, 01328 Dresden, Germany*
[3] *Japan Synchrotron Radiation Research Institute (JASRI), 1-1-1 Kouto, Sayo, Hyogo 679-5198, Japan*
[4] *Center for Science and Technology under Extreme Conditions, Graduate School of Engineering Science, Osaka University, 1-3 Machikaneyama, Toyonaka, Osaka 560-8531, Japan*
[5] *V. L. Ginzburg Center for High-Temperature Superconductivity and Quantum Materials, Moscow, 119991 Russia*
[6] *National Research University Higher School of Economics, Moscow 101000, Russia*
[7] *Dipartimento di Fisica, Sapienza Università di Roma, Piazzale Aldo Moro 5, 00185 Roma, Italy*
[8] *Department of Materials and Environmental Chemistry, Stockholm University, SE-10691 Stockholm, Sweden*
[9] *Department of Electrotechnology, Faculty of Electrical Engineering, Czech Technical University in Prague, Czech Republic*
[10] *Elettra - Sincrotrone Trieste S.C.p.A., Strada Statale 14 - km 163,5 in AREA Science Park, 34149 Basovizza, Trieste, Italy*
[11] *A.V. Shubnikov Institute of Crystallography of the Kurchatov Complex of Crystallography and Photonics (KKKiF), 59 Leninsky Prospekt, Moscow 119333, Russia*
[12] *Shanghai Advanced Research in Physical Sciences (SHARPS), 68 Huatuo Rd, Bldg 3, Pudong, Shanghai 201203, P.R. China*
[13] *Center for High Pressure Science and Technology Advanced Research (HPSTAR), Bldg. 6, 1690 Cailun Rd., Pudong, Shanghai 201203, P.R. China*

*Corresponding authors: Dmitrii V. Semenok (dmitrii.semenok@hpstar.ac.cn), Di Zhou (di.zhou@hpstar.ac.cn), and Viktor V. Struzhkin (viktor.struzhkin@hpstar.ac.cn).

†These authors contributed equally.

# Contents

# I. Methods

*Structure prediction*

Crystal structure predictions were performed using the evolutionary algorithm as implemented in the USPEX 10.5 code[1,2]. The search was conducted for the Ba–Si–H system at pressure of 50 GPa. The fitness of each candidate structure was evaluated based on its enthalpy of formation. The initial generation consisted of 80 random structures, generated with symmetry constraints (space groups 2–230) and subject to minimum interatomic distance constraints (Ba–Ba/Si: 1.2 Å, Ba–H: 0.8 Å, Si–Si: 1.2 Å, Si–H: 0.8 Å, H–H: 0.5 Å). A variable-composition approach was employed, allowing the number of atoms per unit cell to vary between 24 and 36. The subsequent population size was fixed at 40 structures per generation. To evolve the population, a suite of variation operators was applied with the following probabilities: heredity: 50%, soft mutation: 20% (atomic positions) & 10% (lattice vectors), rotation: 40%, atom permutation: 10%, random symmetry-guided generation: 20%; composition-specific operators: 20%.

The enthalpy of each candidate structure was computed via density functional theory (DFT) calculations using the VASP package[3,4]. The calculations employed the PBE exchange-correlation functional and PAW pseudopotentials. A plane-wave energy cutoff of 600 eV and a k-point mesh with a resolution of 0.12–0.04 2π/Å (progressively refined during the search) were used to ensure convergence of total energies.

*Selection of a structural solution*

A search for the structures that best explain complex experimental X-ray powder diffraction patterns has been performed using the XRpostprocessing code[5]. In particular, we used the script powder_xrd_screening.py, which screens a collection of structures to identify those that best match experimental powder XRD data. The input parameters of the powder_xrd_screening.py script include a set of candidate structures in the POSCAR format, an experimental powder diffraction pattern in the form of angles and intensities and an array of coefficients k (for instance, from 0.6 to 1.4 with a step of 0.001) for performing isotropic deformations of the unit cell. For each structure, the script does the following:

1. Multiplies the lattice vectors by the first provided value of *k*.

2. Computes the theoretical XRD pattern of the so-obtained crystal structure.

3. Computes a fitness value $F'$, denoting the degree of disagreement between theoretical and experimental powder diffraction patterns.

4. Goes back to the point #1 for the next value of *k*, until the last *k* is reached.

We define a *match* between a theoretical and an experimental peak the case when the two peaks are less than *match_tol* degrees apart, where *match_tol* is a user-defined parameter. For peak intensities normalized to 100, $F'$ is given by

$$F' = \left\{ \sum_{i,j} \left( \frac{h_{i,j}^{exp} - h_{i,j}^{th}}{100} \right)^2 \left( \frac{h_{i,j}^{exp}}{100} \right)^2 \right\}_{match} + \left\{ \sum_{i} \left( \frac{h_i^{exp}}{100} \right)^2 \right\}_{exp\,residue} + \left\{ p^{th} \cdot \sum_{i} \left( \frac{h_i^{th}}{100} \right)^2 \right\}_{th\,residue}, \quad (S1)$$

where the $h_i$ are diffraction intensities. The first term of the fitness is a sum over the matched peaks, and each addend of this sum will be smaller the more the intensities of the two matched peaks are close to each other. The second term is a sum over the experimental peaks that are left after our matching, we call these peaks experimental rest. The third term, analogously, is a sum over the theoretical rest. The user-defined coefficient $p_{th} > 1$ which appears in the last term introduces an additional penalty for non-matching theoretical peaks. It is clear that a low $F'$ gives a good agreement between calculated and experimental PXRD patterns.

We define the fitness $F$ of each candidate structure as the minimum value of $F'$ across all values of $k$

$$F = \min_k F'(k). \quad (S2)$$

A low value of $F'$, together with the corresponding value of $k$, enables a rapid and automatic screening of structures even when the experimental volume of cell or pressure differs from the pressure/volume of the candidate structures.

*Experiment*

In most XRD experiments, we used a BX-90 mini, as well as Mao-type symmetrical DACs and diamond anvils with culet's diameters of 150-300 μm. 300 μm tick tungsten and tantalum sheets were used as gaskets. Transport measurements were performed in asymmetric BeCu DACs with a diameter of 25 mm. The composite insulating gasket was composed of $CaF_2$/epoxy/$NH_3BH_3$ and supported by a non-magnetic steel gasket with a chamber of 300-350 μm in diameter.

**Table S1.** Summary of experiments performed in the Ba–Si–H system. Experiments are grouped by DACs, with corresponding pressure ranges, techniques, and main findings.

| DAC/Experiment | Starting composition | Pressure, GPa | Performed studies | Results |
|---|---|---|---|---|
| BaSi precursor | $Cmcm$-BaSi $I4_2/mnm$-$Ba_3Si_4$ | 0 | Powder XRD, SEM, EDS | Mechanochemical synthesis of BaSi precursor; multiphase material with near-equimolar Ba:Si ratio suitable for high-pressure hydrogenation |
| DAC BS-1 | BaSi/AB | 31 → 0 | XRD, full decompression at 300 K | $Fm\bar{3}m$-$BaSiH_8$ $hP$-$BaSiH_x$ ($x \approx 8$)<br><br>Successful synthesis of cubic $BaSiH_8$ at 31 GPa; structure preserved upon full decompression to ambient pressure |
| DAC BS-2 | BaSi/AB | 18 → 0 | XRD, full decompression at 300 K | $Fm\bar{3}m$-$BaSiH_8$ $Fm\bar{3}m$-$BaSiH_{4-x}$<br><br>$BaSiH_8$ synthesized at 18 GPa; coexistence with lower-H phase; $BaSiH_8$ remains stable to 0 GPa, while $BaSiH_{4-x}$ destabilizes < ~3.5 GPa |
| DAC BS-3 | BaSi/AB | 142 | Transport measurements (four-probe van der Pauw) | Superconducting transition observed at $T_c \approx 9$ K; multistep transition suggests phase mixture; upper critical field $B_{c2}(0) \approx 13$–16 T |
| DAC BS-4 | BaSi/AB | 30 → 40 → 48 | Transport measurements, low-T XRD, detailed XRD map at 300 K | $Fm\bar{3}m$-$BaSiH_{8-x}$ $hP$-$BaSiH_x$<br><br>Semiconducting behavior at ~30 GPa evolving into poor metallicity above ~40 GPa; negative magnetoresistance, PPC, photovoltaic effect at 48 GPa |
| DAC BS-NMR | BaSi/AB | 28 → 42** | $^1$H NMR, XRD | $P2/m$-$BaSiH_{x\approx 8}$<br><br>Long spin–lattice relaxation times consistent with degenerate semiconductor |

| | | | | | or poor metal with low density of states at the Fermi level |
|---|---|---|---|---|---|
| DAC BS-5 | BaSi/AB | 55 → 21* | Transport measurements before and after laser heating, three-probe vdP scheme | | $dR/dT > 0$, metallic behavior above 78 K |
| DAC BS-6 | BaSi/AB | 10 | XRD at 300 K, SC XRD | | $P1$-BaSiH$_{6-8}$ Formation of distorted, molecular polyhydride BaSiH$_{6-8}$ at low pressure; exact hydrogen content difficult to determine |

*Pressure dropped after laser heating
**Pressure increased after NMR and XRD experiments

*Photovoltaic effect*

The Shockley equation for a diode[6] in the case of photocurrent and infinite load resistance leads to the following equation for the photo-EMF of a semiconducting sample:

$$\mathcal{E}_{ph} \propto \frac{k_B T}{e} \ln\left(1 + \frac{I_{ph}}{I_0}\right), \quad (S3)$$

where $k_B$ – is the Boltzmann constant, $T$ – is temperature, e – is the electron charge, $I_0$ - is the dark saturation current, $I_{ph} = e\eta\Phi = e\eta IA/h\nu$ – is the photocurrent as the product of quantum efficiency (η), illumination intensity ($I$), and semiconductor area ($A$), divided by the photon energy ($h\nu$).

In the case of low light intensity ($I_{ph} \ll I_0$), formula (S3) becomes linearly dependent on the irradiation intensity:

$$\mathcal{E}_{ph} \propto \frac{k_B T}{e} \ln\left(1 + \frac{I_{ph}}{I_0}\right) \approx \frac{k_B T}{e I_0} I_{ph} \propto \frac{k_B T}{h\nu} IA. \quad (S4)$$

*Weak localization*

According to Refs. [7,8], the change in conductivity Δσ due to weak localization in the three-dimensional case is expressed by the following formula

$$\Delta\sigma = \frac{e^2}{2\pi^2 \hbar}\sqrt{\frac{eB}{\hbar c}} f_{3D}\left(\frac{\hbar c}{4eBL_{Th}^2}\right) = \frac{e^2}{2\pi^2 \hbar}\sqrt{\frac{eB}{\hbar c}} \left( \sum_{n=0}^{\infty} 2\sqrt{n+1+\frac{\hbar c}{4eBL_{Th}^2}} - 2\sqrt{n+\frac{\hbar c}{4eBL_{Th}^2}} - \frac{1}{\sqrt{n+0.5+\frac{\hbar c}{4eBL_{Th}^2}}} \right), \quad (S5)$$

where $B$ – this is the magnetic field induction, while $L_{Th} = \sqrt{\tau_\varphi \tau_{tr} V_F^2/3}$ – is characteristic inelastic scattering distance. Asymptotic expressions for the quantum correction to the conductivity are

$$\Delta\sigma = \begin{cases} 0.605 \frac{e^2}{2\pi^2 \hbar}\sqrt{\frac{eB}{\hbar c}}, & \frac{\hbar c}{4eBL_{Th}^2} \ll 1 \text{ (high field)} \\ \frac{e^2}{96\pi^2 \hbar}\sqrt{\frac{eB}{\hbar c}}\left(\frac{\hbar c}{4eBL_{Th}^2}\right)^{-3/2}, & \frac{\hbar c}{4eBL_{Th}^2} \gg 1 \text{ (low field)} \end{cases}. \quad (S6)$$

In these formulas, the phase decoherence time $\tau_\varphi$ gives us the temperature dependence of the magnetoresistance, which is determined mainly by phonon scattering at high temperatures.

*Quantum ESPRESSO calculations*

Phonon and electron phonon calculations were performed using Density Functional Theory and Density Functional Perturbation Theory as implemented in Quantum ESPRESSO. We employed Optimized Norm-Conserving Vanderbilt pseudopotentials and the Perdew-Burke-Ernzerhof exchange correlation functional. To compute the ground-state charge density, the Kohn-Sham wavefunctions were expanded in plane waves up to an energy cutoff of 100 Ry, and integrals over the Brillouin zone were performed on a 16x16x16 k-grid, while treating partial occupancies with a Methfessel-Paxton smearing of 0.015 Ry. The phonon properties were computed by calculating the dynamical matrices on a 8x8x8 q-grid. Dispersions were then Fourier interpolated. The integral of the electronic part of the electron-phonon matrix elements was performed over a 24x24x24 k-grid, computed non-self-consistently, with a Gaussian smearing of 0.015 Ry.

## II. Optical microscopy

In this section, we present optical microscopy data for BaSi/BaSiH$_x$ samples that were not presented in the main text or in Supporting Information, along with results from other measurements.

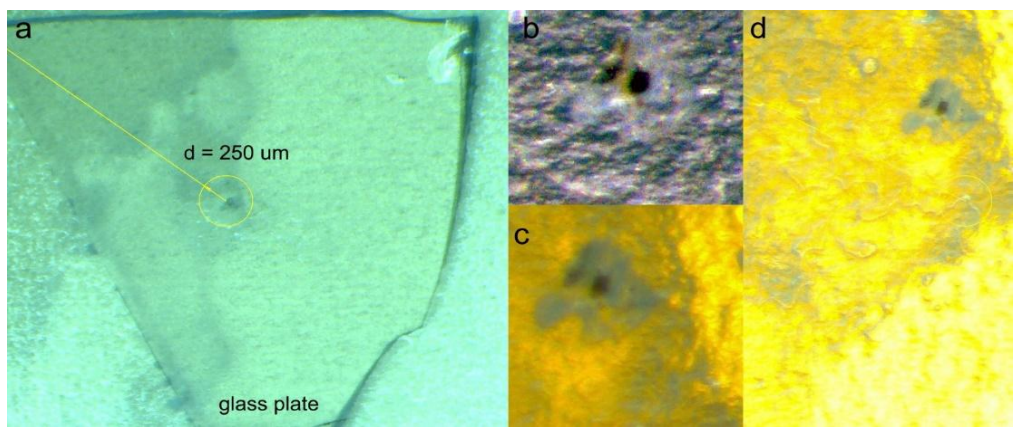

**Figure S1.** Optical photographs (a-d) at different magnifications of the cubic BaSiH$_8$ sample after opening the DAC BS-1 and transferring the particle to a glass plate in an Ar glove box. The sample (dark spot in the center) shows no visible signs of decomposition, remains opaque, but possibly reacts with an organic adhesive used to fix the sample. Single-crystal X-ray diffraction study of this sample reveals the presence of cubic Si and Si$_3$N$_4$ [9] (see section IV. Single-crystal X-ray diffraction data).

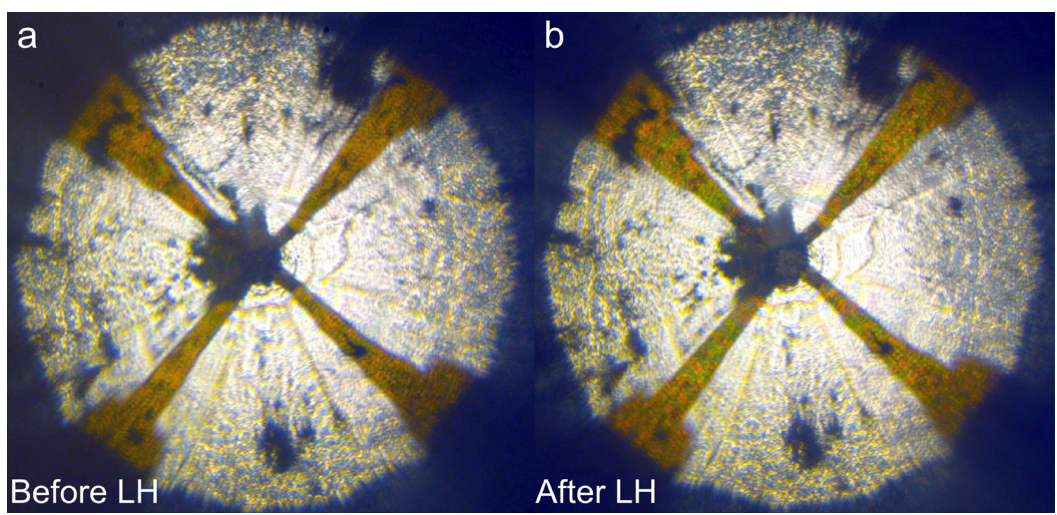

**Figure S2.** Optical photograph of a DAC BS-3 with 50 µm culet diameter: (a) before laser heating at 148 GPa, and (b) after laser heating. The pressure after laser heating dropped to 142 GPa.

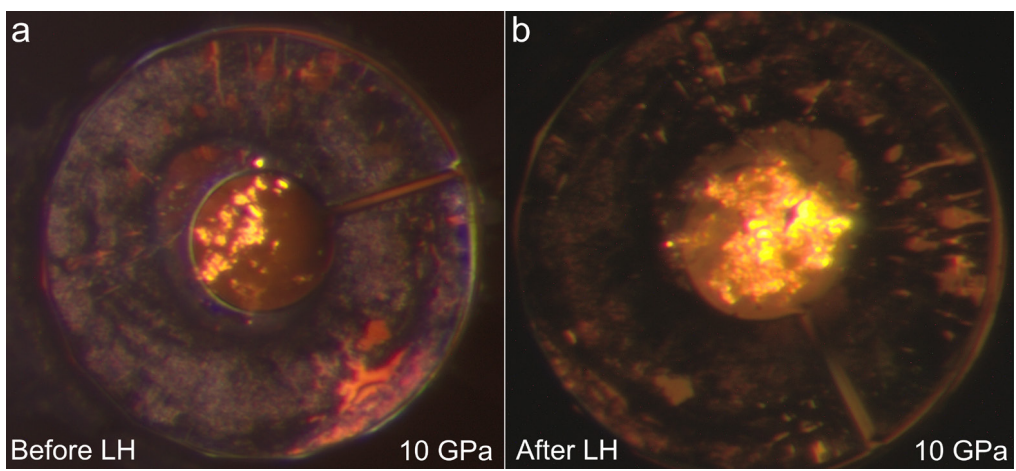

**Figure S3.** Optical photograph of the DAC BS-6 sample at 10 GPa (a) before laser heating (BiSi/AB) and (b) after laser heating. The Ta/Au Lenz lens and the irregularly shaped sample (dark), which becomes almost completely transparent after laser heating, are visible.

# III. Additional powder X-ray diffraction data

Powder XRD analysis of the starting BaSi material, performed at BL12B2 (SPring-8, 2023), confirms the presence of *Cmcm*-BaSi as the main product of the reaction. The diffraction pattern is complex (Figure S4), indicating the presence of other barium silicides (possibly $BaSi_2$, and $Ba_3Si_4$) as well as the oxidation product, $Ba_2SiO_4$, also detected by single-crystal diffraction methods (see section IV).

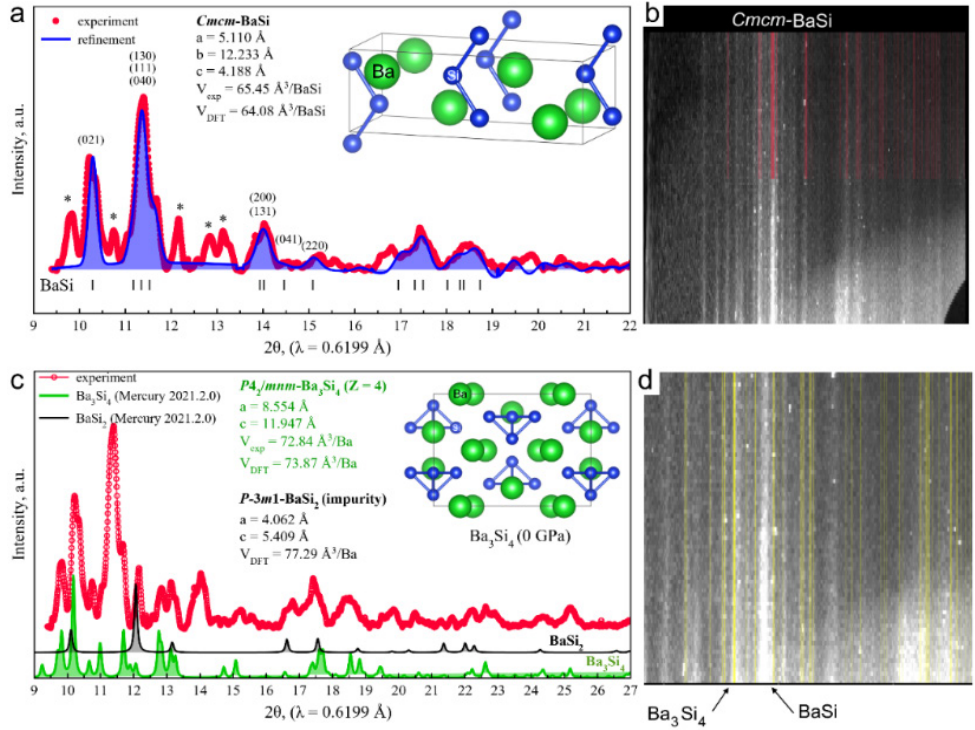

**Figure S4.** X-ray diffraction analysis and crystal structure of BaSi starting material at ambient conditions. (a) Complex powder X-ray diffraction pattern of the parent BaSi mixture showing experimental data (red) and Le Bail refinement (blue) for the *Cmcm*-BaSi phase (a = 5.110 Å, b = 12.233 Å, c = 4.188 Å, $V_{exp}$ = 65.45 Å$^3$/BaSi, $V_{DFT}$ = 64.08 Å$^3$/BaSi). Miller indices are labeled for major reflections, with tick marks indicating calculated peak positions. Asterisks (*) denote peaks from impurity or unidentified phases. Inset: crystal structure of orthorhombic *Cmcm*-BaSi showing Ba atoms (green) and Si atoms (blue) in the unit cell. (b) Raw diffraction image ("cake") of starting BaSi material with red lines highlighting the primary diffraction of *Cmcm*-BaSi. (c) The same powder X-ray diffraction pattern revealing presence of by-products: experimental data (red curve) compared to the calculated patterns for $Ba_3Si_4$ (green spectrum, $P4_2/mnm$-$Ba_3Si_4$ with Z = 4, a = 8.554 Å, c = 11.947 Å, $V_{exp}$ = 72.84 Å$^3$/Ba, $V_{DFT}$ = 73.87 Å$^3$/Ba) and $BaSi_2$ (black spectrum, $P$-$3m1$-$BaSi_2$ impurity phase, a = 4.062 Å, c = 5.409 Å). Inset: crystal structure of $P4_2/mnm$-$Ba_3Si_4$. (d) 2D raw diffraction image showing distinct diffraction lines from $Ba_3Si_4$ (yellow lines) and *Cmcm*-BaSi phases.

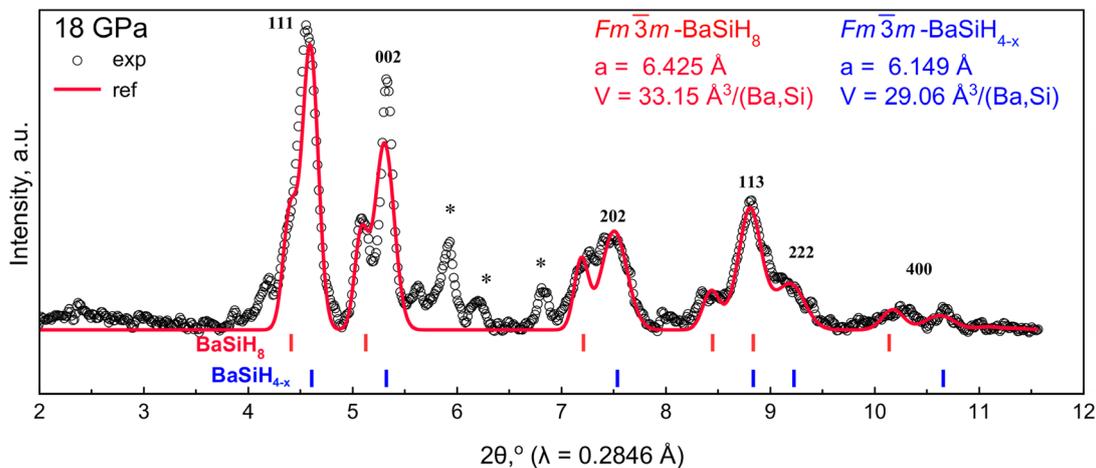

**Figure S5.** Experimental diffraction pattern and Le Bail refinement of the unit cell parameters of cubic $BaSiH_8$ and $BaSiH_{4-x}$ at 18 GPa in DAC BS-2. Black circles represent experimental data, and the red curve represents the refinement. Asterisks mark reflections from $Ba_2SiO_4$.

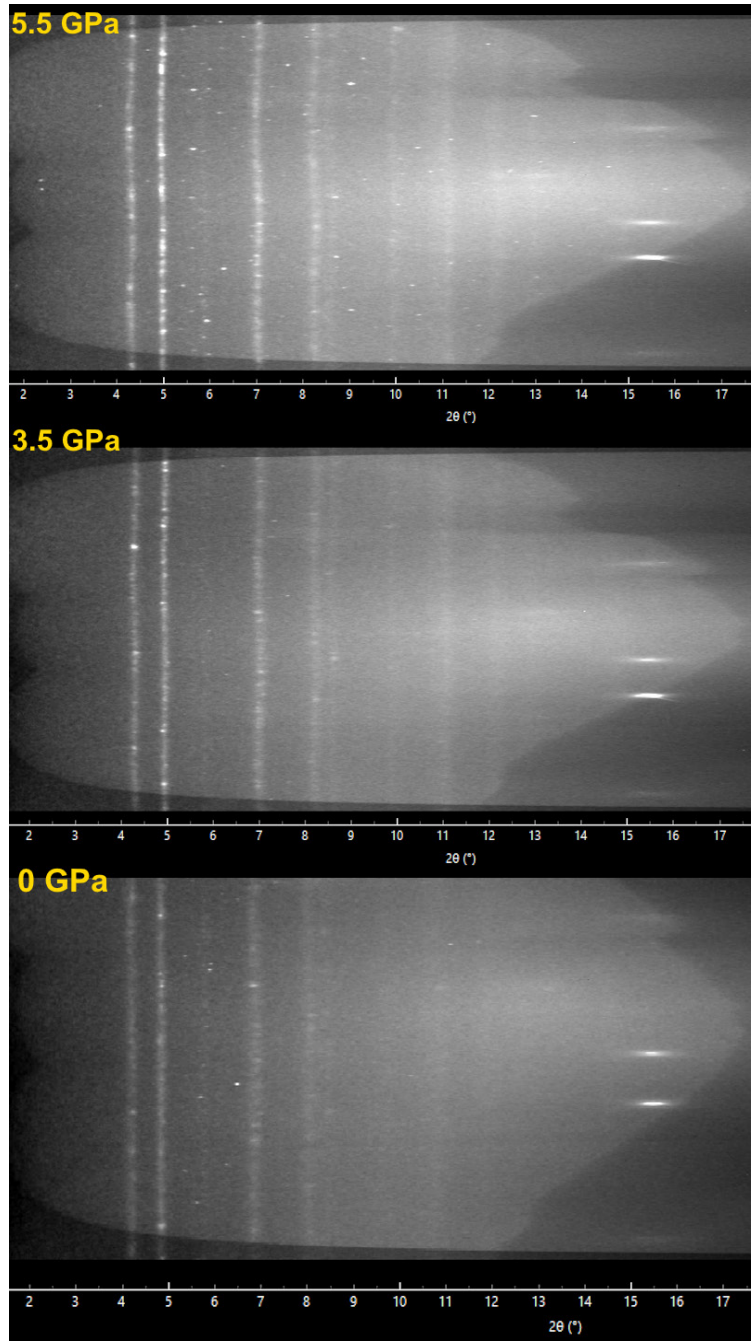

**Figure S6**. Raw X-ray diffraction images ("cake", λ = 0.2846 Å) of cubic BaSiH$_8$ in the DAC BS-2 obtained during decompression of the cell at 5.5, 3.5, and 0 GPa. Diffraction patterns do not qualitatively change during decompression, which speaks in favor of a fairly high stability of the $Fm\bar{3}m$-BaSiH$_8$.

**Table S2.** Unit cell parameter and volume of cubic BaSiH$_8$, as well as BaSiH$_{4-x}$ (refined as $Fm\bar{3}m$).

| Pressure, GPa | a, Å | V(BaSiH$_8$), Å$^3$/(Ba,Si) | V(BaSiH$_{4-x}$), Å$^3$/(Ba,Si) |
|---|---|---|---|
| 31 | 6.236 | 30.31 [a] | – |
| 30 | 6.226 | 30.17 | – |
| 18 | 6.475 | 33.93 [b] | 29.28 |
| 10 | *distorted* | 36.67 – 37.17 [d] | – |
| 7 | 6.722 | 37.97 | – |
| 5.5 | 6.715 | 37.85 [c] | 33.25 |
| 3.8 | 6.745 | 38.36 [c] | 34.91 |
| 0 | 6.945 | 41.87 [c] | – |
| 0 | 6.906 | 41.89 | – |

[a] SSRF experiment in 2023; [b] ID 11 ESRF in 2024; [c] PETRA III in 2024; [d] SPring-8 2025;

**Table S3.** Ideal structural models of cubic BaSiH$_6$ and pseudocubic $C$2-BaSiH$_4$ found by evolutionary search using USPEX code.

| Formula (Pressure) | POSCAR |
|---|---|
| BaSiH$_6$ (50 GPa) 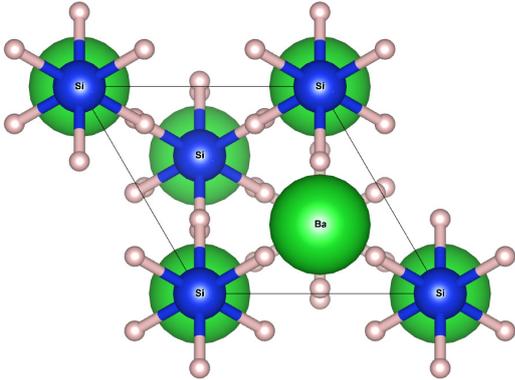 BaSi sublattice is $Fm\bar{3}m$ | EA57    4.179  4.180  4.181 61.11 118.89 118.89<br>Sym.group:  166<br>1.0<br>   4.179147   -0.006276    0.042149<br>  -2.014347    3.662944   -0.006450<br>  -2.053007    1.182073    3.445162<br> Ba  Si  H<br>  1   1   6<br>Direct<br>   0.725813    0.882053    0.635283<br>   0.225822    0.382041    0.135204<br>   0.120288    0.487417    0.700647<br>   0.791130    0.276570    0.029963<br>   0.660569    0.487425    0.240521<br>   0.331173    0.276833    0.569669<br>   0.331544    0.816789    0.029758<br>   0.120372    0.947386    0.240665 |
| BaSiH$_4$ (20 GPa) 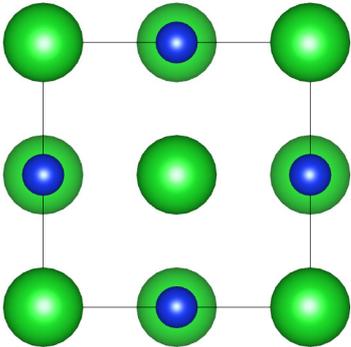 BaSi sublattice is $Fm\bar{3}m$ | $C$2-Ba2Si2H8<br>  1.0<br>   4.41350   0.00340  -0.10013<br>   0.10364   4.41073   0.08271<br>  -0.11929   0.14896   6.18429<br>  Ba  Si  H<br>   2   2   8<br>Direct<br>  0.31174 0.13819 0.24834<br>  0.81485 0.64115 0.74842<br>  0.29773 0.12402 0.74836<br>  0.81121 0.63762 0.24843<br>  0.10748 0.67351 0.11374<br>  0.62684 0.81802 0.06751<br>  0.33687 0.42385 0.86454<br>  0.78259 0.33835 0.10958<br>  0.59713 0.16333 0.63188<br>  0.98984 0.45135 0.42936<br>  0.84779 0.93344 0.38353<br>  0.51159 0.60950 0.38695 |

During XRD experiment with DAC BS-1 we also found $P6/mmm$-Si (Si-V) with a = 2.476 Å, c = 2.332 Å, V = 12.38 Å$^3$/atom [10].

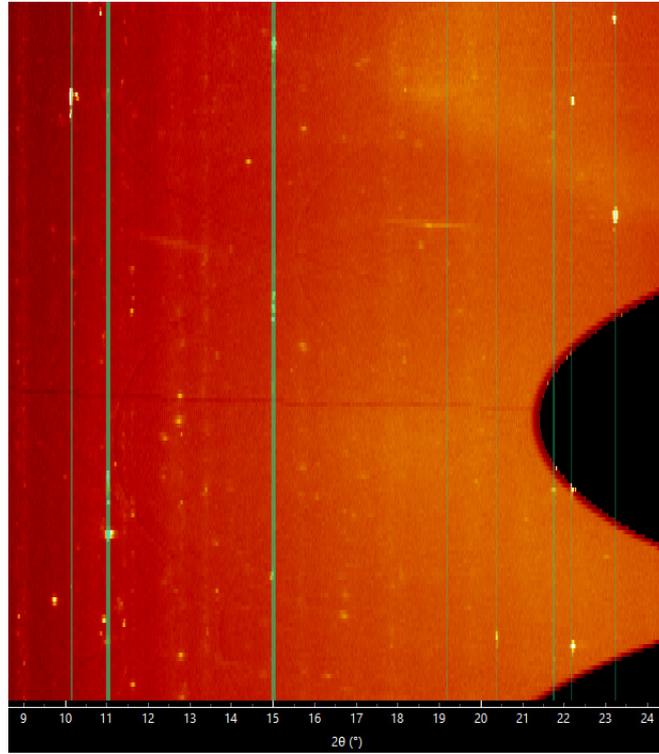

**Figure S7.** Raw X-ray diffraction image of a primitive hexagonal silicon (Si-V) [10,11] impurity in the Ba-Si hydrides mixture at 30 GPa. The image displays raw XRD data collected from the sample DAC BS-1 after laser heating. Bright, discrete diffraction spots, characteristic of single-crystal-like grains, are observed. The green vertical lines highlight the positions of these Si-V diffraction peaks along the 2θ axis. Wavelength is 0.4124 Å.

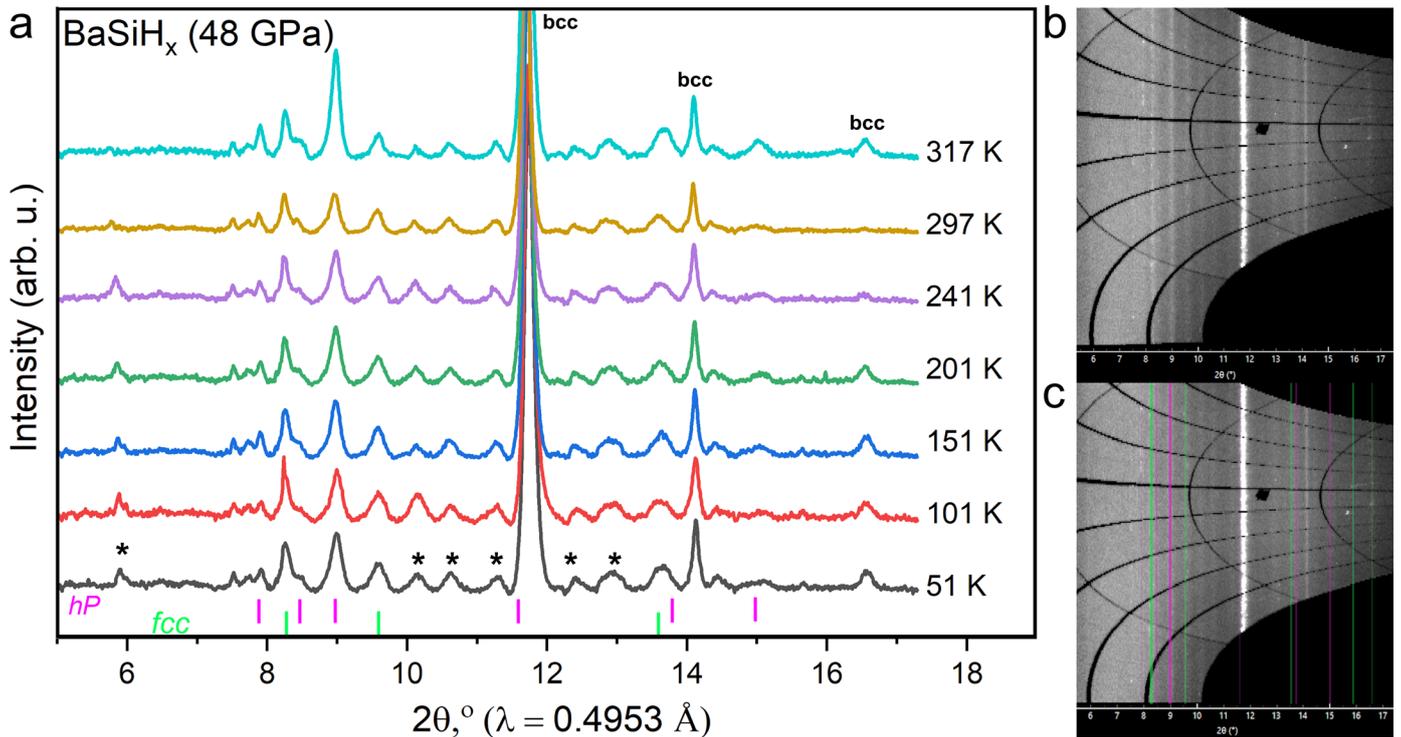

**Figure S8.** Low-temperature X-ray diffraction study of the DAC BS-4 sample in the range from 317 K to 51 K at 48 GPa using a wide X-ray beam ($d = 50$ μm). Measurements were performed at Elettra synchrotron. (a) Stack of diffraction patterns recorded at different temperatures containing reflections from the cubic (*fcc*) and hexagonal (*hP*) phases, which have a unit cell volume close to $BaSiH_8$. No phase transitions are observed in the range of 51–317 K. The asterisk marks uninterpreted reflections; "*bcc*" corresponds to the metal components of the DAC and/or tantalum electrodes. (b, c) Raw diffraction images with (c) and without (b) marking of phases used for integration. Green lines indicate the cubic phase, probably $BaSiH_8$; purple lines indicate the hexagonal phase $hP$-$BaSiH_x$.

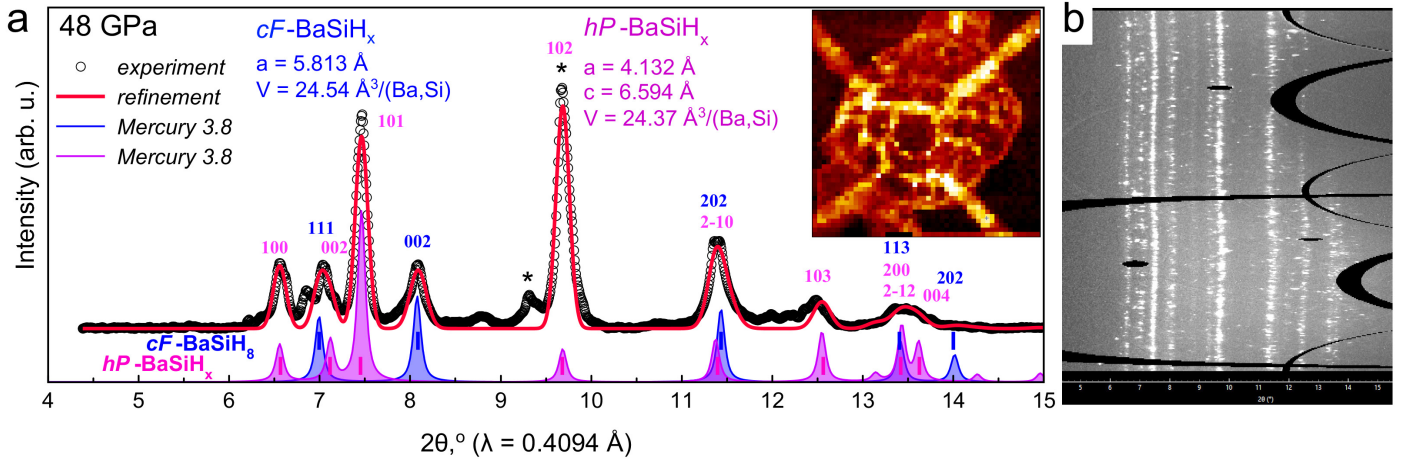

**Figure S9.** X-ray diffraction analysis and spatial XRD mapping of BaSiH$_x$ sample at 48 GPa in DAC BS-4. (a) Le Bail refinement of the *cF*-BaSiH$_x$ and *hP*-BaSiH$_x$ unit cell parameters. Experimental data (black open circles), refinement fit (red line), and calculated XRD patterns (by Mercury 3.8 software) for cubic *cF*-BaSiH$_x$ (x = 6-8, blue) and hexagonal *hP*-BaSiH$_x$ (refined as *P*6$_3$/*mmc*, magenta) are displayed. The cubic phase has refined lattice parameter a = 5.813 Å with unit cell volume V = 24.54 Å$^3$/(Ba,Si) which is about 2-3 Å$^3$/(Ba,Si) lower than expected for BaSiH$_8$. The hexagonal phase (*hP*-BaSiH$_x$) shows lattice parameters a = 4.132 Å and c = 6.594 Å with V = 24.37 Å$^3$/(Ba,Si). Asterisks (*) mark unexplained reflection. The wavelength in this experiment was λ = 0.4094 Å. Top right inset: XRD intensity map 100 ×100 μm$^2$ showing the spatial distribution of the X-ray diffraction signal at 2θ = 10-11° (golden electrodes) with a color scale from dark red/black (low intensity) through orange and yellow to white (highest intensity). (b) Two-dimensional raw X-ray diffraction image used for integration.

**Table S4.** Different BaSiH$_x$ phases found experimentally in DACs BS-1 and 4. When calculating the volumes of the unit cell, we assumed that in all hydrides the ratio of Ba and Si atoms was 1:1.

| Pressure, GPa | a, Å  | c, Å  | V, Å$^3$/(Ba,Si) | Comment         |
| ---           | ---   | ---   | ---              | ---             |
| 48            | 4.132 | 6.594 | 24.37            | *hP*, DAC BS-4  |
| 48            | 5.813 | -     | 24.54            | *cF*, DAC BS-4  |
| 31            | 4.420 | 7.430 | 31.4             | *hP*-I, DAC BS-1  |
| 31            | 4.460 | 7.176 | 30.9             | *hP*-II, DAC BS-1 |

In the SPring-8 (BL10XU) experiment in 2025 at 10 GPa, we studied a BaSi/AB sample before and after laser heating. The DAC was equipped a Ta/Au Lenz lens and a Ta/W/Ta$_2$O$_5$ gasket. XRD mapping a 100×100 μm$^2$ region with a 16×16 grid (256 points) reveals the presence of a Ta lens/gasket and Ta$_2$O$_5$ signals at the edge of the studied region (Supporting Figure S11). XRD pattern of Ta$_2$O$_5$ is in agreement with literature [12]. The complex mixture of the initial barium silicides ("BaSi") located in the center of the DAC's chamber was not analyzed in detail. A series of sequential IR laser pulses (wavelength 1.04 μm, pulses of 5, 7, 10 W, the pulse duration was about 0.5 seconds) leads to the formation of possibly low-symmetry coarse-crystalline phase *P*1-BaSiH$_x$ (x = 2 ±1, Supporting Figure S12), distorted *P*1-BaSiH$_{6-8}$ (edge of the heated region, Supporting Figure S13, Figure S15), as well as other unidentified products, and the oxide *Pnma*-Ba$_2$SiO$_4$ (Supporting Figure S14). In general, laser heating was ineffective immediately after the first pulse due to the formation of a transparent window: additional laser pulses of higher power no longer affect the composition of the reaction mixture studied by XRD. For this reason, the SPring-2025 experiment was completed by intensely heating the entire diamond cell, followed by re-mapping a 100×100 μm$^2$ region with the same 16×16 grid.

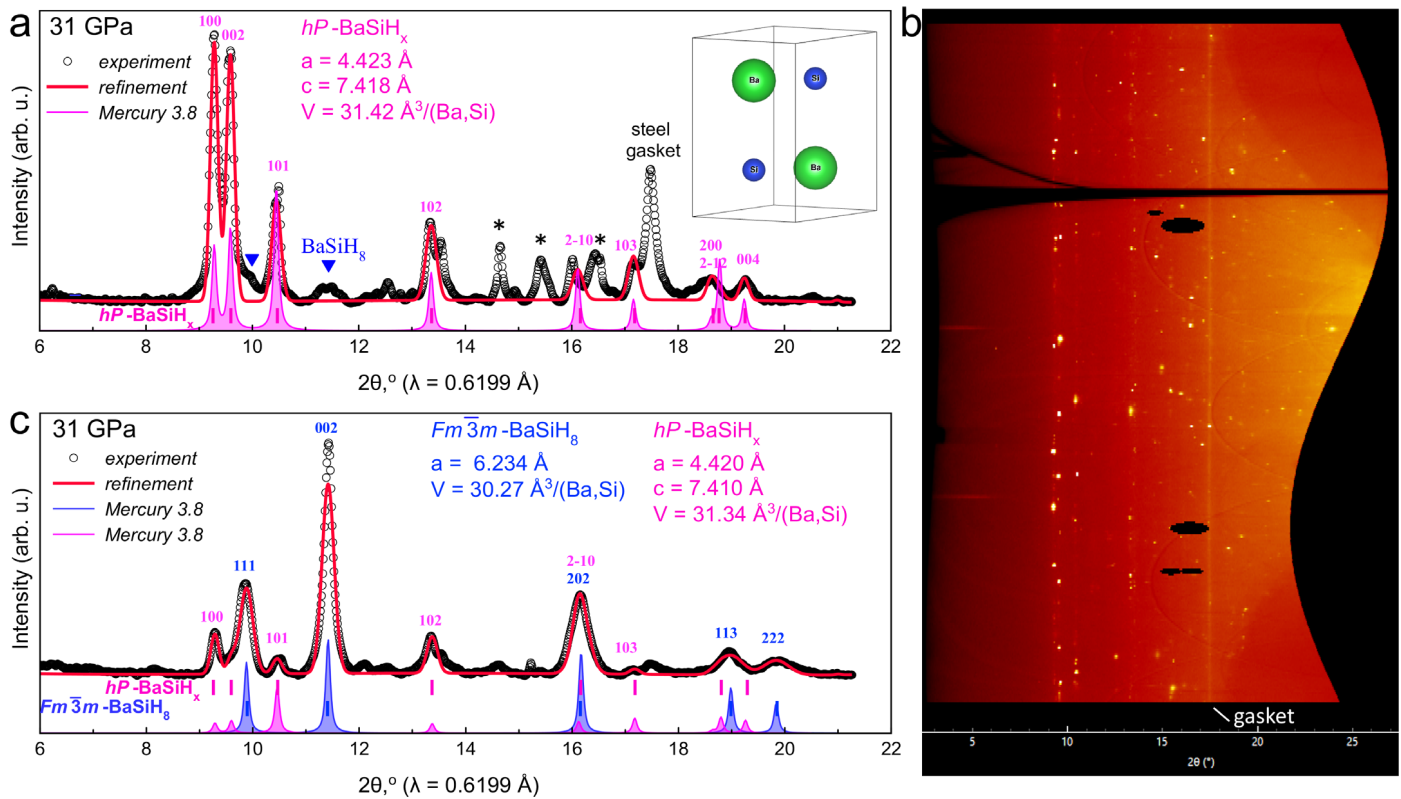

**Figure S10.** High-pressure X-ray diffraction analysis of hexagonal BaSiH$_x$ phase at 31 GPa (DAC BS-1). (a) Powder X-ray diffraction pattern at 31 GPa showing experimental data (black circles) and Le Bail refinement (red line) for the $P6_3/mmc$ ($hP$) BaSiH$_x$ phase (magenta, a = 4.423 Å, c = 7.418 Å, V = 31.42 Å$^3$/(Ba,Si)). Calculated pattern from Mercury 3.8 (pink) is shown for comparison. Miller indices label the main reflections of $hP$-BaSiH$_x$. The pattern reveals residual BaSiH$_8$ phase (blue triangle marker) and impurity peaks from steel gasket, as well as unknown phases (*, asterisks). Inset: possible structure of Ba-Si sublattice of BaSiH$_x$, which may explain increased intensity of reflections 100 and 002. Ba atoms are green, and Si atoms are blue. (b) Two-dimensional raw diffraction image ("cake") at 31 GPa. (c) Experimental XRD and Le Bail refinement at different point of DAC BS-1, comparing experimental data (black circles) with two hydride phases: cubic BaSiH$_8$ (blue spectrum, a = 6.234 Å, V = 30.27 Å$^3$/(Ba,Si)) and $hP$-BaSiH$_x$ (magenta, a = 4.420 Å, c = 7.410 Å, V = 31.34 Å$^3$/(Ba,Si)). The Le Bail refinement (red line) and calculated patterns from Mercury 3.8 demonstrate that these two structural models can mostly explain the observed diffraction peaks.

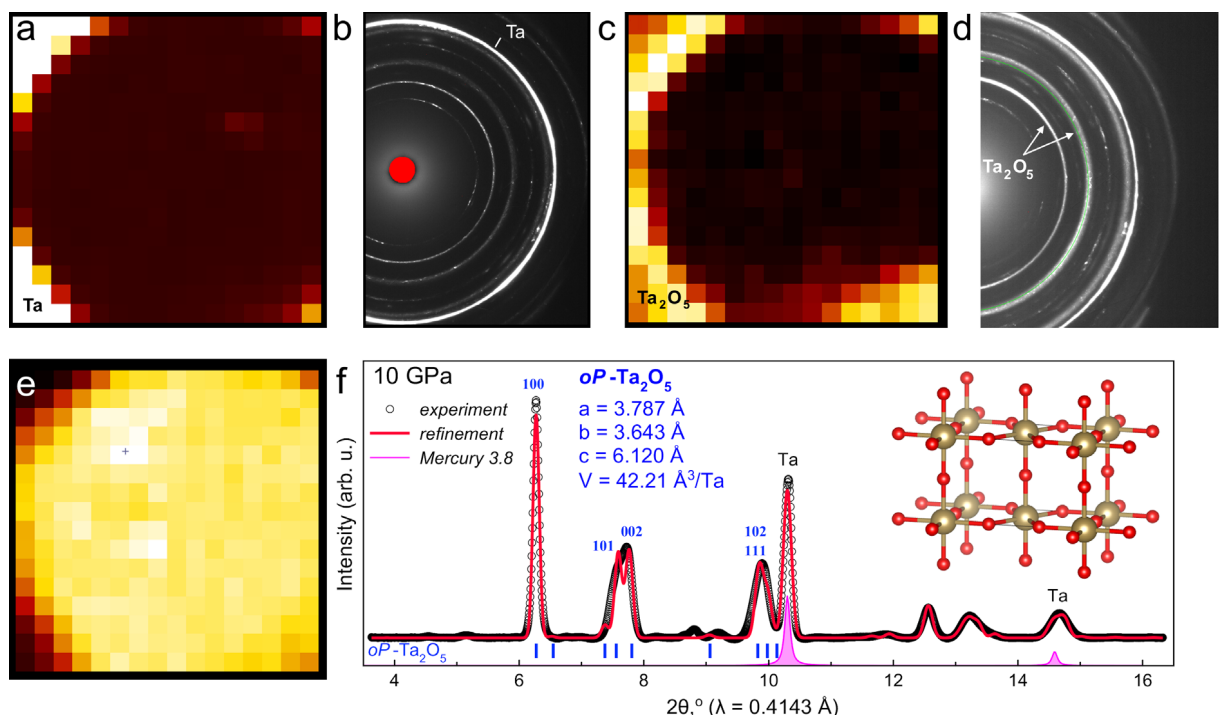

**Figure S11.** X-ray diffraction mapping of the gasket material at 10 GPa in DAC BS-6. (a) XRD map showing the spatial distribution of tantalum (Ta, 110 reflection) in the chamber area, with warm colors indicating higher Ta concentration

and dark colors indicating lower concentration. (b) Two-dimensional raw X-ray diffraction pattern of the Ta metal (gasket and Lenz lens material), showing characteristic Debye-Scherrer rings. (c) XRD map showing the spatial distribution of $Ta_2O_5$ phase in the sample chamber using the 100 reflection, with distinct regions of higher concentration (bright) and lower concentration (dark). (d) Two-dimensional raw X-ray diffraction pattern showing diffraction rings from the $oP$-$Ta_2O_5$ phase [12], with the $Ta_2O_5$ main XRD rings labeled. The pattern demonstrates the polycrystalline nature of the oxide phase. (e) Qualitative XRD mapping of the culet's center, showing bright regions (yellow-white) corresponding to high diffraction intensity from BaSi starting material, and darker regions (red-brown) with lower BaSi concentration. (f) Le Bail refinement of the integrated XRD pattern for orthorhombic $oP$-$Ta_2O_5$ around 10 GPa. Experimental data (black open circles), refinement fit (red line), and calculated pattern from Mercury 3.8 (for Ta only, magenta) are shown. Major diffraction peaks of $oP$ structure are indexed. Right inset: possible crystal structure of $oP$-$Ta_2O_5$, showing tantalum atoms (brown spheres) coordinated by oxygen atoms (red spheres). The wavelength used was $\lambda = 0.4143$ Å.

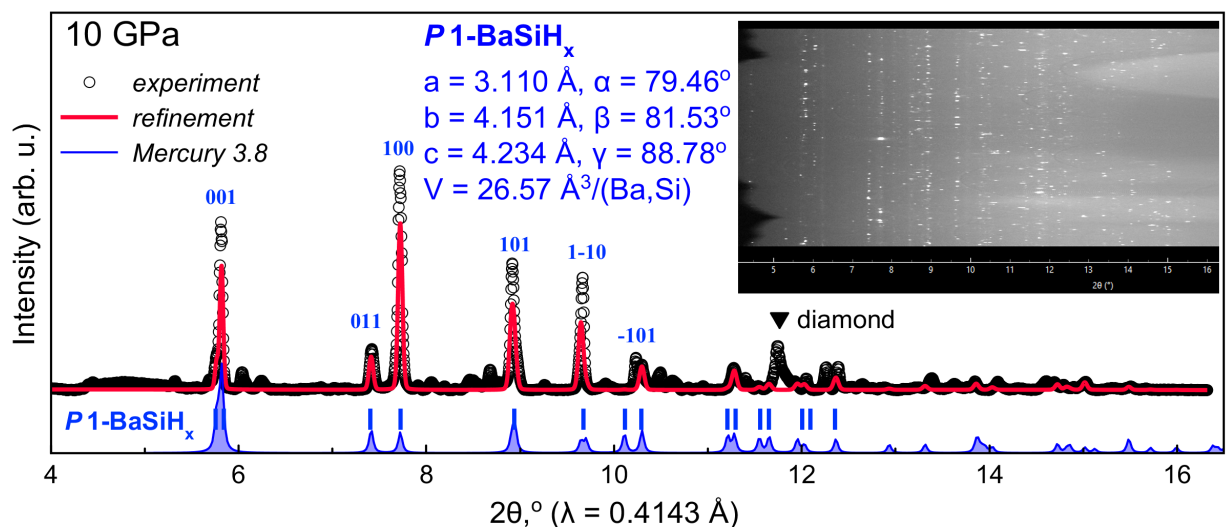

**Figure S12.** X-ray diffraction pattern and Le Bail refinement of $P1$-$BaSiH_x$ unit cell at 10 GPa (DAC BS-6). The experimental data (black open circles) and Le Bail refinement fit (red line, Jana2006) are shown. The calculated pattern from Mercury 3.8 software is displayed as blue vertical lines below the experimental data. Key diffraction peaks are indexed. A possible diamond diffraction peak was also marked by an inverted triangle. The refined lattice parameters are a = 3.110 Å ($\alpha = 79.46°$), b = 4.151 Å ($\beta = 81.53°$), c = 4.234 Å ($\gamma = 88.78°$), with a unit cell volume of V = 26.57 Å³/(Ba,Si). Inset: raw diffraction image used for integration. The wavelength used for XRD data collection was $\lambda = 0.4143$ Å.

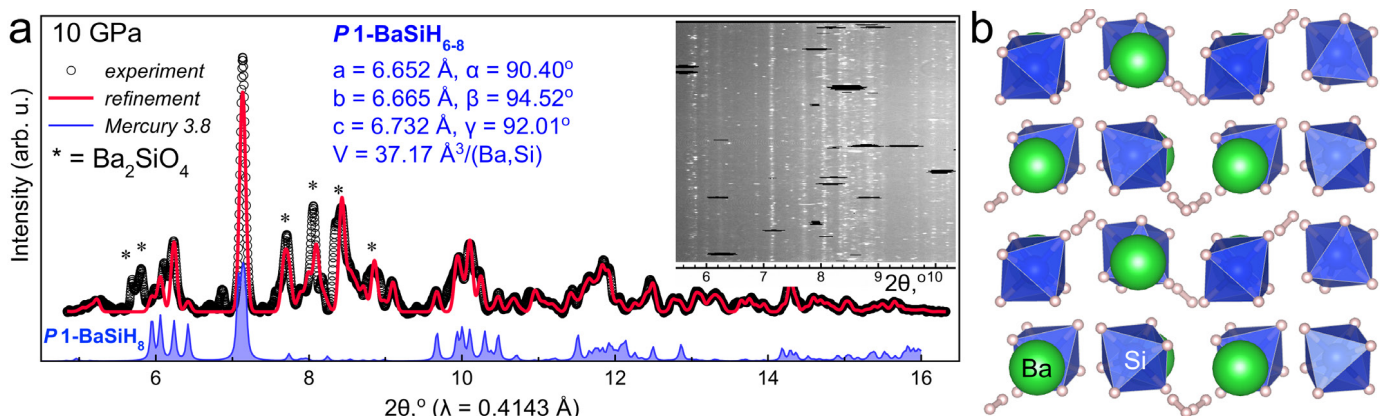

**Figure S13.** X-ray diffraction pattern, Le Bail refinement, and possible crystal structure of $P1$-$BaSiH_{6-8}$ at 10 GPa (DAC BS-6). (a) Experimental X-ray diffraction pattern (black open circles) with Le Bail refinement fit (red line) and the calculated pattern from Mercury 3.8 (blue line below) for the $P1$-$BaSiH_{6-8}$ phase. The refined lattice parameters are a = 6.652 Å ($\alpha = 90.40°$), b = 6.665 Å ($\beta = 94.52°$), c = 6.732 Å ($\gamma = 92.01°$), with unit cell volume V = 37.17 Å³/(Ba,Si). This volume corresponds to $BaSiH_x$ composition with x = 6–8. Asterisks (*) mark diffraction peaks from the oxidation product $Ba_2SiO_4$ which presents in a significant amount and explains X-ray reflections at 7.5-9.2°. The X-ray wavelength used was $\lambda = 0.4143$ Å. Inset: complex two-dimensional raw X-ray diffraction image used for integration. (b) Crystal structure of $P1$-$BaSiH_8$ showing the arrangement of Ba atoms (green spheres), Si atoms (indicated in blue polyhedra labeled "Si"), and H atoms (pink spheres). The Si atoms are coordinated by hydrogen in polyhedral arrangements (blue polyhedra, [$SiH_6$]), while Ba atoms occupy interstitial positions together with unknown number of $H_2$ molecules

**Table S5.** Structural model of distorted $P1$-BaSiH$_8$ used for the Le Bail refinement and found by fixed composition search via USPEX code at 25 GPa.

| Formula (Pressure) | POSCAR |
|---|---|
| $P1$-BaSiH$_8$ (25 GPa)<br><br>BaSi sublattice is also $P1$<br><br>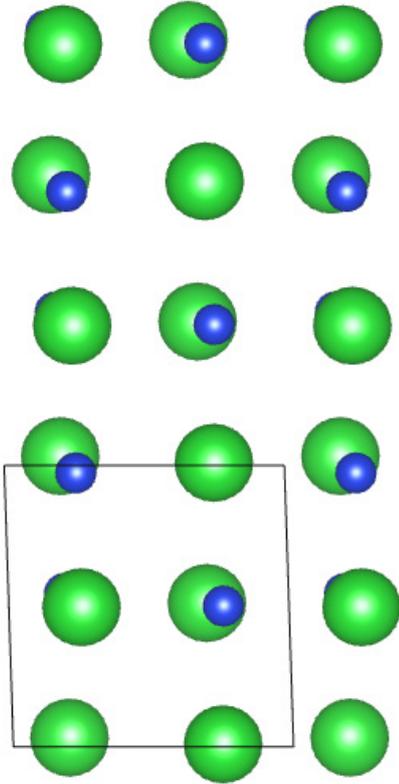<br><br>Ba – green spheres<br>Si – blue spheres | BaSiH8_1_p25gpa<br>1.0<br>  6.447242437  -0.297468349  0.250026901<br>  0.321280125  6.446439584  -0.234343068<br>  0.063326154  0.021861527  6.385609017<br>Ba Si H<br>4 4 32<br>Direct<br>  0.7546358532  1.0322468559  0.2012160093<br>  0.7778513296  0.5104554013  0.7070583535<br>  0.2545525816  0.0086439300  0.7484594463<br>  0.2789996629  0.4954814768  0.2589986333<br>  0.2389351146  0.9725430366  0.2546623405<br>  0.2926887810  0.4996743921  0.7658875225<br>  0.7259463574  -0.0164173753  0.6890715520<br>  0.7534191655  0.5446122298  0.2001940394<br>  0.1410977772  0.3916106873  0.6110215108<br>  0.6597216505  0.1557957715  0.8529768098<br>  0.4390273301  0.6063586586  0.9184549203<br>  0.3770179929  0.1469244860  0.3600236673<br>  0.3906769529  0.6461337372  0.5966104259<br>  0.0631236275  0.0148786800  0.4207727528<br>  0.1172761909  0.6532159292  0.8523419549<br>  0.0973244828  0.8109491475  0.1434866134<br>  0.6570277123  0.3799266142  0.3631331515<br>  0.7860347443  0.8241045572  0.5129831421<br>  0.9093129918  0.6228136505  0.3546148213<br>  0.1436156064  0.1402765105  0.1157680632<br>  0.4528037025  0.3309806570  0.6778876017<br>  0.3435680997  0.8289369197  0.4054815386<br>  0.6109502986  0.4405642782  0.0452998017<br>  0.8950113962  -0.1040209867  0.8357844367<br>  0.5855640961  0.6988005766  0.2945902983<br>  0.8860405258  0.1395313187  0.6015045374<br>  0.5635661334  0.0677623107  0.5435813945<br>  0.2081092340  0.3482237485  0.9273658007<br>  0.8258241427  0.6992871186  0.0334559112<br>  0.5778757240  0.8290237464  0.8009060396<br>  0.9278738932  0.3947846826  0.1118456830<br>  0.4124644639  0.9062270549  0.0838033638<br>  -0.0041147064  0.2698683344  0.4378578705<br>  -0.0458651286  0.3288223661  0.3467235013<br>  0.1020157558  0.6998380010  0.5859515158<br>  1.0357723293  0.7631973483  0.5152818879<br>  0.4165635854  0.1794754247  0.0727154264<br>  0.4861291645  0.2501270322  0.0069184609<br>  0.9282113511  0.1834907453  0.8550514301<br>  0.9883340333  0.2514979462  0.9314477694 |

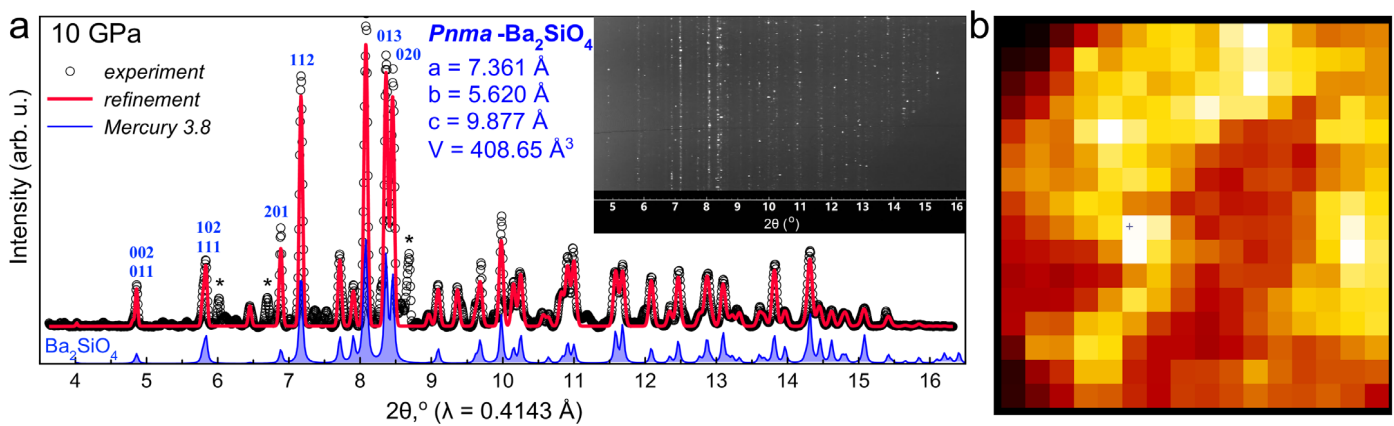

**Figure S14.** X-ray diffraction analysis of impurity of $Pnma$-Ba$_2$SiO$_4$ in DAC BS-6 at 10 GPa. (a) Le Bail refinement of the integrated X-ray diffraction pattern for orthorhombic $Pnma$-Ba$_2$SiO$_4$. Experimental data (black open circles), refinement fit (red line), and calculated by Mercury 3.8 XRD pattern (blue line below) are shown. Major diffraction peaks are indexed. Asterisks (*) mark peaks from unidentified phases. The refined lattice parameters are a = 7.361 Å, b = 5.620 Å, c = 9.877 Å, with unit cell volume V = 408.65 Å³. The wavelength used was λ = 0.4143 Å. Inset: two-dimensional XRD image ("cake") showing Debye-Scherrer rings. (b) XRD mapping (100×100 μm² region with a 16×16

grid, 256 points) showing the Ba$_2$SiO$_4$ distribution across the sample area. The color map ranges from dark red (low intensity of 013, 020 reflections) through yellow to white (high intensity of 013, 020 reflections). The cross (+) symbol corresponds to XRD pattern given in panel (a).

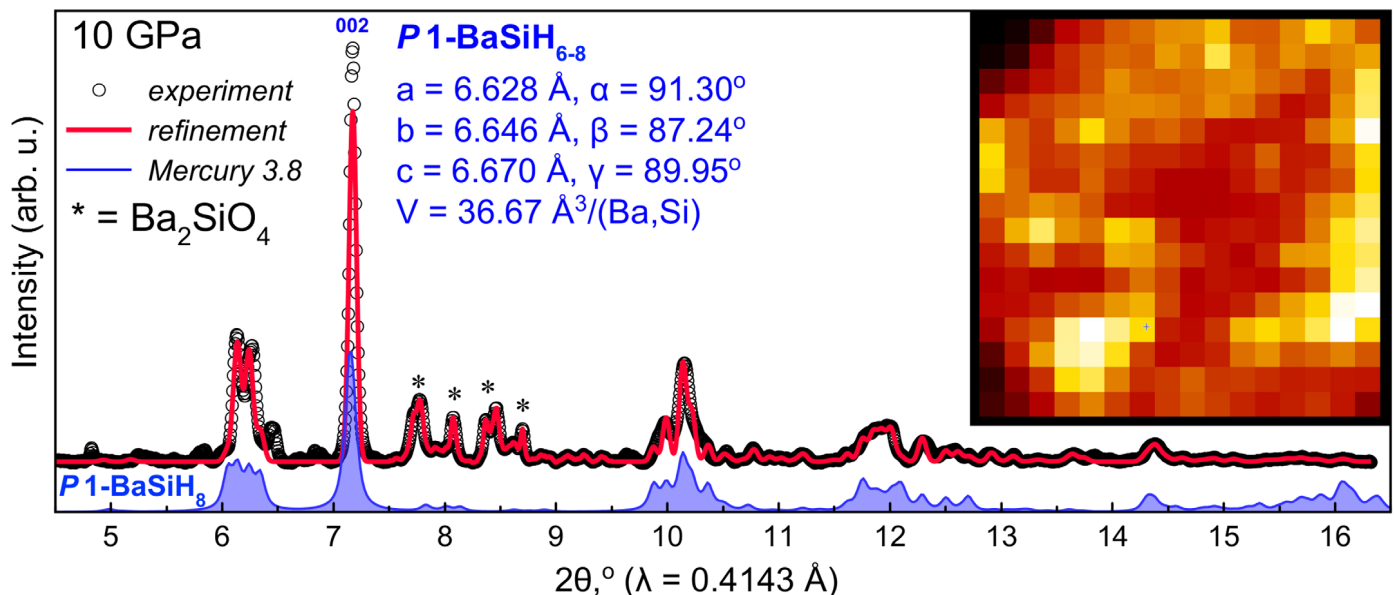

**Figure S15.** X-ray diffraction analysis and spatial mapping of $P$1-BaSiH$_{6-8}$ at 10 GPa in DAC BS-6. Le Bail refinement of the integrated XRD pattern for $P$1-BaSiH$_{6-8}$. Experimental data (black open circles), refinement fit (red line), and calculated pattern from Mercury 3.8 (blue filled pattern) are displayed. The refined lattice parameters are a = 6.628 Å (α = 91.30°), b = 6.646 Å (β = 87.24°), c = 6.670 Å (γ = 89.95°), with unit cell volume V = 36.67 Å$^3$/(Ba,Si). Asterisks (*) mark diffraction peaks from the Ba$_2$SiO$_4$ impurity phase present in the sample. The wavelength used was λ = 0.4143 Å. The triclinic structure is nearly pseudocubic, as indicated by the similar lattice parameters and angles close to 90°. Inset: XRD intensity mapping (100×100 μm$^2$ region with a 16×16 grid, 256 points) showing the spatial distribution of intensity of 002 reflection across the sample area. The color scale ranges from dark red/black (low intensity) through orange and yellow to white (highest intensity).

The presence of a significant amount of Ba$_2$SiO$_4$ in the SPring-8 experiment in 2025 is explained by the aging of the starting BaSi material: despite storing the BaSi in an Ar glovebox for 2 years (2023-2025), it gradually oxidizes, adsorbs oxygen, which, when laser heated, produces a stable Ba$_2$SiO$_4$ phase (Supporting Figure S14).

Final laser heating of the DAC BS-6 sample at 10 GPa leads to the formation of not only Ba$_2$SiO$_4$ and distorted $P$1-BaSiH$_{6-8}$, but also a large number of unknown complex compounds whose structures are difficult to decipher. We will provide just one example: the possible formation of $C$2/$m$-BaSi$_2$H$_4$ containing barium-stabilized disilene (Si$_2$H$_4$) in its structure (Supporting Figure S16). Symmetrical disilene Si$_2$H$_4$ cannot be obtained by chemical synthesis and was only detected in a supersonic molecular beam through the discharge products of silane [13]. In this case, a localized region with high concentration of BaSi$_2$H$_4$ was found in the vicinity of the lens/gasket material (Figure S16b, this may lead to the inclusion of Ta and O in the compound), with a characteristic diffraction pattern of $C$2/$m$-BaSi$_2$H$_{4+x}$ and an unknown impurity with a $Cmmm$ structure (Figure S17, marked by *).

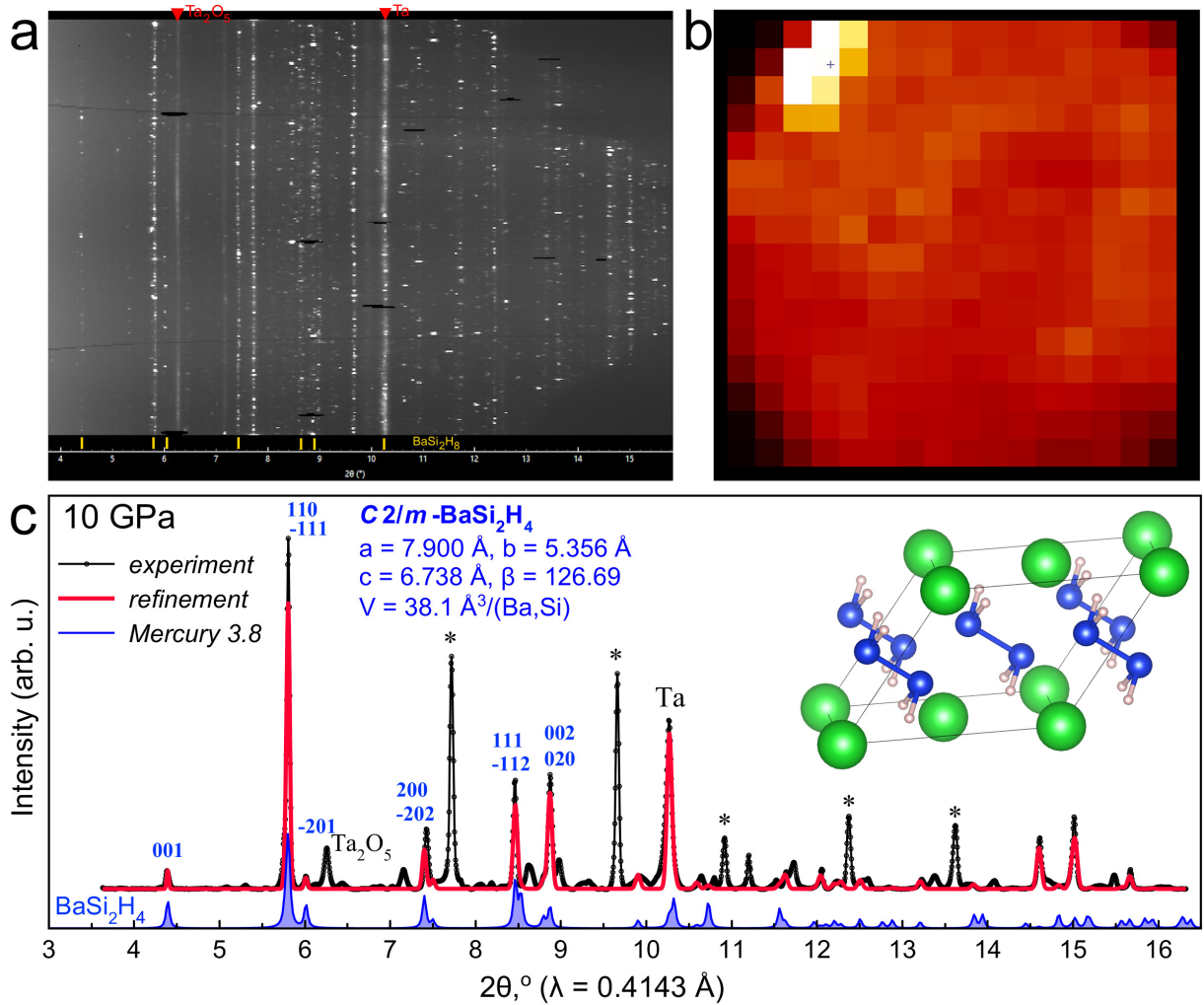

**Figure S16.** X-ray diffraction analysis, spatial mapping, and possible crystal structure of $C2/m$-BaSi$_2$H$_4$ at 10 GPa (DAC BS-6). (a) Two-dimensional raw X-ray diffraction image ("cake"). Yellow tick marks at the bottom indicate the 2θ positions of the main reflections for BaSi$_2$H$_4$, with red triangles at the top marking Ta/Ta$_2$O$_5$ (fine-grained polycrystalline material). (b) XRD intensity mapping displaying the spatial distribution of 110 diffraction reflection across the mapped sample area. The color scale ranges from black (no signal) through dark red and orange to white (highest intensity of 110 reflection). The cross (+) marks the refinement point (panel c). (c) Le Bail refinement of the integrated X-ray diffraction pattern for monoclinic $C2/m$-BaSi$_2$H$_4$. Experimental data (black line), refinement fit (red line), and calculated pattern from Mercury 3.8 (blue filled pattern below) are shown. Asterisks (*) mark peaks from possible impurity of $Cmmm$ phase. The refined lattice parameters are a = 7.900 Å, b = 5.356 Å, c = 6.738 Å, β = 126.69°, with unit cell volume V = 38.1 Å$^3$/(Ba,Si) at 10 GPa. The wavelength used was λ = 0.4143 Å. Right inset: possible crystal structure of $C2/m$-BaSi$_2$H$_4$ showing Ba atoms (large green spheres) and Si atoms (blue spheres) with H atoms (small pink/white spheres) coordinating the Si centers.

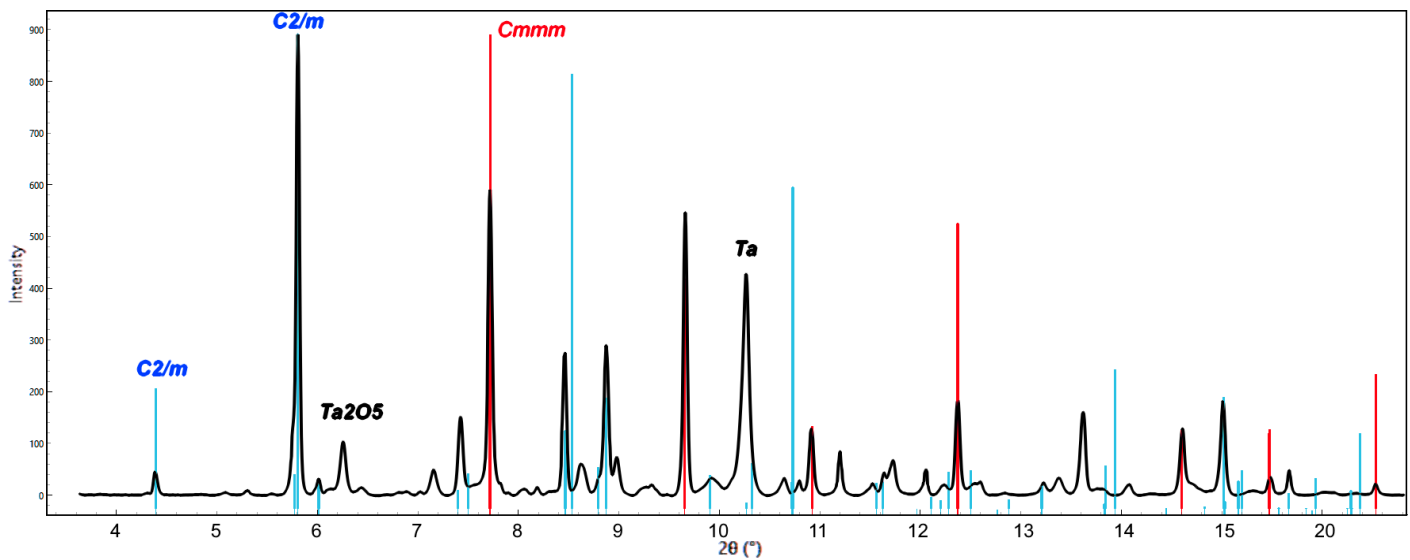

**Figure S17.** X-ray diffraction pattern of DAC BS-6 and its qualitative interpretation via $C2/m$-BaSi$_2$H$_8$ and $Cmmm$ impurity phase at 10 GPa. The integrated diffraction pattern (black line) shows multiple phases present in the sample. Vertical colored lines indicate the calculated peak positions for two structures: blue lines mark reflections from the $C2/m$-BaSi$_2$H$_4$ phase (labeled at $2\theta \approx 4.4°$ and near $5.8°$), red lines correspond to the $Cmmm$ impurity (labeled at $2\theta \approx 7.75°$). Additional peaks belong to Ta$_2$O$_5$ (at $2\theta \approx 6.2°$) and Ta metal from the lens/gasket (Ta, at $2\theta \approx 10.3°$). The wavelength used was $\lambda = 0.4143$ Å.

However, this interpretation has certain problems: DFT relaxation of $C2/m$-BaSi$_2$H$_4$ at 10 GPa leads to significantly different unit cell parameters and, accordingly, to a different volume of the compound. In turn, the $Cmmm$ prototype (BaSiH$_{13}$), although it perfectly describes the seven residual reflections (Figure S17), contains Ba and Si atoms that are too closely spaced (Supporting Table S6).

**Table S6.** Structural model of $C2/m$-BaSi$_2$H$_4$, and $Cmmm$-BaSiH$_x$ found by evolutionary search using USPEX code at 10 GPa.

| Formula (Pressure) | POSCAR |
|---|---|
| $C2/m$-BaSi$_2$H$_4$ (10 GPa) 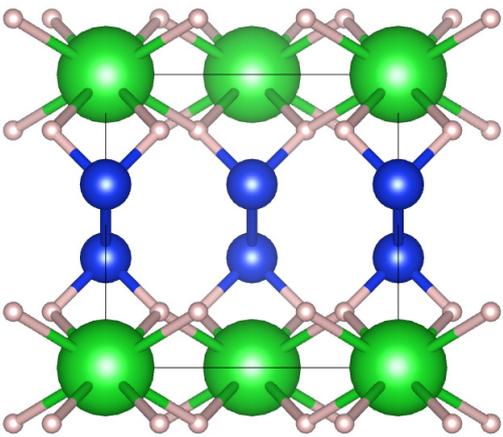 | Ba2 Si4 H8<br>1.0<br>    7.8997998238    0.0000000000    0.0000000000<br>    0.0000000000    5.3565001488    0.0000000000<br>   -4.0258771662    0.0000000000    5.4031754878<br>Ba  Si  H<br>2  4  8<br>Direct<br>   0.000000000    0.000000000    0.000000000<br>   0.500000000    0.500000000    0.000000000<br>   0.180059999    0.000000000    0.625069976<br>   0.819939971    0.000000000    0.374930024<br>   0.680060029    0.500000000    0.625069976<br>   0.319940001    0.500000000    0.374930024<br>   0.203041002    0.184052005    0.809354007<br>   0.796958983    0.815948009    0.190645993<br>   0.796958983    0.184052005    0.190645993<br>   0.203041002    0.815948009    0.809354007<br>   0.703041017    0.684051991    0.809354007<br>   0.296958983    0.315948009    0.190645993<br>   0.296958983    0.684051991    0.190645993<br>   0.703041017    0.315948009    0.809354007 |
| $Cmmm$-BaSiH$_x$ (10 GPa) | $Cmmm$-BaSi<br>1.0<br>    6.1469998360    0.0000000000    0.0000000000<br>    0.0000000000    2.6809999943    0.0000000000<br>    0.0000000000    0.0000000000    3.0850000381<br>Ba  Si<br>2  2 |

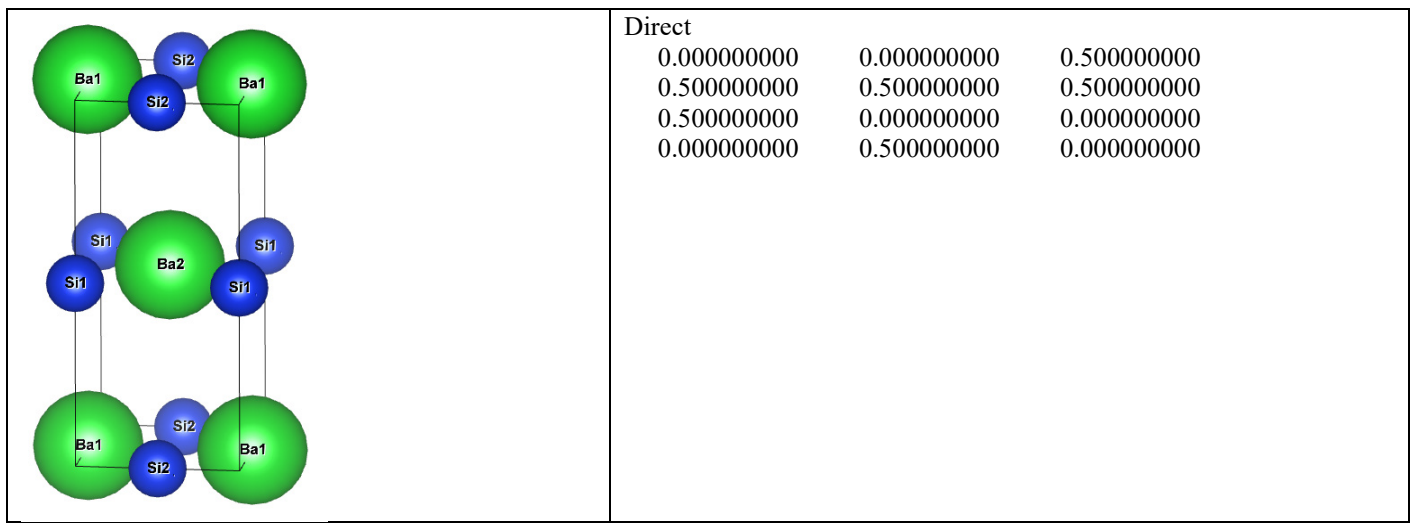

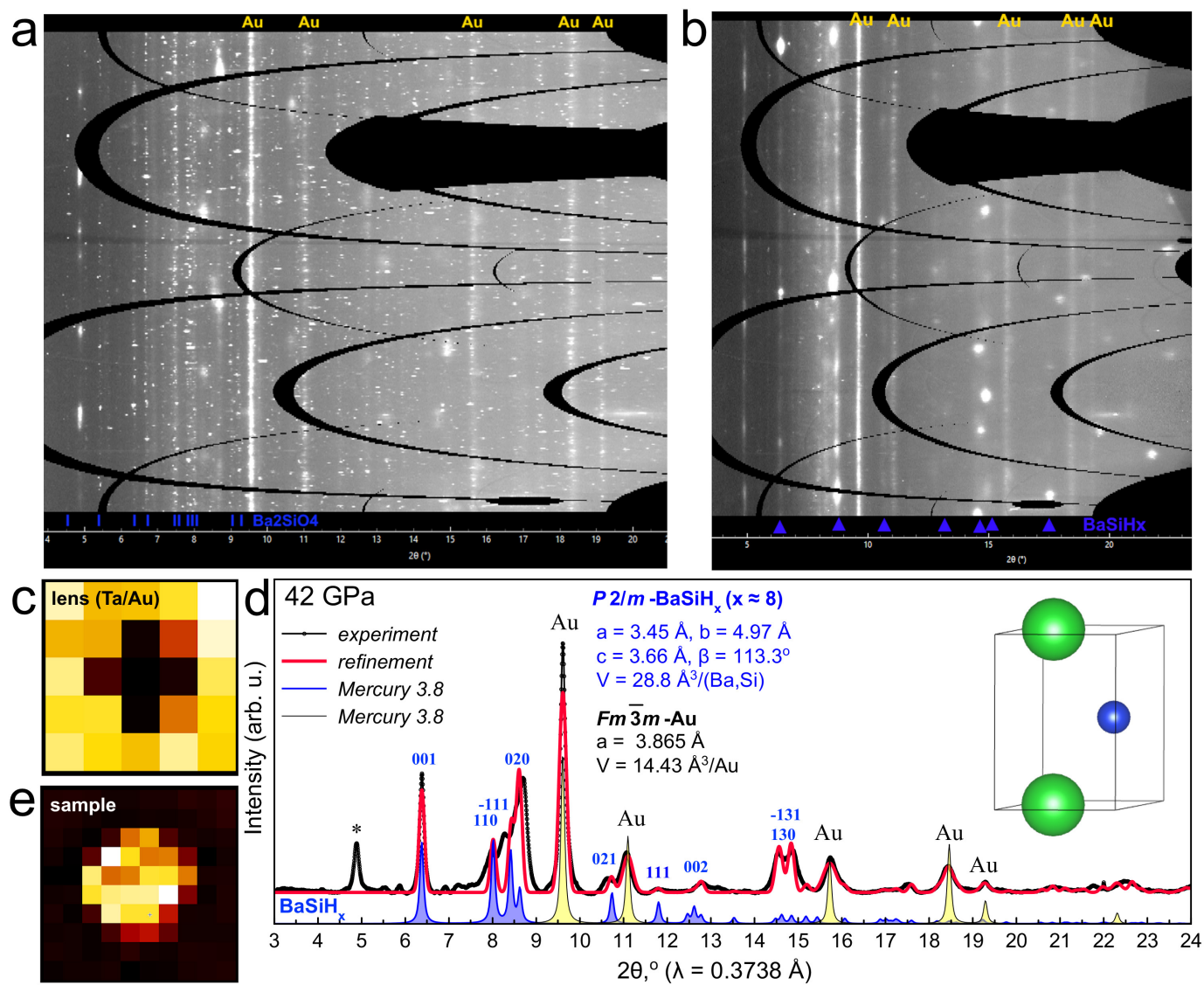

**Figure S18.** Synchrotron X-ray diffraction analysis and mapping of BaSiH$_x$ sample in diamond anvil cell BS-NMR at ~42 GPa. (a) Two-dimensional raw X-ray diffraction image of the Ba$_2$SiO$_4$-enriched region. Gold (Au) reflections from the Ta/Au lens are labeled at the top. Vertical lines indicate Ba$_2$SiO$_4$ reflections (blue tick marks at bottom). (b) Two-dimensional raw XRD image from a BaSiH$_x$-enriched region, showing similar Au reflections and BaSiH$_x$ peaks (labeled with blue arrows at bottom). (c) XRD intensity map of the gasket/sample area on a 5×5 grid (25 points, 20 μm step). The color scale ranges from yellow (high intensity from Ta highest peak) through orange and brown to black (low intensity of the Ta signal). (d) Le Bail refinement of integrated XRD pattern at 42 GPa (Au pressure scale). Experimental data (black line), refinement fit (red line), and calculated patterns from Mercury 3.8 for $P2/m$-BaSiH$_x$ (x ≈ 8, blue color)

and *fcc* Au (gray, bottom) are shown. The *P2/m*-BaSiH$_x$ phase has refined unit cell parameters: a = 3.45 Å, b = 4.97 Å, c = 3.66 Å, β = 113.3°, V = 28.8 Å$^3$/(Ba,Si). Gold reflections are labeled "Au" (a = 3.865 Å, V = 14.43 Å$^3$/Au). An asterisk (*) marks uninterpreted reflection. Inset: Crystal structure of *P2/m*-BaSiH$_x$ showing Ba atoms (large green spheres) and Si atoms (small blue sphere) in the monoclinic unit cell. Hydrogen positions are unknown. (e) Higher-resolution XRD intensity map on a 10×10 grid with 10 μm step focused on the sample region.

**Table S7.** Structural model of *P2/m*-BaSiH$_x$ (x ≈ 8) found by XRpostprocessing code[5] at 42 GPa.

| Formula (Pressure) | POSCAR |
|---|---|
| *P2/m*-BaSiH$_x$ (42 GPa) | BaSiHx<br>1.0<br>    3.4552381039    0.0000000000    0.0000000000<br>    0.0000000000    4.9706749916    0.0000000000<br>   -1.4475865111    0.0000000000    3.3597897493<br>  Ba  Si<br>  1   1<br>Direct<br>    0.265177995    0.000000000    0.448702991<br>    0.765177965    0.500000000    0.448702991 |

# IV. Single-crystal X-ray diffraction data

**Table S8.** Crystallographic data for γ-Si$_3$N$_4$ at 1 bar. The refinement was done with the Olex2 software.[1]

|  |  | γ-Si$_3$N$_4$ |  |
|---|---|---|---|
| Pressure (bar) |  | 1 |  |
| Space group, # |  | $Fm\bar{3}m$ |  |
| Z |  | 8 |  |
| a (Å) |  | 7.7777(7) |  |
| V (Å$^3$) |  | 470.49(13) |  |
|  |  |  |  |
| **Refinement details** |  |  |  |
| Wavelength (λ, Å) |  | 0.4099 |  |
| μ (mm$^{-1}$) |  | 0.860 |  |
| # measured/independent reflections (I ≥ 2σ) |  | 202 / 55 (37) |  |
| (sin θ/λ)$_{max}$ (Å$^{-1}$) |  | 1.034 |  |
| R$_{int}$ (%) |  | 15.80 |  |
| R[F$^2$ > 2σ(F$^2$)] (%) |  | 10.98 |  |
| wR(F2) (%) |  | 28.35 |  |
| S |  | 1.163 |  |
| No. of parameters |  | 6 |  |
| Δρ$_{min}$, Δρ$_{max}$ (eÅ$^{-3}$) |  | -0.747, 1.764 |  |
|  |  |  |  |
| **Atomic positions** |  |  |  |

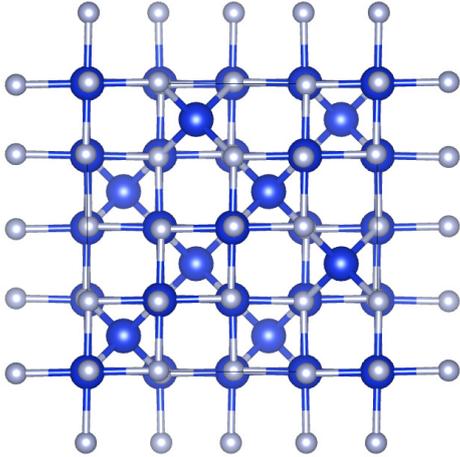

N (grey), Si (blue)

| Atom | Wyckoff position | Fractional atomic coordinates (x; y; z) | U$_{iso}$ (Å$^2$) |
|---|---|---|---|
| Si1 | 16c | 0; 0; 0 | 0.021(2) |
| Si2 | 8b | 3/8; 3/8; 3/8 | 0.023(3) |
| N1 | 32e | 0.2451(8); 0.2451(8); 0.2451(8) | 0.006(3) |

**Table S9.** Crystallographic data for Si at 1 bar. The refinement was done with the Olex2 software.[1]

|  | **Si** |
|---|---|
| Pressure (bar) | 1 |
| Space group, # | $Ia\bar{3}$ |
| Z | 16 |
| $a$ (Å) | 6.6679(19) |
| $V$ (Å$^3$) | 296.5(3) |
|  |  |
| ***Refinement details*** |  |
| Wavelength ($\lambda$, Å) | 0.4099 |
| $\mu$ (mm$^{-1}$) | 0.355 |
| # measured/independent reflections (I ≥ 2σ) | 183 / 47 (41) |
| $(\sin\theta/\lambda)_{max}$ (Å$^{-1}$) | 0.662 |
| $R_{int}$ (%) | 10.21 |
| $R[F^2 > 2\sigma(F^2)]$ (%) | 8.47 |
| $wR(F2)$ (%) | 20.25 |
| S | 1.010 |
| No. of parameters | 4 |
| $\Delta\rho_{min}, \Delta\rho_{max}$ (eÅ$^{-3}$) | -1.036, 0.732 |
|  |  |
| **Atomic positions** | 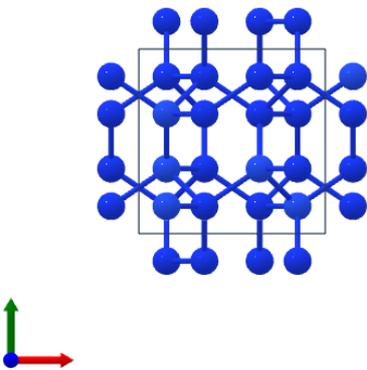 |

| Atom | Wyckoff position | Fractional atomic coordinates (x; y; z) | $U_{iso}$ (Å$^2$) |
|---|---|---|---|
| Si1 | 16c | 0.1006(4); 0.1006(4); 0.1006(4) | 0.037(2) |

**Table S10.** Crystallographic data for $Ba_2SiO_4$ at 18 GPa. The refinement was done with the Olex2 software.[1]

| | $Ba_2SiO_4$ |
|---|---|
| Pressure (GPa) | 18 |
| Space group, # | *Pnma* |
| Z | 4 |
| *a* (Å) | 7.152(5) |
| *b* (Å) | 5.52(3) |
| *c* (Å) | 9.431(2) |
| *V* (Å$^3$) | 372 ± 2 |
| | |
| ***Refinement details*** | |
| Wavelength (λ, Å) | 0.3738 |
| μ (mm$^{-1}$) | 3.151 |
| # measured/independent reflections (I ≥ 2σ) | 170 / 125 (109) |
| (sin θ/λ)$_{max}$ (Å$^{-1}$) | 0.751 |
| $R_{int}$ (%) | 1.58 |
| $R[F^2 > 2σ(F^2)]$ (%) | 7.35 |
| $wR(F2)$ (%) | 18.54 |
| S | 1.093 |
| No. of parameters | 20 |
| Δρ$_{min}$, Δρ$_{max}$ (eÅ$^{-3}$) | -1.664, 1.747 |
| | |
| **Atomic positions** | 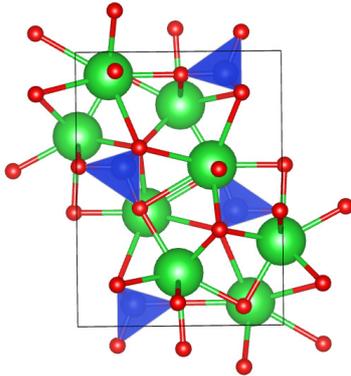<br>Ba (green), O (red), Si (blue) |

| Atom | Wyckoff position | Fractional atomic coordinates (x; y; z) | $U_{iso}$ (Å$^2$) |
|---|---|---|---|
| Ba1 | 4*c* | 0.4800(3); 1/4; 0.68604(13) | 0.0092(11) |
| Ba2 | 4*c* | 0.1529(3); 1/4; 0.40668(15) | 0.0122(12) |
| Si1 | 4*c* | 0.2336(14); 1/4; 0.0796(7) | 0.0105(17) |
| O1 | 4*c* | 0.305(4); 1/4; 0.921(2) | 0.018(6) |
| O2 | 4*c* | 0.014(4); 1/4; 0.094(3) | 0.031(8) |
| O3 | 8*d* | 0.3216(19); 0.017(8); 0.1508(12) | 0.009(3) |

# V. Theoretical calculations and the equation of state

**Table S11.** Results of a brief USPEX search (13 gen.) for the most thermodynamically stable phases in the Ba-Si-H system at 50 GPa.

| Number (EA) | Ba | Si | H | Enthalpy, eV/atom | Volume, Å$^3$/atom | Fitness, eV/block | Space group | Convex hull (X, eV/block) | Convex hull (Y, eV/block) |
|---|---|---|---|---|---|---|---|---|---|
| 38 | 1 | 1 | 26 | -1.7814 | 3.9433 | 0.0 | 1 | 0.963 | -0.1287 |
| 59 | 1 | 1 | 6 | -1.1642 | 6.6198 | 0.0 | 2 | 0.857 | -0.4734 |
| 60 | 2 | 2 | 2 | 1.022 | 12.7672 | 0.0 | 63 | 0.5 | -0.5179 |
| 500 | 0 | 0 | 8 | -2.0203 | 2.9629 | 0.0 | 7 | 1.0 | 0.0 |
| 610 | 7 | 7 | 0 | 3.061 | 18.0684 | 0.0 | 1 | -0.0 | 0.0 |
| 535 | 1 | 1 | 16 | -1.6438 | 4.5225 | 5.0E-4 | 2 | 0.941 | -0.1992 |
| 632 | 0 | 0 | 8 | -2.0193 | 2.9936 | 9.0E-4 | 1 | 1.0 | 9.0E-4 |
| 595 | 1 | 1 | 20 | -1.7115 | 4.2139 | 0.0027 | 1 | 0.952 | -0.1605 |
| 133 | 0 | 0 | 8 | -2.0172 | 3.0036 | 0.003 | 4 | 1.0 | 0.003 |
| 63 | 2 | 2 | 5 | -0.1388 | 9.2559 | 0.0066 | 8 | 0.714 | -0.4846 |
| 57 | 1 | 1 | 6 | -1.1562 | 6.6179 | 0.0091 | 166 | 0.857 | -0.4643 |
| 336 | 0 | 0 | 12 | -2.0101 | 3.0351 | 0.0101 | 1 | 1.0 | 0.0101 |
| 490 | 1 | 1 | 26 | -1.7663 | 3.9709 | 0.0156 | 1 | 0.963 | -0.1131 |
| 428 | 0 | 0 | 10 | -2.0032 | 3.0419 | 0.0171 | 1 | 1.0 | 0.0171 |
| 548 | 1 | 1 | 30 | -1.7944 | 3.7784 | 0.0174 | 1 | 0.968 | -0.0947 |
| 585 | 1 | 1 | 12 | -1.5172 | 4.9834 | 0.0186 | 2 | 0.923 | -0.24 |
| 102 | 1 | 1 | 16 | -1.6266 | 4.5688 | 0.0187 | 1 | 0.941 | -0.181 |
| 58 | 1 | 1 | 10 | -1.4342 | 5.368 | 0.0197 | 1 | 0.909 | -0.2845 |
| 122 | 12 | 12 | 0 | 3.0713 | 17.8655 | 0.0206 | 4 | -0.0 | 0.0206 |
| 656 | 1 | 1 | 20 | -1.6929 | 4.2153 | 0.0222 | 1 | 0.952 | -0.141 |
| 442 | 1 | 1 | 6 | -1.1447 | 6.6718 | 0.0223 | 2 | 0.857 | -0.4512 |
| 62 | 1 | 1 | 8 | -1.3132 | 5.7419 | 0.0264 | 2 | 0.889 | -0.3436 |
| 646 | 1 | 1 | 22 | -1.7131 | 4.1421 | 0.0283 | 1 | 0.957 | -0.1214 |
| 483 | 1 | 1 | 8 | -1.3097 | 5.8777 | 0.0303 | 1 | 0.889 | -0.3397 |
| 157 | 1 | 1 | 28 | -1.7677 | 3.9377 | 0.0306 | 1 | 0.966 | -0.0892 |
| 200 | 1 | 1 | 26 | -1.7495 | 3.9839 | 0.0331 | 1 | 0.963 | -0.0956 |
| 283 | 0 | 0 | 16 | -1.9848 | 3.0439 | 0.0354 | 1 | 1.0 | 0.0354 |
| 593 | 1 | 1 | 14 | -1.5618 | 4.6956 | 0.0367 | 1 | 0.933 | -0.1885 |

**Table S12.** List of POSCARS of the most thermodynamically stable phases in the Ba-Si-H system at 50 GPa.

```
EA38   5.530  4.203  5.705 111.50 65.43 90.16 Sym.group:   1
1.0
    5.520162   -0.318015   -0.064177
    0.231232    4.194427    0.127900
    2.294144   -2.363498    4.657623
  Ba  Si  H
   1   1  26
Direct
   0.967375   0.990502   0.029851
   0.466468   0.495141   0.029993
   0.248222   0.947582   0.520336
   0.700437   0.027776   0.532408
   0.004150   0.734094   0.538719
   0.301411   0.520584   0.607206
   0.258335   0.835108   0.207069
   0.670361   0.157407   0.857847
   0.671275   0.664680   0.861590
   0.259381   0.327460   0.199406
   0.030397   0.586395   0.225892
   0.326482   0.404121   0.834172
   0.900654   0.391413   0.838034
   0.601903   0.590997   0.230273
   0.506775   0.194013   0.315788
   0.859569   0.819195   0.637330
   0.650198   0.501333   0.457414
   0.051690   0.150182   0.438283
   0.652779   0.421016   0.554205
   0.236693   0.520869   0.514435
   0.484847   0.041137   0.212982
   0.624132   0.933994   0.453818
   0.429443   0.832755   0.732772
   0.451156   0.925491   0.860336
   0.043116   0.386536   0.820847
```

```
EA428  4.338  2.148  3.568 73.04 102.76 104.34 Sym.group:   1
1.0
    4.327276    0.059746   -0.295150
   -0.561410    2.073817    0.008352
   -0.570292    0.910118    3.402807
  H
  10
Direct
   0.257735   0.678079   0.781122
   0.441171   0.406902   0.507173
   0.403737   0.713521   0.367193
   0.592367   0.492707   0.047536
   0.016321   0.814541   0.300970
   0.211595   0.870027   0.891991
   0.056662   0.145908   0.163328
   0.693389   0.253688   0.061633
   0.857267   0.327488   0.604009
   0.782391   0.975492   0.662273
EA548  5.061  4.820  5.712 72.59 66.54 76.51 Sym.group:   1
1.0
    5.056802    0.004796    0.199885
    1.129520    4.680834   -0.211886
    2.072344    1.490684    5.109532
  Ba  Si  H
   1   1  30
Direct
   0.910580   0.066788   0.084375
   0.523012   0.415219   0.491836
   0.278382   0.656852   0.500738
   0.202371   0.120254   0.310365
   0.039201   0.916852   0.686113
   0.762468   0.158415   0.490793
   0.641469   0.552641   0.019606
```

```
  0.889978   0.596157   0.236114
  0.929776   0.290751   0.517375
  0.303830   0.068238   0.611948
EA59   4.208  3.561  3.688 73.80 87.82 91.92 Sym.group:  2
1.0
  4.207608   0.011346   0.012592
 -0.128634   3.558234   0.011762
  0.127190   1.022620   3.540763
 Ba  Si  H
  1  1  6
Direct
  0.459987   0.402038   0.182632
  0.960054   0.901944   0.682614
  0.939529   0.324334   0.456373
  0.635930   0.779532   0.560304
  0.138763   0.828537   0.348111
  0.781281   0.975418   0.017102
  0.980624   0.479523   0.908763
  0.284250   0.024369   0.804878
EA60   8.085  3.257  2.959 90.00 79.49 90.00 Sym.group:  63
1.0
  8.084555  -0.009721  -0.006472
  0.003892   3.256589   0.029502
  0.542300  -0.027000   2.908890
 Ba  Si  H
  2  2  2
Direct
  0.720811   0.546182   0.062986
  0.370760   0.046290   0.240720
  0.102673   0.546213   0.873187
  0.988892   0.046193   0.429786
  0.473143   0.546358   0.688705
  0.618459   0.046230   0.614925
EA500   2.057  3.510  3.284 90.02 91.08 89.62 Sym.group:  7
1.0
  2.054179   0.015902  -0.104649
 -0.004942   3.509878  -0.026516
  0.104827   0.024004   3.282022
 H
 8
Direct
  0.822345   0.392760   0.185653
  0.057568   0.244512   0.129359
  0.544065   0.866001   0.235638
  0.336128   0.776085   0.075758
  0.555836   0.113393   0.731052
  0.837817   0.585487   0.699807
  0.326355   0.209017   0.583566
  0.051335   0.729974   0.610317
EA610   5.840  8.983  4.954 85.16 81.50 80.42 Sym.group:  1
1.0
  5.822415   0.340936  -0.289406
  0.961461   8.926537  -0.294352
  0.947144   0.478544   4.839065
 Ba  Si
  7  7
Direct
  0.128625   0.727064   0.971714
  0.940991   0.309319   0.935532
  0.665123   0.064567   0.183262
  0.664790   0.627512   0.971960
  0.965649   0.008057   0.626073
  0.655710   0.416901   0.472400
  0.147050   0.542979   0.507547
  0.383559   0.386880   0.014795
  0.210551   0.238571   0.367798
  0.216789   0.025247   0.094942
  0.372990   0.813329   0.384287
  0.757501   0.765606   0.438519
  0.419487   0.175321   0.724852
  0.488750   0.914969   0.737729
EA535   4.711  4.038  4.863 89.77 62.30 95.59 Sym.group:  2
1.0
  4.711255   0.003972   0.019021
 -0.396632   4.017696  -0.097377
  2.242739   0.345630   4.301175
 Ba  Si  H
  1  1  16
Direct
  0.444008   0.404031   0.821356
  0.941657   0.904622   0.825303
  0.589719   0.918328   0.928976
  0.098115   0.935092   0.181658
  0.616335   0.522467   0.667480
  0.446885   0.314598   0.302241
  0.386509   0.846118   0.822462
  0.683815   0.893420   0.869687
  0.436563   0.899656   0.098043
  0.476249   0.005501   0.442534
  0.909665   0.378386   0.640163
  0.068150   0.340983   0.627906
  0.549384   0.850979   0.482867
  0.378051   0.819782   0.239044
  0.697230   0.031066   0.761714
  0.680936   0.542275   0.880616
  0.421418   0.269371   0.936599
  0.211528   0.573411   0.223076
  0.332134   0.707264   0.898565
  0.119936   0.540706   0.824783
  0.054731   0.192135   0.380316
  0.173350   0.972348   0.653793
  0.836739   0.792108   0.539715
  0.057971   0.557046   0.248141
  0.998922   0.748835   0.496677
  0.012596   0.572166   0.953588
  0.404441   0.302929   0.062398
  0.747379   0.594452   0.242341
  0.339884   0.203902   0.725081
EA585   4.650  4.454  3.723 101.40 110.36 95.32 Sym.group:  2
1.0
  4.648533   0.032146  -0.129967
 -0.447539   4.429536  -0.124755
 -1.194576  -0.763380   3.442706
 Ba  Si  H
  1  1  12
Direct
  0.669060   0.763244   0.152185
  0.169025   0.263271   0.152190
  0.340846   0.403795   0.594209
  0.872196   0.494435   0.691742
  0.177882   0.942680   0.228192
  0.481283   0.242650   0.109286
  0.160054   0.583608   0.075773
  0.162475   0.847558   0.691117
  0.997013   0.122874   0.709873
  0.175329   0.679263   0.614214
  0.465654   0.032746   0.612868
  0.640488   0.094213   0.720550
  0.856537   0.283555   0.194553
  0.497650   0.431735   0.583606
EA102   4.459  4.145  4.993 79.33 114.86 92.79 Sym.group:  1
1.0
  4.457577   0.111152   0.005572
 -0.304679   4.133864   0.015299
 -2.124176   0.753633   4.455163
 Ba  Si  H
  1  1  16
Direct
  0.228161   0.297069   0.789278
  0.745608   0.789029   0.830326
  0.463252   0.773572   0.920591
  0.409573   0.911738   0.273186
  0.367588   0.813268   0.152370
  0.558634   0.603815   0.574130
  0.787825   0.195319   0.235761
  0.371613   0.387902   0.341168
  0.940424   0.968407   0.077088
  0.186670   0.406436   0.238227
  0.092570   0.901616   0.454617
  0.025961   0.798171   0.740929
  0.269655   0.928409   0.482567
  0.722962   0.133781   0.353599
  0.753068   0.614958   0.392645
  0.849472   0.637712   0.298119
  0.842986   0.438542   0.006792
  0.634674   0.130384   0.625447
EA58   3.846  4.412  3.927 90.49 91.06 104.74 Sym.group:  1
1.0
  3.845077   0.040414  -0.055848
 -1.167574   4.254486   0.005387
 -0.015390  -0.044124   3.926690
 Ba  Si  H
  1  1  10
Direct
  0.668676   0.229535   0.369170
  0.967362   0.831322   0.910345
```

```
   0.247591    0.898883    0.191364
   0.964978    0.681654    0.575572
   0.575477    0.186416    0.252525
   0.974576    0.357120    0.371938
   0.918657    0.131225    0.072500
   0.917732    0.455244    0.273596
   0.637644    0.908111    0.450944
   0.293325    0.889453    0.720148
   0.784829    0.880903    0.468659
   0.394813    0.169168    0.360636
   0.303856    0.622849    0.396961
   0.484079    0.636890    0.287554
   0.850490    0.581220    0.034066
   0.031880    0.228920    0.615936
EA632   3.274  2.052  3.565 89.21 89.83 90.20 Sym.group:    1
1.0
   3.254534    0.354393   -0.067030
  -0.225673    2.033813    0.150817
   0.105843   -0.202705    3.557483
 H
 8
Direct
   0.126707    0.665928    0.476905
   0.662310    0.892669    0.654568
   0.153772    0.405472    0.337517
   0.601010    0.434812    0.018888
   0.101396    0.198931    0.844165
   0.181375    0.943528    0.969125
   0.692038    0.667462    0.149425
   0.606564    0.139873    0.509704
EA595   4.950  5.143  4.036 108.87 81.18 107.17 Sym.group:    1
1.0
   4.949790   -0.013195   -0.053815
  -1.506647    4.914623   -0.156247
   0.657930   -1.042270    3.843432
 Ba Si  H
  1  1 20
Direct
   0.833271    0.844684    0.754327
   0.529504    0.364086    0.079031
   0.977296    0.451912    0.331171
   0.120758    0.503603    0.403387
   0.095621    0.731286    0.163761
   0.207565    0.868762    0.212363
   0.801136    0.600901    0.150998
   0.014358    0.100789    0.322139
   0.246034    0.150578    0.044574
   0.433738    0.499958    0.457937
   0.411985    0.544594    0.935633
   0.337090    0.887682    0.796744
   0.322524    0.731597    0.694172
   0.557803    0.947673    0.347731
   0.116612    0.400455    0.871761
   0.287298    0.224335    0.544794
   0.969816    0.372332    0.812957
   0.658730    0.244800    0.704202
   0.511459    0.786252    0.285510
   0.664624    0.194573    0.209567
   0.313803    0.092440    0.558757
   0.869967    0.127334    0.381289
EA133   2.045  3.576  3.285 89.86 90.15 89.16 Sym.group:    4
1.0
   2.045072    0.000349    0.035228
   0.053183    3.575245   -0.076847
  -0.064971    0.079453    3.283527
 H
 8
Direct
   0.059524    0.983495    0.474699
   0.944759    0.518224    0.969179
   0.074489    0.654106    0.822891
   0.455816    0.021319    0.970312
   0.565594    0.157121    0.817037
   0.462662    0.344552    0.325208
   0.978888    0.849795    0.312789
   0.575016    0.490766    0.464539
EA63   4.756  3.731  4.755 89.89 80.82 89.88 Sym.group:    8
1.0
   4.755384   -0.038735   -0.054241
   0.037770    3.731019   -0.020201
   0.812322    0.026374    4.685291
 Ba Si  H
  2  2  5
   0.349914    0.304165    0.855379
   0.107983    0.956800    0.270090
   0.259398    0.577838    0.527016
   0.612968    0.738343    0.111577
   0.267110    0.596763    0.334685
   0.013450    0.508421    0.946765
   0.350152    0.927946    0.787757
   0.918825    0.156876    0.863045
   0.453705    0.475517    0.881600
   0.763381    0.742168    0.575959
EA122   9.267  8.434  5.516 90.00 95.92 90.00 Sym.group:    4
1.0
   9.266455    0.016635   -0.024975
  -0.015308    8.433890   -0.053737
  -0.554421    0.034000    5.487625
 Ba Si
 12 12
Direct
   0.858849    0.939579    0.591016
   0.132698    0.056195    0.938667
   0.645055    0.439588    0.997472
   0.371226    0.556209    0.649795
   0.915967    0.806317    0.097492
   0.967983    0.214034    0.300127
   0.814908    0.566916    0.496444
   0.315394    0.386083    0.107973
   0.587958    0.306321    0.490995
   0.535954    0.714057    0.288375
   0.689018    0.066925    0.092049
   0.188536    0.886069    0.480493
   0.424042    0.144008    0.825783
   0.414426    0.873952    0.886494
   0.403577    0.052703    0.238779
   0.552337    0.981784    0.582448
   0.079887    0.643983    0.762699
   0.089500    0.373928    0.702035
   0.100341    0.552663    0.349713
   0.951568    0.481778    0.006045
   0.261454    0.219155    0.523239
   0.637882    0.762887    0.817035
   0.242481    0.719154    0.065233
   0.866037    0.262870    0.771476
EA656   4.837  4.162  4.914 80.30 72.58 91.37 Sym.group:    1
1.0
   4.824051   -0.025333   -0.346752
  -0.072058    4.160227    0.079325
   1.802640    0.773267    4.505682
 Ba Si  H
  1  1 20
Direct
   0.702746    0.179952    0.555431
   0.527296    0.780262    0.161797
   0.145778    0.358168    0.686089
   0.734900    0.722948    0.329211
   0.324300    0.292527    0.965254
   0.436664    0.639313    0.752336
   0.990250    0.029482    0.125714
   0.337826    0.982109    0.377786
   0.143082    0.926214    0.705562
   0.027938    0.332549    0.831076
   0.107376    0.032664    0.197955
   0.025787    0.866170    0.845731
   0.446869    0.317493    0.025936
   0.344005    0.460805    0.343991
   0.720622    0.608560    0.919779
   0.970364    0.535141    0.090547
   0.725703    0.113626    0.032228
   0.933055    0.670555    0.601962
   0.101593    0.521080    0.142099
   0.408306    0.654025    0.608709
   0.345963    0.855992    0.969860
   0.025966    0.699576    0.445276
EA442   4.359  3.795  3.425 109.49 91.61 90.19 Sym.group:    2
1.0
   4.358884   -0.015536   -0.063495
  -0.005511    3.768091   -0.446907
  -0.050535   -0.754625    3.339951
 Ba Si  H
  1  1  6
Direct
   0.718423    0.039892    0.659937
   0.224641    0.532403    0.156643
   0.911002    0.638814    0.989722
```

```
Direct
  0.709452  0.814947  0.718311
  0.144578  0.816393  0.153557
  0.246656  0.344623  0.633642
  0.624617  0.344721  0.255461
  0.451068  0.328574  0.989026
  0.628793  0.922732  0.240682
  0.980560  0.328476  0.460028
  0.232066  0.922678  0.637842
  0.930565  0.319835  0.939094
EA57   4.179 4.180 4.181 61.11 118.89 118.89 Sym.group:  166
1.0
   4.179147  -0.006276   0.042149
  -2.014347   3.662944  -0.006450
  -2.053007   1.182073   3.445162
 Ba Si  H
  1  1  6
Direct
  0.725813  0.882053  0.635283
  0.225822  0.382041  0.135204
  0.120288  0.487417  0.700647
  0.791130  0.276570  0.029963
  0.660569  0.487425  0.240521
  0.331173  0.276833  0.569669
  0.331544  0.816789  0.029758
  0.120372  0.947386  0.240665
EA336  3.528 3.551 3.564 99.23 119.60 100.83 Sym.group:  1
1.0
   3.528323  -0.012498  -0.011319
  -0.654616   3.489751   0.068504
  -1.754043  -0.968712   2.946971
  H
  12
Direct
  0.634621  0.518682  0.343783
  0.593381  0.040729  0.724273
  0.011291  0.509576  0.033051
  0.434994  0.525285  0.763759
  0.461404  0.887102  0.503045
  0.677116  0.375839  0.475580
  0.260782  0.376794  0.765051
  0.015867  0.428707  0.212056
  0.132429  0.863543  0.826210
  0.182277  0.031701  0.008282
  0.721194  0.962949  0.205866
  0.958626  0.956988  0.330095
EA490  5.095 5.315 5.002 74.48 62.21 96.03 Sym.group:  1
1.0
   5.094488  -0.082958   0.022283
  -0.472772   5.292580   0.096928
   2.337940   1.476411   4.168240
 Ba Si  H
  1  1 26
Direct
  0.305839  0.448733  0.108664
  0.994696  0.913209  0.987201
  0.287504  0.104155  0.904617
  0.697155  0.738811  0.064998
  0.869497  0.629506  0.755585
  0.461930  0.093458  0.425268
  0.789662  0.032617  0.239084
  0.252521  0.784683  0.357961
  0.804713  0.470785  0.126228
  0.188647  0.771665  0.763197
  0.989206  0.120238  0.714011
  0.187093  0.509954  0.665779
  0.848585  0.593854  0.630302
  0.639609  0.145019  0.022625
  0.652089  0.901337  0.770753
  0.512456  0.098328  0.992678
  0.070541  0.992478  0.474763
  0.551910  0.558141  0.528817
  0.540872  0.912403  0.709987
  0.154897  0.358154  0.698964
  0.983432  0.705561  0.267771
  0.828247  0.337823  0.101255
  0.008849  0.985592  0.364405
  0.588306  0.702320  0.519909
  0.432032  0.196622  0.507872
  0.741050  0.331171  0.504152
  0.419130  0.792112  0.301129
  0.801175  0.265796  0.610052
  0.529333  0.423996  0.320994
  0.410561  0.820026  0.052948
  0.239802  0.271726  0.715810
  0.202743  0.790658  0.589823
  0.027922  0.241219  0.260371
EA62   4.600 3.266 4.082 89.94 91.80 69.52 Sym.group:  2
1.0
   4.599270  -0.033375   0.039666
   1.164735   3.051488   0.013731
  -0.162746   0.048064   4.078621
 Ba Si  H
  1  1  8
Direct
  0.491086  0.008643  0.400215
  0.991238  0.008441  0.900327
  0.841658  0.345625  0.631462
  0.570602  0.524270  0.899362
  0.262489  0.159143  0.893370
  0.110787  0.697864  0.610800
  0.412637  0.495231  0.900565
  0.141000  0.670903  0.168756
  0.871700  0.318723  0.189619
  0.719233  0.859239  0.907236
EA646  4.561 4.508 5.171 104.27 75.59 88.85 Sym.group:  1
1.0
   4.559949  -0.075041  -0.068788
   0.163080   4.503615  -0.105988
   1.340557  -1.210220   4.844958
 Ba Si  H
  1  1 22
Direct
  0.376701  0.234668  0.659434
  0.943591  0.704311  0.544984
  0.979567  0.655742  0.230748
  0.153965  0.101594  0.297890
  0.958409  0.356228  0.495100
  0.519744  0.087867  0.036132
  0.253508  0.952520  0.257094
  0.714692  0.827926  0.231082
  0.653720  0.990691  0.931367
  0.769783  0.496129  0.922454
  0.383013  0.756915  0.895898
  0.704900  0.294956  0.251168
  0.285933  0.702056  0.500194
  0.918037  0.743211  0.852137
  0.632531  0.456759  0.309577
  0.404907  0.415652  0.131415
  0.546321  0.810670  0.240525
  0.774073  0.323898  0.897979
  0.995814  0.027647  0.986933
  0.898966  0.050364  0.603862
  0.600572  0.705056  0.604902
  0.224844  0.734910  0.974074
  0.236500  0.456802  0.156752
  0.236419  0.184937  0.078528
EA483  4.090 3.712 4.286 115.40 90.37 88.89 Sym.group:  1
1.0
   4.069275  -0.342843  -0.234554
   0.373169   3.688748  -0.174025
   0.058750  -1.669726   3.947458
 Ba Si  H
  1  1  8
Direct
  0.265612  0.022757  0.324959
  0.701247  0.767646  0.730357
  0.914338  0.442660  0.098907
  0.053032  0.631703  0.683174
  0.517432  0.489386  0.849261
  0.783211  0.084388  0.604655
  0.386208  0.027513  0.826298
  0.611429  0.516836  0.346513
  0.063539  0.375132  0.978996
  0.768379  0.006149  0.109367
```

*Equation of state and superconductivity of cubic BaSiH$_8$*

The unit-cell calculations are done in the $Fm\bar{3}m$-BaSiH$_8$ phase (space group 225) with lattice parameter $a$. Si occupies the Wyckoff position 4a (0, 0, 0), Ba the 4b (1/2, 1/2, 1/2), and the H atoms the 32f ($x, x, x$). The lattice parameter $a$ and the Wyckoff parameter $x$ are shown in Figure S19 over the pressure range from 0 GPa to 100 GPa.

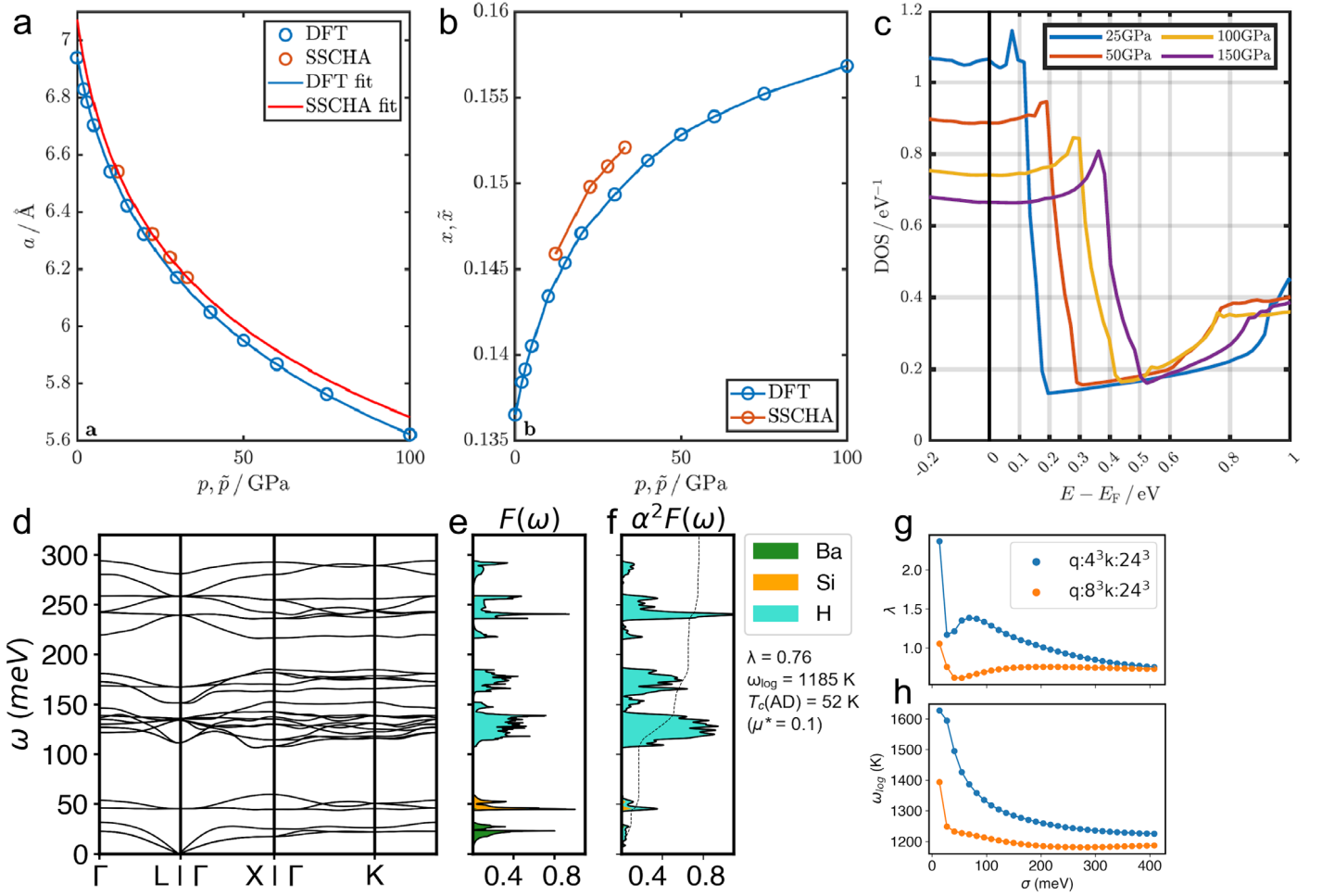

**Figure S19.** Structural parameters, phonon and superconducting properties for $Fm\bar{3}m$-BaSiH$_8$ within harmonic DFT and anharmonic SSCHA: (a) Lattice parameter (circles) and equation-of-state fit (solid line) over pressure calculated within harmonic DFT (blue) and SSCHA (red). See also Ref. [14] for details of calculations. (b) Hydrogen Wyckoff parameter ($x$). (c) Stepwise density of electron states near the Fermi level in $Fm\bar{3}m$-BaSiH$_8$ at different pressures. Relatively small doping of BaSiH$_8$ can lead to a sharp drop in the density of electron states by several times. (d) Phonon band structure at 142 GPa in the harmonic approximation. (d) Corresponding phonon density of states and (e) the Eliashberg function at 142 GPa. (g, h) Dependence of the electron-phonon interaction parameter ($\lambda$) and the logarithmically averaged phonon frequency ($\omega_{\log}$) on the Gaussian broadening ($\sigma$) in the framework of calculations in Quantum Espresso for two phonon q-grids: 4×4×4 and 8×8×8.

We fitted the Birch–Murnaghan equation of state

$$p(V) = (3B_0)/2\left[(V_0/V)^{(7/3)} - (V_0/V)^{(5/3)}\right]1 + 3/4(B_0' - 4)\left[(V_0/V)^{(2/3)} - 1\right], \quad (S5)$$

for the *fcc* unit-cell volumes $V(p)$ and the lattice parameters $a(p)$ (related via $V = a^3/4$) shown in Figure S19, and obtained the following fit parameters: $B_{0;DFT} = 39.4$ GPa, $B'_{0;DFT} = 4.3$, ($V_{0;DFT} = 83.6$ Å$^3$ fixed from DFT), and $B_{0;SSCHA} = 29.2$ GPa, $B'_{0;SSCHA} = 5.0$, $V_{0;SSCHA} = 88.3$ Å$^3$. Here $B_0$ is the bulk modulus, and $B_0'$ its the derivative with respect to pressure.

# VI. Raman spectroscopy

A systematic study of Raman scattering of $BaSiH_x$ samples is beyond the scope of this work. In this section, we only aim to show qualitatively what kind of Raman spectra are detected from typical BaSi/AB and $BaSiH_x$ samples before and after laser heating. Since BaSi, and $BaSiH_x$ are bad metals or semiconductors, their Raman signal is rather weak and cannot be confidently separated from other Ba silicides, and Ba-Si oxides (e.g., $Ba_2SiO_4$). Raman spectra of the samples were recorded using 532 nm excitation laser of the Renishaw Raman inVia system with 60-120 s exposition time at 1-2 mW of power, and focusing lens with x20 magnification. Raman spectra are available on the github: *https://github.com/mark6871/BaSiHx-project/tree/Raman*

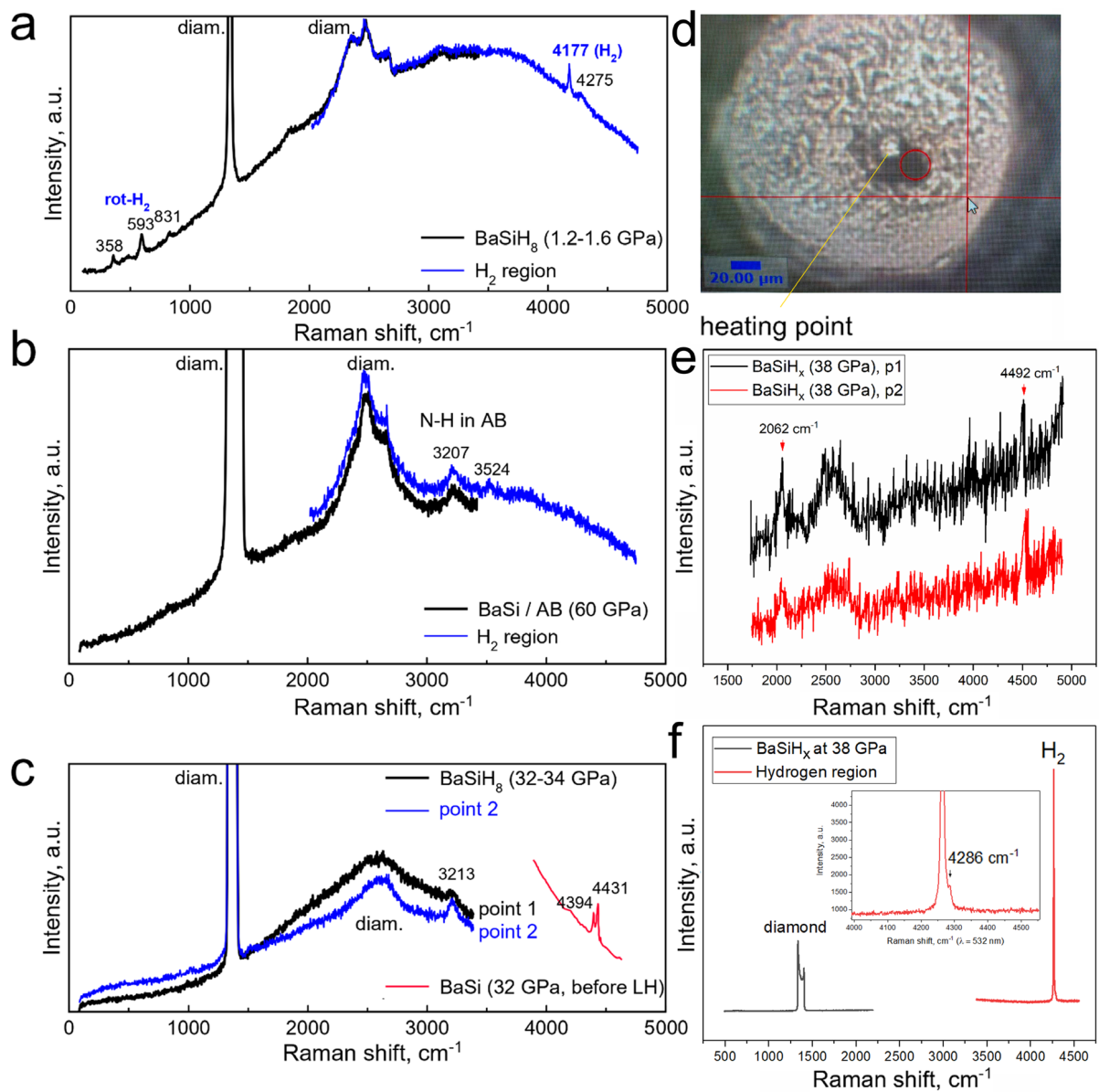

**Figure S20.** Raman spectra of different BaSi/AB and $BaSiH_x$ samples at various pressures. (a) Raman spectra of $BaSiH_x$ after laser heating and decompression to about 1-1.5 GPa. The excess hydrogen is observed with clearly detectable $H_2$ vibrons and rotons. There is also an additional peak of $H_2$ at 4275 cm$^{-1}$. (b) Raman spectra of BaSi/AB sample before heating at 60 GPa. There are signals in the region of N-H vibrations, possibly related to $NH_3BH_3$. (c) Raman spectra of BaSiHx sample after laser heating at 32-34 GPa. In this DAC, we observed a double signal around 4394-4431 cm$^{-1}$ before laser heating. After laser heating this signal is no longer observed, but there is a signal at 3213 cm$^{-1}$. (d) Microscopy of the BaSi/AB sample after laser heating at 38 GPa (diamond pressure scale[15]). The studied region is shown by red circle, and is a bit away of the laser heating spot. (e) Raman spectrum of the sample after laser heating in two points: p1 and p2. There are peaks at 4492 cm$^{-1}$, broad signal at 2600-2750 cm$^{-1}$, and around 2062 cm$^{-1}$. (f) Raman spectrum of the sample at 38 GPa (diamond scale). There is a large peak from molecular hydrogen, which, however, corresponds to a much lower pressure (24 GPa, $H_2$-scale). There is also a small shoulder at the hydrogen peak at 4286 cm$^{-1}$.

S28

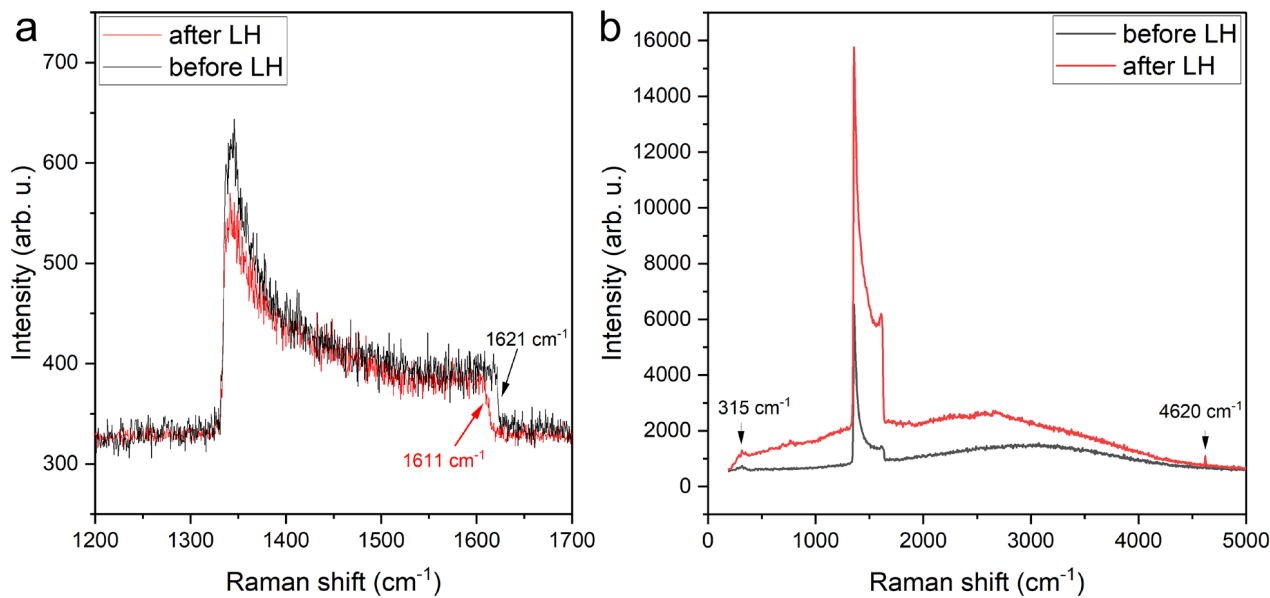

**Figure S21.** Raman spectroscopy of the DAC BS-3 sample before and after laser heating (LH). (a) Diamond Raman edge (1200-1700 cm$^{-1}$) showing the change of pressure after laser heating. (b) Full-range Raman spectra comparing the sample before and after laser heating. The spectrum shows possible weak Raman peaks only at 315 cm$^{-1}$ and near 4620 cm$^{-1}$ probably due to metallization of the sample.

# VII. Additional transport measurements

The BS-5 DAC with a BaSi/AB sample was laser heated at 55 GPa. in this DAC we studied the temperature dependence of electrical resistance of the BaSi sample before laser heating and found typical metallic behavior with $dR/dT > 0$ upon cooling from 300 to 78 K at 55 GPa (Figure S22). After laser heating, the DAC's pressure dropped significantly to 21 GPa; however, the $R(T)$ remained essentially unchanged ($dR/dT > 0$), and we found no signs of high-temperature superconductivity above 78 K.

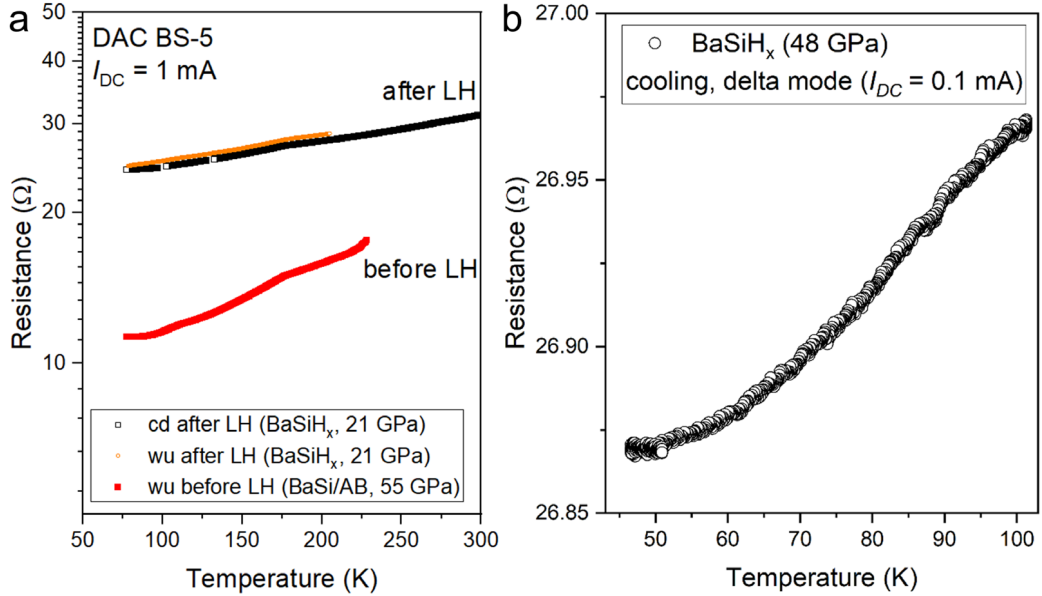

**Figure S22.** Electrical resistance measurements in DAC BS-4 and 5. (a) Electrical resistance of the BaSi/AB sample before and after laser heating at 55 GPa in heating (wu) and cooling (cd) cycles in DAC BS-5. (b) Temperature dependence of the electrical resistance of the DAC BS-4 sample at 48 GPa, measured in a cooling cycle to 45 K. It can be seen that $dR/dT > 0$ as in normal metals. The resistance of BS-4 decreases monotonically with increasing pressure from 30 GPa to 48 GPa.

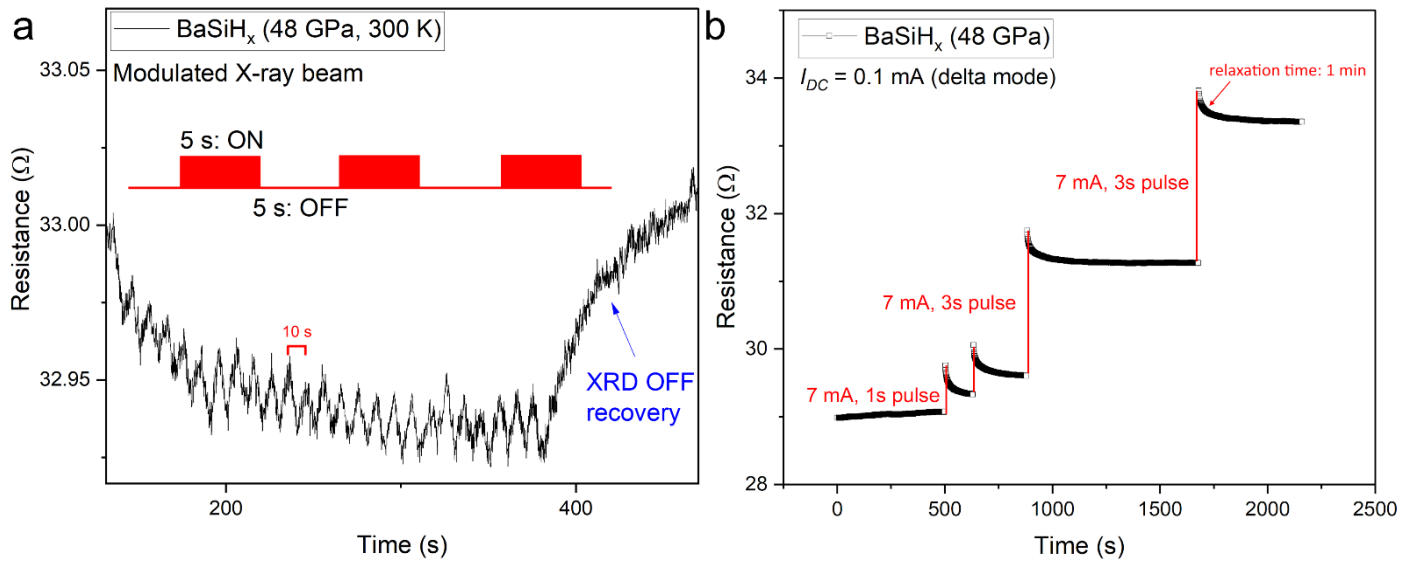

**Figure S23.** (a) Modulation of the electrical resistance of a $BaSiH_x$ sample in DAC BS-4 caused by periodically switching on and off a 25 keV X-ray beam (about $10^{10}$ photons/s) with a period of 10 seconds. Due to the sample's photoconductivity, a periodic sawtooth response is observed in the sample's electrical resistance. (b) A sudden and irreversible increase in the resistance of the $BaSiH_x$ sample when passing 7 mA DC current pulses through the sample for 1-3 s. The exponential relaxation part of $R(t)$ ($\tau \sim 60$ sec) likely corresponds to the relaxation of the ohmic heat released during the passage of electric current through the sample.

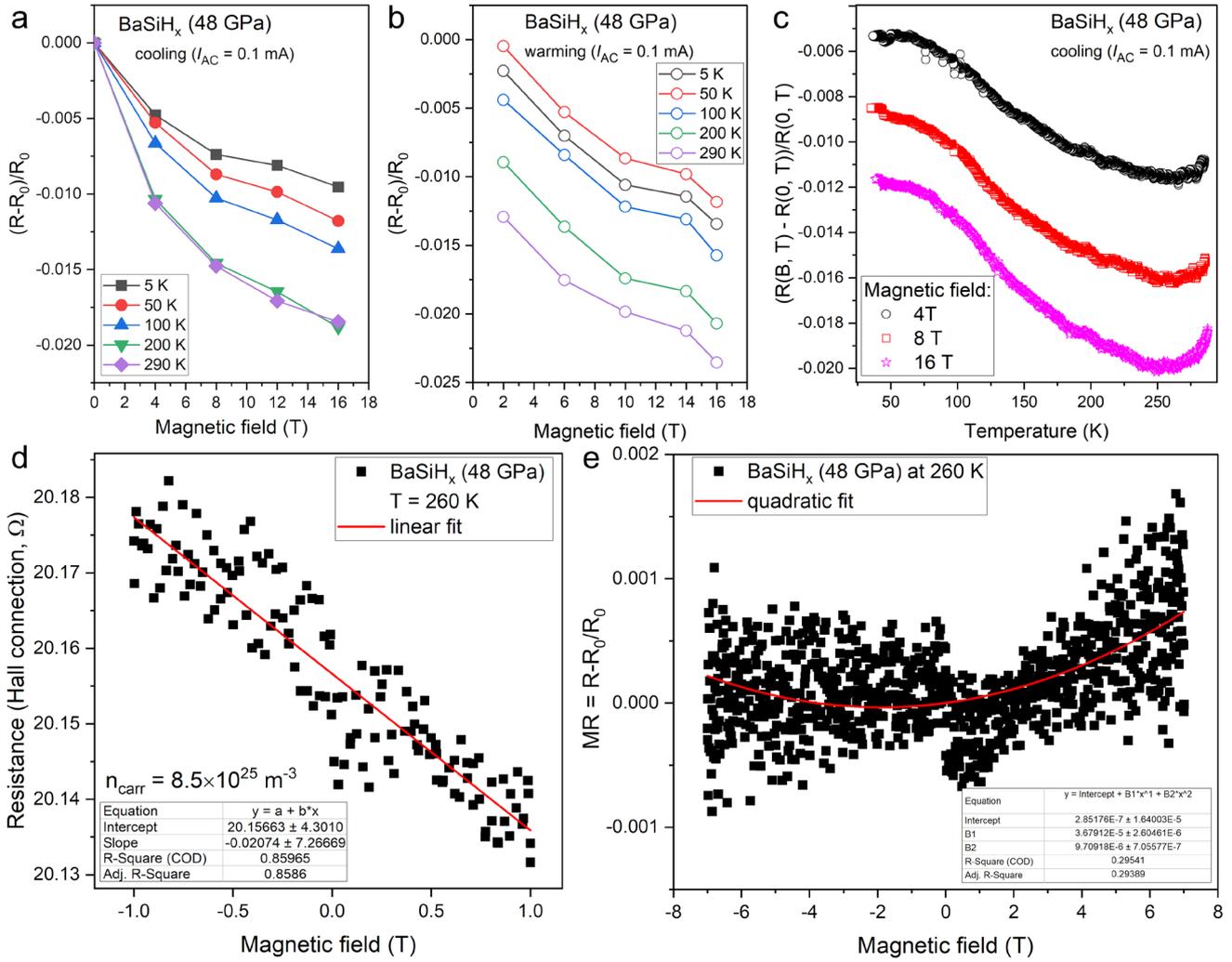

**Figure S24.** Negative magnetoresistance of BaSiH$_x$ at ≈40 GPa in DAC BS-4. (a) Negative magnetoresistance during cooling cycle measured with AC current $I_{AC}$ = 0.1 mA. The relative resistance change (R–R$_0$)/R$_0$ is plotted as a function of magnetic field (0-16 T) at five different temperatures: 5 K, 50 K, 100 K, 200 K, and 290 K. The negative MR effect is strongest at the highest temperature (290 K) with approximately –2% change at 16 T, and systematically weakens with decreasing temperature, becoming minimal at 5 K. (b) Negative magnetoresistance during warming cycles at the same temperatures and measurement conditions. (c) Temperature dependence of magnetoresistance at three fixed magnetic fields (4 T, 8 T, and 16 T) during cooling. The data shows that *(R(B,T) – R(0,T))/R(0,T)* becomes more negative with increasing temperature, with the effect being most pronounced around ~250 K. (d) Hall resistance as a function of magnetic field. The linear fit (red line) yields a calculated carrier density of $n_e$ = 8.5×10$^{25}$m$^{-3}$ assuming only one type of carriers (electrons). (e) Normalized magnetoresistance (MR = (R–R$_0$)/R$_0$) measured up to 7 T at 260 K in DAC BS-4. Between panels (a-c) and (e) about 5 months passed. Magnetoresistance appeared to be very small, but positive, which indicates dynamic processes occurring in the sample. The data is approximated by a quadratic fit (red curve).

Estimation of carrier relaxation time (τ) using the Drude formula for conductivity

$$\tau_{min} = \frac{m^*}{n_e^{max} e^2 \rho},$$

and maximum carrier concentration ($n_e^{max}$), obtained from the Hall effect assuming the presence of only one type of carrier (electrons) with mass $m^* = m_e$, gives a very short relaxation time $\tau_{min}$ = 7.54×10$^{-16}$ s = 0.75 fs. At the same time, the free electron model allows us to estimate the maximum Fermi wavenumber from above $k_F^{max} = (3\pi^2 n_e^{max})^{1/3} = 1.36 \times 10^9 \ m^{-1}$, and the maximum Fermi velocity $V_F^{max} = \hbar k_F^{max}/m_e = 1.57 \times 10^5 \ m/s$ and the Fermi energy $E_F^{max} = 70 \ meV$. This is 2-3 times smaller than the typical Fermi velocity in superhydrides (2.5–3.5×10$^5$ m/s) [16]. The resulting Fermi energy is close in magnitude to ω$_{ph}$ in polyhydrides (~1000-2000 K) that makes the Migdal parameter ω$_{ph}$/$E_F$ ~ 1 close to unity and excludes the application of the Migdal-Eliashberg theory to the evaluation of superconducting properties.

The obtained estimations of the relaxation time and Fermi velocity lead to the mean free path of electrons $l_e = V_F^{max}\tau_{min} = 1.18$ Å. This unrealistically small value, close to the interatomic distance in a hydrogen molecule, shows that both electrons and holes participate in the Hall effect.

More realistic estimates can be made based on a small field-dependent magnetoresistance in the DAC BS-4 sample, studied five months after (Feb. 2026, Figure S24e) the discovery of negative magnetoresistance (Sept. 2025). The resulting mobility value $\Delta R/R \propto (\mu B)^2$ is $\mu = 3.11 \times 10^{-3}$ m²/(V·s). From the Drude model, we then obtain $\tau = \frac{\mu m_e}{e} = 1.77 \times 10^{-14}$ s $= 17.7$ fs, the carrier concentration $n_e = 3.6 \times 10^{24}$ m⁻³, the Fermi energy $E_F = 8.6$ meV (close to the band gap value $E_g \sim 1$ meV obtained from resistive measurements), the Fermi velocity $V_F = 5.5 \times 10^4$ m/s, and mean free path $l_e = 9.7$ Å.

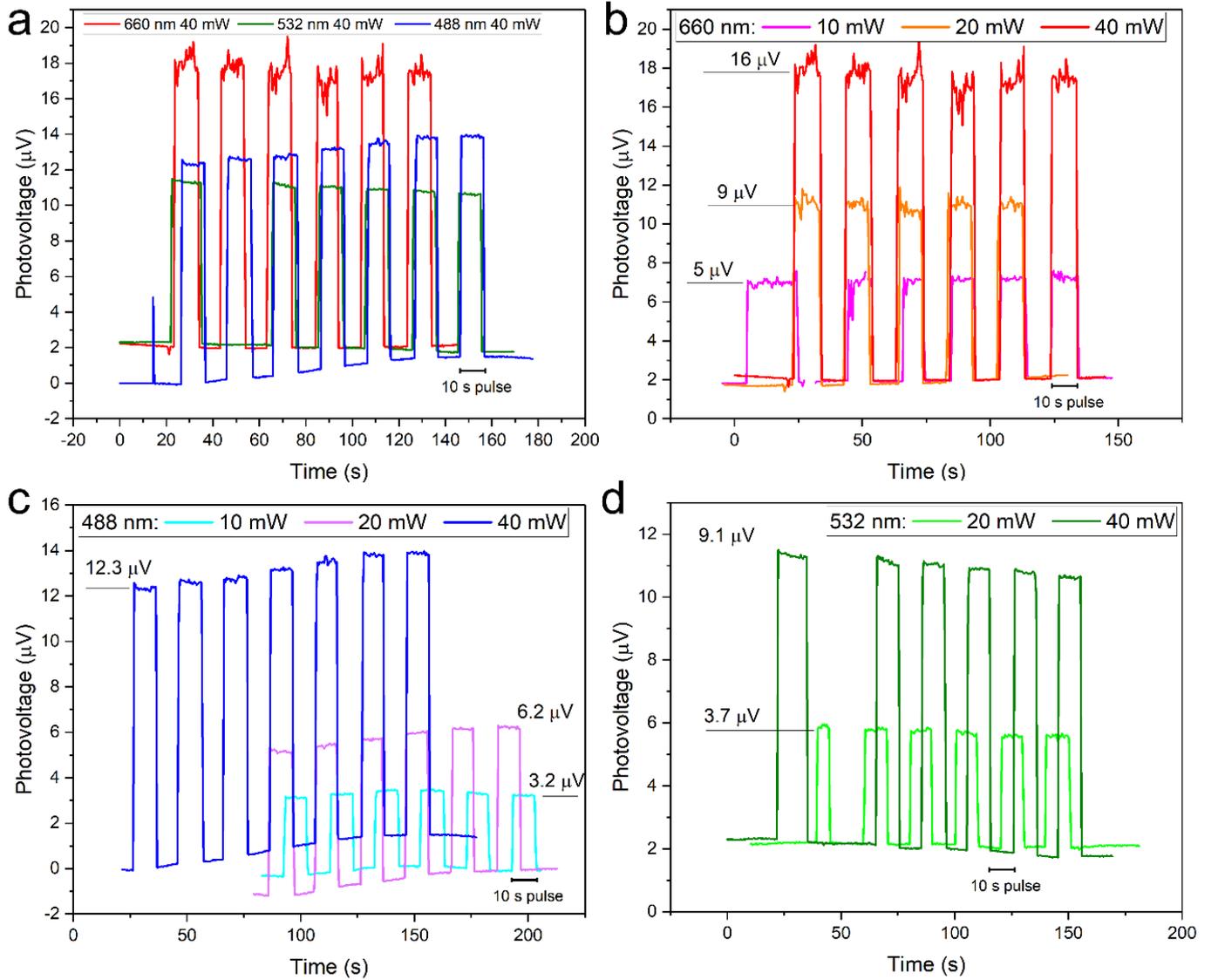

**Figure S25**. Photovoltaic response of BaSiH$_x$ at 48 GPa under illumination at different laser wavelengths and power levels. (a) Photovoltage response to pulsed illumination (10 s Π-pulses) at three different wavelengths with constant 40 mW power: 660 nm (red), 532 nm (green), and 488 nm (blue). The photovoltage non-monotonously depends on the wavelength. All three wavelengths show reproducible square-wave response over multiple on-off cycles, demonstrating reversible photovoltaic effect. (b) Wavelength-dependent response at 660 nm for three different laser powers: 10 mW (magenta), 20 mW (orange), and 40 mW (red). The photovoltage linearly scales with power, showing values of approximately 5, 9, and 16 μV for 10, 20, and 40 mW, respectively. (c) Power-dependent photovoltage at 488 nm wavelength for 10 (cyan), 20 (pink), and 40 mW (blue) illumination. The photovoltage increases linearly with power: 3.2 (10 mW), 6.2 (20 mW), and 12.3 μV (40 mW). (d) Photovoltage response at 532 nm for two power levels: 20 (light green) and 40 mW (dark green). The response shows 3.7 and 9.1 μV for 20 and 40 mW, respectively.

# VIII. Nuclear magnetic resonance

The DAC BS-NMR with a BaSi/AB sample at 28 GPa, equipped with two Lenz lenses [17] (Supporting Figure S26) and two trimmer capacitors (10 nF), was heated using a pulsed infrared laser (Nd:YAG, 1 μm). After the chemical reaction, the BaSiH$_x$ sample was examined using nuclear magnetic resonance spectroscopy in a magnetic field of 7.25 T (the resonance frequency of $^1$H nuclei was 308.8 MHz) at two temperature points: 150 and 100 K. Failure of the DACs electrical circuit at low temperature and loss of resonance prevented the NMR study below 100 K. The spin-echo detection mode was used, in which a 90º pulse had a duration of 2 μs, optimal RF power is 28 % out of 250 W, and optimal delay time ($d_1$) before 180º pulse was about 100 ms. Nutation 2D scan of 90º-pulse duration from 1 μs to 8 μs with a step of 0.5 μs, and pulse power from 10 to 30 % with a step of 5 % indicates that 90º spin-echo conditions are achieved at 6.5 μs pulse duration and 25 % of power. However, considering the total time required for experiment, we mostly used shorter pulses of 2 μs with higher power of 28 %.

As a result, $^1$H NMR spectra of BaSiH$_x$ were obtained (Supporting Figure S27). As can be seen, the main signal from the sample has a width of 44–60 kHz (142–194 ppm) and corresponds to a chemical shift of –39-43 ppm. The NMR signal consists of two components, where the smaller signal (a shoulder) corresponds to the chemical shift of about + 50 ppm. Moreover, a $d_1$-scan at 150 K in the range of $d_1$ from 100 to 2000 ms indicates the presence of another peak at around – 400 ppm (Figure S27d) with a significantly longer spin-lattice relaxation time ($T_1$) than it was measured for the main reaction product.

A later X-ray diffraction study at ESRF ID27 (in 2023) revealed an increase in the DAC's pressure to 42 GPa (according to the equation of state for gold). The complex diffraction pattern was unresolved, revealing Ba$_2$SiO$_4$ and, possibly, *P*2/*m*-BaSiH$_8$ (Supporting Figure S18), along with other uninterpreted phases.

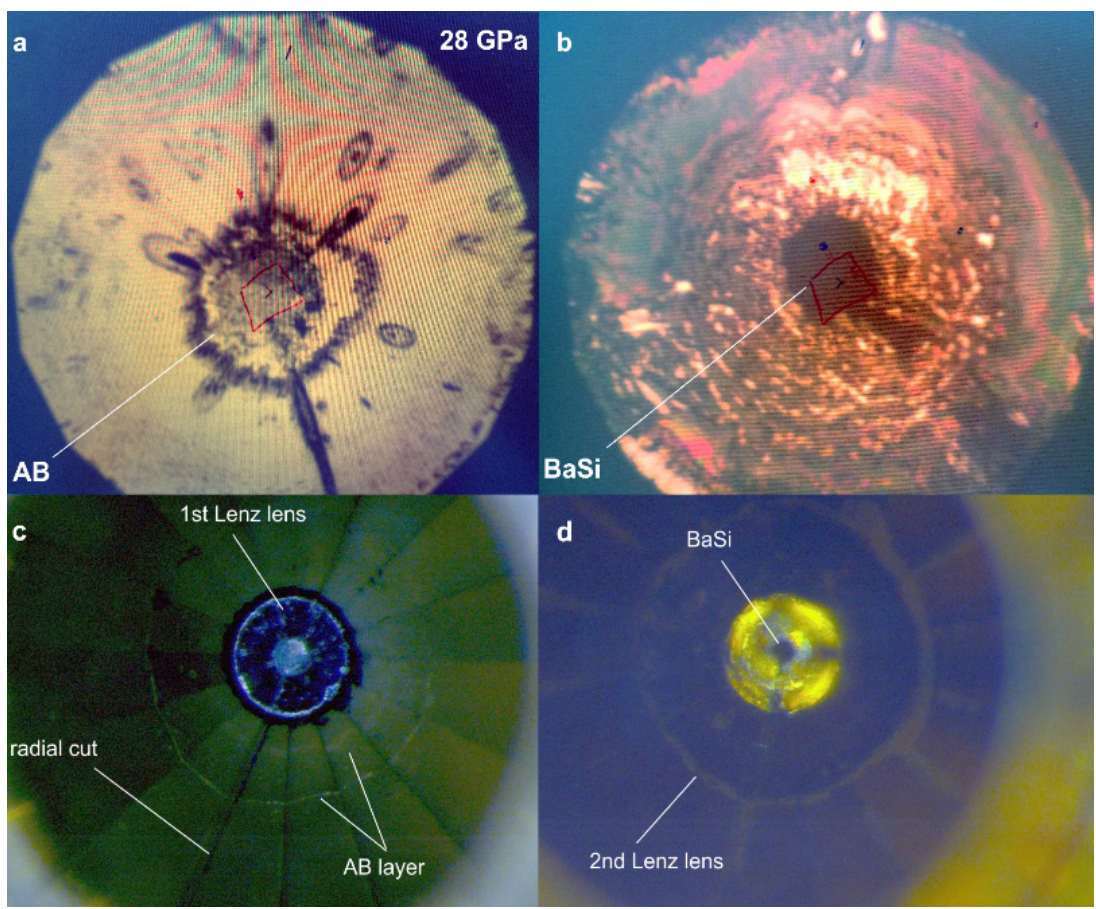

**Figure S26.** Optical photographs of a diamond cell DAC BS-NMR culet with a BaSi/AB sample under a pressure of 28 GPa with different magnifications and illumination way (from piston/cylinder/reflected or transmitted lights). The sample is opaque to visible light. The Lenz lens is made of copper and marked ("radial cut", "2$^{nd}$ Lenz lens"). Gasket material is tungsten. Insulating layer is Ta$_2$O$_5$ obtained by oxidation of thin layer of Ta sputtered over W.

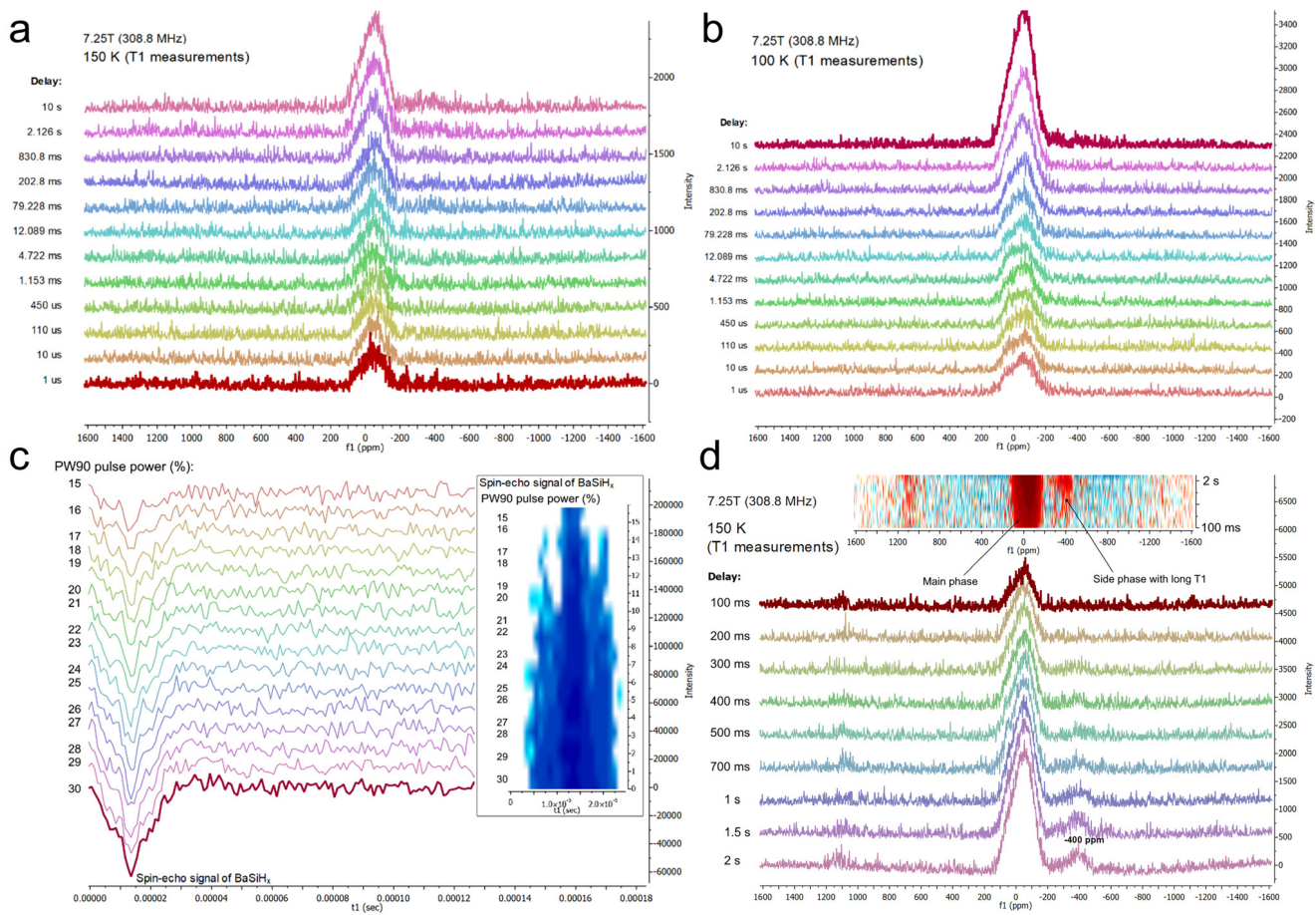

**Figure S27.** $^1$H NMR (308.8 MHz, 7.25 T) study of BaSiH$_x$ sample at 28 GPa in DAC BS-NMR. (a) Inversion recovery experiment performed in order to determine the spin-lattice relaxation time $T_1$ at 150 K. Delays (d$_1$) from 1 μs to 10 s were used. (b) The same for 100 K. (c) Spin-echo nutation experiment with using different 90° pulse power (pw90) from 15 to 30 % (where 100 % is 250 W) at the constant RF pulse duration (2 μs). Sample is BaSiH$_x$ at 150 K. Inset: colormap of intensity of the spin-echo signal. (d) The inversion recovery experiment at 150 K with a larger number of accumulation cycles. Inset: intensity colormap which indicates appearance of an additional signal with longer $T_1$.

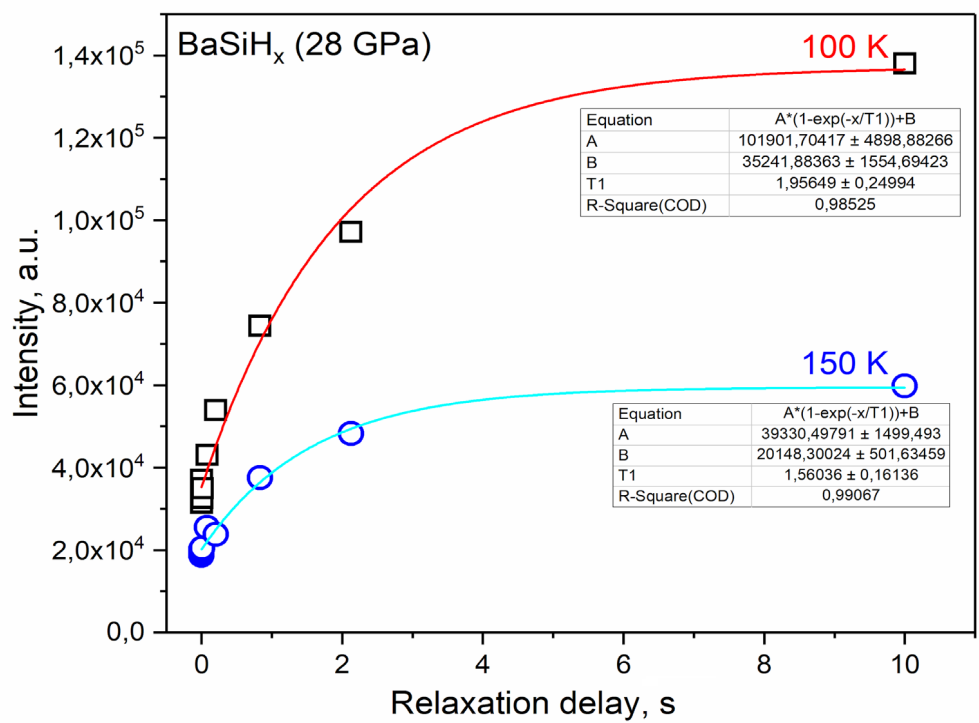

**Figure S28.** Interpolation of the integrated NMR signal intensity of the main signal depending on the delay (d$_1$) carried out using an exponential decay function at 100 and 150 K.

Thus, the sample contains several phases, which is confirmed by the complex nature of the X-ray diffraction spectra (Supporting Figure S18) and by the NMR as well. Estimations of the spin-lattice relaxation time $T_1$ for the main phase show that the $T_1$ increases from 1.5 to 2 seconds when the sample is cooled from 150 to 100 K (Figure S28). It should be noted that since full saturation of the signal was not reached even at the relaxation delay 10 s, we cannot be completely sure of the $T_1$ values. Such long relaxation times also serve as an additional argument in favor of the (degenerated) semiconducting character of the obtained barium-silicon hydrides. The main conclusion is that there is a pronounced broad hydrogen signal, typical of solid-state NMR, with spin-lattice relaxation time $T_1$ significantly lower than in the $NH_3BH_3$, but significantly higher than it should be for the metallic hydrogen sublattice.